\documentclass[epj,final]{svjour}

\usepackage{latexsym}
\usepackage{graphics}

\usepackage[utf8]{inputenc}
\usepackage[english]{babel}
\usepackage[square,numbers]{natbib} 
\usepackage[nottoc]{tocbibind}

\usepackage{amsmath}
\usepackage{amssymb}
\usepackage{graphicx}
\usepackage{bbm}
\usepackage{caption}
\usepackage{subcaption}
\usepackage{epsfig}
\usepackage{braket}
\usepackage{slashed}
\usepackage{bm}
\usepackage[usenames,dvipsnames]{color}
\usepackage{color}
\usepackage{float}
\usepackage{hyperref}
\usepackage{calligra}

\usepackage[abs]{overpic}

\begin{document}

\title{Simulating Lattice Gauge Theories within Quantum Technologies}

\author{M.C. Ba{\~n}uls\inst{1,2} \and R. Blatt\inst{3,4} \and J. Catani\inst{5,6,7} \and A. Celi\inst{3,8} \and J.I. Cirac\inst{1,2} \and M. Dalmonte\inst{9,10} \and L. Fallani\inst{5,6,7} \and K. Jansen\inst{11} \and M. Lewenstein\inst{8,12,13} \and S. Montangero\inst{7,14} \thanks{simone.montangero@unipd.it} \and C.A. Muschik\inst{3} \and B. Reznik\inst{15} \and E. Rico\inst{16,17} \thanks{enrique.rico.ortega@gmail.com} \and L. Tagliacozzo\inst{18} \and K. Van Acoleyen\inst{19} \and F. Verstraete\inst{19,20} \and U.-J. Wiese\inst{21} \and M. Wingate\inst{22} \and J. Zakrzewski\inst{23,24} \and P. Zoller\inst{3}}                     

\institute{Max-Planck-Institut f{\"u}r Quantenoptik, Hans-Kopfermann-Stra{\ss}e 1, 85748 Garching, Germany \and Munich Center for Quantum Science and Technology (MCQST), Schellingstr. 4, 80799 Muenchen, Germany \and Institut f{\"u}r Quantenoptik und Quanteninformation, {\"O}sterreichische Akademie der Wissenschaften, Technikerstra{\ss}e  21a, 6020 Innsbruck, Austria \and Institut f{\"u}r Experimentalphysik, Universit{\"a}t Innsbruck, Technikerstra{\ss}e  25, 6020 Innsbruck, Austria \and LENS and Dip. di Fisica e Astronomia, Universit{\`a} di Firenze, 50019 Sesto Fiorentino, Italy \and CNR-INO, S.S. Sesto Fiorentino, 50019 Sesto Fiorentino, Italy \and INFN Istituto Nazionale di Fisica Nucleare, I-50019 Sesto Fiorentino, Italy \and Departament de Fisica, Universitat Autonoma de Barcelona, E-08193 Bellaterra, Spain \and SISSA, Via Bonomea 265, I-34136 Trieste, Italy \and Abdus Salam ICTP, Strada Costiera 11, I-34151 Trieste, Italy \and NIC, DESY, Platanenallee 6, D-15738 Zeuthen, Germany \and ICFO - Institut de Ci{\`e}ncies Fot{\`o}niques, The Barcelona Institute of Science and Technology, 08860 Castelldefels (Barcelona), Spain \and ICREA, Pg.  Llu{\'i}s Companys 23, 08010 Barcelona, Spain \and Dipartimento di Fisica e Astronomia G. Galilei, Universit{\`a} degli Studi di Padova, I-35131 Padova, Italy \and School of Physics and Astronomy, Raymond and Beverly Sackler Faculty of Exact Sciences, Tel Aviv University, Tel-Aviv 69978, Israel \and Department of Physical Chemistry, University of the Basque Country UPV/EHU, Apartado 644, 48080 Bilbao, Spain \and IKERBASQUE, Basque Foundation for Science, Maria Diaz de Haro 3, E-48013 Bilbao, Spain \and Departament de F{\'i}sica Qu{\`a}ntica i Astrof{\'i}sica and Institut de Ci{\`e}ncies del Cosmos (ICCUB), Universitat  de  Barcelona,  Mart{\'i} i Franqu{\`e}s 1, 08028 Barcelona, Spain \and Department of Physics and Astronomy, Ghent University, Krijgslaan 281, S9, 9000 Gent, Belgium \and Vienna Center for Quantum Science and Technology, Faculty of Physics, University of Vienna, Boltzmannga{\ss}e 5, 1090 Vienna, Austria \and Albert Einstein Center for Fundamental Physics, Institute for Theoretical Physics, University of Bern, Sidlerstra{\ss}e 5, CH-3012 Bern, Switzerland \and Department of Applied Mathematics and Theoretical Physics, University of Cambridge, Cambridge CB3 0WA, UK \and Instytut Fizyki imienia Mariana Smoluchowskiego, Uniwersytet Jagiellonski, Lojasiewicza 11, 30-348 Krakow, Poland \and Mark Kac Complex Systems Research Center, Jagiellonian University, Lojasiewicza 11, 30-348 Krakow, Poland}

\date{\today}

\abstract{Lattice gauge theories, which originated from particle physics in the context of Quantum Chromodynamics (QCD), provide an important intellectual stimulus to further develop quantum information technologies. While one long-term goal is the reliable quantum simulation of currently intractable aspects of QCD itself, lattice gauge theories also play an important role in condensed matter physics and in quantum information science. In this way, lattice gauge theories provide both motivation and a framework for interdisciplinary research towards the development of special purpose digital and analog quantum simulators, and ultimately of scalable universal quantum computers.
In this manuscript, recent results and new tools from a quantum science approach to study lattice gauge theories are reviewed. Two new complementary approaches are discussed: first, tensor network methods are presented - a classical simulation approach - applied to the study of lattice gauge theories together with some results on Abelian and non-Abelian lattice gauge theories. Then, recent proposals for the implementation of lattice gauge theory quantum simulators in different quantum hardware are reported, e.g., trapped ions, Rydberg atoms, and superconducting circuits. Finally, the first proof-of-principle trapped ions experimental quantum simulations of the Schwinger model are reviewed.
} 

\maketitle

\section{Introduction}
\label{intro}

In the last few decades, quantum information theory has been fast developing and consequently its application to the real world has spawned different technologies that -- as for classical information theory -- encompass the fields of communication, computation, sensing, and simulation \cite{RevModPhys.86.419,RevModPhys.89.035002,RevModPhys.86.153}. To date, the technological readiness level of quantum technologies is highly diverse: some quantum communication protocols are ready for the market, while, e.g., universal quantum computers -- despite experiencing an incredibly fast development -- are still at the first development stage\cite{feynman1986quantum,Ladd:2010fq}.  

Some particularly interesting and potentially disruptive applications of quantum information theory and of quantum technologies lay within different scientific fields, such as high-energy, nuclear, condensed matter physics or chemistry\cite{trabesinger2012quantum}. Indeed, in the last years, it became increasingly clear that concepts and tools from quantum information can unveil new directions and will most probably provide new tools to attack long-standing open problems such as the study of information scrambling in black holes\cite{susskind2016computational}, the solution of complex chemical or nuclear systems\cite{PhysRevLett.120.210501}, or the study of lattice gauge theories (LGTs) -- the main subject of this review.  

LGTs are characterised by an extensive number of symmetries, that is, conservation laws that hold at every lattice site. They describe an incredibly vast variety of different phenomena that range from the fundamental interactions of matter at high energies \cite{kronfeld2012twenty} -- the standard model of particle physics -- to the low-energy behaviour of some materials with normal and/or topological order in condensed matter physics\cite{altland2010condensed,fradkin2013field}. Moreover, recently it has been shown that most of the hard problems in computer science can be recast as a LGT\cite{de2009unifying,10.3389/fphy.2014.00005}. The connection passes through the recasting of the classical problem in Hamiltonian form, which generally assumes the form of an Ising Hamiltonian with long-range disordered interactions. This class of Hamiltonians can be mapped exactly into two-dimensional LGT\cite{Lechnere1500838}. 

For all the aforementioned scenarios, quantum science provided two novel pathways to analyse them. The first one has its root in Feynman's first intuition \cite{feynman1982simulating} of quantum computers: having quantum hardware able to precisely reproduce another physical quantum model, allows a powerful investigation tool for computing the observables of the model, and to verify or compare its prediction with the physical system. Today, the research frontier is at the edge of having universal quantum computers and quantum simulators able to perform such investigations beyond proof of principle analysis. Thus, detailed studies and proposals have been put forward to perform quantum simulations of LGT in the near and mid-term\cite{preskill2018quantum}. The second pathway exploits a class of numerical methods -- tensor network methods (TNM) -- which have been developed in the condensed matter and quantum information communities to study strongly correlated many-body quantum systems\cite{RevModPhys.77.259}. Indeed, as it has been shown recently, TNM can be exploited to study LGT going in regimes where standard approaches are severely limited\cite{Dalmonte2016,banuls2019review}. 

Lattice gauge theory was originally constructed by Wilson in order to define Quantum Chromodynamics (QCD) --- the relativistic $SU(3)$ gauge field theory that describes the strong interaction between quarks and gluons --- beyond perturbation theory. For this purpose, he introduced a hyper-cubic space-time lattice as a gauge invariant regulator of ultraviolet divergences, with quark fields residing on lattice sites and gluons fields residing on links connecting nearest-neighbour sites. This framework makes numerous important physical quantities accessible to first principles Monte Carlo simulations using classical computers. These include static properties, like masses and matrix elements, of baryons (such as protons and neutrons) and mesons (such as pions). Properties of the high-temperature quark-gluon plasma in thermal equilibrium are accessible as well. This includes, e.g., the critical temperature of the phase transition in which the chiral symmetry of the quarks, which is spontaneously broken at low temperatures, gets restored. After more than four decades of intensive research, lattice QCD has matured to a very solid quantitative tool that is indispensable for correctly interpreting a large variety of experiments, including the ones at the high-energy frontier of the Large Hadron Collider (LHC) at CERN. 

However, there are other important aspects of the QCD dynamics, both at high baryon density (such as in the core of neutron stars) and for out-of-equilibrium real-time evolution (such as the various stages of heavy-ion collisions), where importance-sampling-based Monte Carlo simulations fail due to very severe sign or complex action problems. In these cases, reliable special purpose quantum simulators or universal quantum computers may be the only tools to successfully address these grand-challenge problems. While immediate results with quantitative impact on particle physics are unrealistic to hope for, a long-term investment in the exploration of quantum technologies seems both timely and most interesting. Lattice gauge theory has a very important role to play in this endeavour, because, besides fully-fledged lattice QCD, it provides a large class of simpler models, in lower dimensions, with simpler Abelian or non-Abelian gauge groups, or with a modified matter content, which often are interesting also from a condensed matter perspective. The real-time evolution of all these models is as inaccessible to classical simulation as the real-time evolution of QCD itself. Hence, learning how to tackle with these challenges in simpler models is a necessary and very promising step towards the ultimate long-term goal of quantum simulating QCD. Along the way, via a large variety of lattice field theory models, particle physics provides an important intellectual stimulus for the development of quantum information technology.

Validation of quantum simulation experiments is vital for obtaining reliable results. In certain cases, which are limited to equilibrium situations, importance sampling Monte Carlo simulations using classical computers can provide such validation. However, Matrix Product States (MPS) and Tensor Network (TN) calculations are often the more promising method of choice, in particular, because they can even work in some out-of-equilibrium real-time situations. This provides an important stimulus to further develop these techniques. While they work best in one (and sometimes in two) spatial dimensions, an extension to higher dimensions is not at all straightforward, but very well worth to pursue vigorously. Even if these methods should remain limited to lower dimensions, they offer a unique opportunity to gain a deep understanding of the real-time evolution of simple lattice gauge models. By quantitatively validating quantum simulators in out-of-equilibrium situations, even if only in lower dimensions, MPS and TN methods play a very important role towards establishing quantum simulators as reliable tools in quantum physics.

This paper reviews the recent activities along these lines, in particular of the groups that form the QTFLAG consortium, a European project funded under QuantERA with the goal of developing novel quantum science approaches to simulate LGT and study physical processes beyond what could be done via standard tools. First, the main concepts of interest are introduced, the LGT formulation and the tools used to study them: quantum simulators on different hardware and tensor network methods. Then, the recent numerical studies of one-dimensional Abelian and non-Abelian LGTs in and out of equilibrium, at zero and finite temperature are presented. Different theoretical proposals for the implementation of LGTs on digital and analog quantum simulators in trapped ions, Rydberg atoms and in superconducting circuits are reviewed. Finally, the first experimental realisations of these ideas are also briefly mentioned. 

\section{Lattice Field Theory background}
\label{LFT}

Gauge field theories are at the heart of the current theoretical understanding of fundamental processes in nature, both in condensed matter and in high-energy physics. Although their formulation appears to be simple, they potentially give rise to very intriguing phenomena, such as asymptotic scaling, confinement, spontaneous chiral symmetry breaking or (non-trivial) topological properties, which shape the observed physical world around us. Solving gauge theories from first principles has been a major goal for several decades. Their formulation on a discrete Euclidean space-time lattice, originally proposed by Wilson in the seventies \cite{Wilson:1974sk}, has provided very powerful methods to study the non-perturbative regimes of quantum field theories \footnote{See, however, \cite{Hartung:2018usn} for a recent alternative approach using a $\zeta$-regularisation}. A most prominent example is the success of ab-initio Lattice Quantum Chromodynamics (LQCD) calculations. Here, starting from the QCD Lagrangian, the low-lying baryon spectrum could be computed on very large lattices and extrapolated to the continuum limit \cite{Durr:2008zz}. Lattice QCD calculations have also provided most important insights into the structure of hadrons \cite{Constantinou:2015agp,Cichy:2018mum}; they provide information on non-perturbative contributions to electroweak processes \cite{Meyer:2018til} and flavour physics \cite{Juettner:2016atf}; and they are very successful to determine thermodynamic properties \cite{Ding:2015ona}. Today, lattice calculations are performed on large lattices -- presently of sizes around $100^3\times 200$ lattice points -- and directly in physical conditions. By controlling systematic errors, such as discretisation and finite-volume effects, lattice field theory, and in particular lattice QCD, is providing most important input to interpret and guide ongoing and planned experiments world-wide, such as those at the Large Hadron Collider at CERN. These most impressive results became possible by a combined progress on algorithmic and computational improvements as well as the development of new supercomputer architectures. Thus, lattice field theory computations have demonstrated the potential to characterise the most fundamental phenomena observed in nature. 

The standard approach of lattice field theory relies on Monte Carlo-based evaluations of path integrals in Euclidean space-time with positive integrands. Thus it suffers from an essential limitation in scenarios that give rise to a sign problem. These include the presence of a finite baryon density, which is relevant for the early universe and for neutron stars; real-time evolution, e.g., to understand the dynamics of heavy-ion collisions; or topological terms, which could shed light on the matter-anti-matter asymmetry of the universe. There is therefore an urgent quest to find alternative methods and strategies that enable tackling these fundamental open problems in the understanding of nature. 

One such alternative is the application of tensor networks (TN). Originally introduced in the context of condensed matter physics, TN can solve quasi-exactly one dimensional strongly correlated quantum many-body problems for system sizes much larger than exact diagonalisation allows. They are naturally free from the sign problem. In fact, for 1-dimensional systems a number of successful studies have demonstrated the power of TN for lattice gauge theory calculations \cite{banuls2018tensor}. In particular, it has been shown that TN provide accurate determinations of mass spectra and that they can map out a broad temperature region. They can also treat chemical potentials and topological terms and they can be used to study real-time dynamics. TN also allow the study of entanglement properties and the entropy (leading in turn to the determination of central charges) in gauge theories, which brings new aspects of gauge theories into focus. However, applications to higher-dimensional problems remain a challenge presently. There are well-founded theoretical formulations such as projected entangled pair states (PEPS) but their practical application is still rather limited (for a recent review see \cite{RTP19}). New ideas such as the ones developed in \cite{Zohar:2017yxl} could have the potential to overcome these limitations but clearly further studies and developments are necessary in order to turn them into practical tools for addressing gauge theories in higher 
dimensions. 

Ultimately, the intrinsic quantum nature of lattice gauge theories will be a limiting factor for classical calculations, even for TN, e.g., when out-of-equilibrium phenomena of a system are to be studied. In this context, quantum simulation, i.e., the use of another well-controlled quantum system to simulate the physics of the model under study, appears as a more adequate strategy. The idea of quantum simulation, first proposed by Feynman \cite{feynman1982simulating}, is now becoming a reality \cite{Jaksch2005,Bloch2012,blatt2012quantum,Cirac2012}, and very different condensed matter models have already been successfully quantum simulated in cold-atom laboratories around the world \cite{Greif1236362,51atom,entanglement}. Regarding the simulation of LGT, a number of proposals have been put forward in the last years \cite{zohar2012simulating,tagliacozzo2013optical,banerjee2012atomic,zohar2013cold,tagliacozzo2013simulation,banerjee2013atomic,mezzacapo2015non,Wiese:2013uua,zohar2015quantum}, and have even been realised by a few pioneering experiments \cite{martinez2016real}. TN calculations have been instrumental in the definition of many of these proposals. It is in particular the approach of hybrid quantum-classical simulation schemes which can take advantage of these new concepts and there is a great potential to realise them on near-term quantum architectures.

Gauge fields on the lattice manifest themselves as parallel transporters residing on the links that connect neighbouring lattice sites. In Wilson's formulation of lattice gauge theory, the link parallel transporters take values in the gauge group \cite{Wilson:1974sk}. As a consequence, for continuous gauge groups such as the Abelian $U(1)$ gauge group of Quantum Electrodynamics (QED) or the non-Abelian $SU(3)$ gauge group of QCD, the link Hilbert space is infinite dimensional. When gauge fields are treated by TN techniques or they are embodied by ultra-cold matter or quantum circuits, representing an infinite dimensional Hilbert space is challenging, because usually only a few quantum states can be sufficiently well controlled in quantum simulation experiments. There are different approaches to addressing this challenge. First, the link Hilbert space of the Wilson theory can be truncated to a finite dimension in a gauge-covariant manner. Gradually removing the truncation by a modest amount allows one to take the continuum limit.

An alternative approach is provided by quantum link models (also known as gauge magnets)\cite{horn_finite_1981,Orland1990,Chandrasekharan:1996ih} which work with quantum degrees of freedom with a finite-dimen\-sional Hilbert space from the outset. For example, the parallel transporters of a $U(1)$ quantum link model are constructed with quantum spins \cite{Banerjee2013QuantumLinkDeconfined}, which can naturally be embodied by ultra-cold matter. Again, when one moderately increases the spin value, one can reach the continuum limit. Both approaches are actively followed presently and it will be interesting to see in the future, which strategy will be most appropriate to treat gauge theories with TN or on quantum devices. 

Even when one restricts oneself to the smallest spin value $\frac{1}{2}$, interesting gauge theories emerge. For example, when its Gauss' law is appropriately modified, the $U(1)$ quantum link model turns into a quantum dimer model \cite{Rokhsar1988Dimers,Banerjee:2014wpa}, which is used in condensed matter physics to model systems related to high-temperature superconductors. Kitaev's toric code \cite{Kitaev:1997wr} -- a topologically protected storage device for quantum information -- provides an example of a $Z(2)$ lattice gauge theory formulated with parallel transporters consisting of quantum spins $\frac{1}{2}$. Quantum spin chains were among the first systems to be quantum simulated successfully. $SU(N)$ quantum spin ladders, i.e.,\ systems consisting of $n$ transversely coupled spin chains, can be quantum simulated with ultra-cold alkaline-earth atoms in optical lattices \cite{laflamme2016cp}. For moderate values of $n$, these $(2+1)$-dimensional systems dimensionally reduce to $(1+1)$-dimensional $CP(N-1)$ models, which are asymptotically free and thus serve as toy models for QCD. Furthermore, for odd $n$ they have non-trivial topology, very much like non-Abelian gauge theories in four space-time dimensions. Also QCD itself can be formulated as a quantum link model \cite{Brower:1997ha,Wiese:2014rla}. In that case, the parallel transporters are matrices with non-commuting matrix elements, just as quantum spins are vectors with non-commuting components. Alkaline-earth atoms can again be used to encode the QCD color degree of freedom in the nuclear spin of these atoms \cite{banerjee2013atomic}. Lattice gauge theory, either in its gauge covariantly truncated Wilson formulation or in the description of quantum link models, which nicely complement each other, provides a broad framework for upcoming quantum simulation experiments.

Whatever the most effective simulations may be in the future, classical Monte Carlo, tensor network, or quantum simulations for addressing gauge theories, there will remain a big challenge: in the end, all calculations aim at providing input for world-wide experiments, whether the ones in condensed matter physics or the large-scale collider experiments in high-energy physics. As a consequence, all results emerging from theoretical computations based on the underlying Hamiltonian or Lagrangian need to have controlled statistical and systematic errors. This will lead to a substantial, demanding effort for such calculations since many simulations at various values of the lattice spacing and lattice volumes as well as possibly other technical parameters (e.g.,\ the bond dimension in the TN approach) have to be executed. Only by performing a controlled continuum and infinite volume (or infinite bond dimension) limit, it will become possible to rigorously attribute the obtained results to the underlying model. In this way, the underlying model can be thoroughly tested and, in turn, any significant deviation seen in experiment will thus open the door to completely new and unexplored physics. 

\section{Quantum Science and Technologies tools}
\label{QSTtools}

In a seminal paper published in 1982, Feynman discussed in great detail the problems connected with the numerical simulation of quantum systems. He envisaged a possible solution, the so-called universal quantum simulator, a quantum-mechanical version of the usual simulators and computers currently exploited in many applications of the 'classical' world. If realised, such a device would be able to tackle many-body problems with local interactions by using the quantum properties of nature itself. Interestingly, even without the advent of a fully universal quantum computer, the construction of dedicated devices, also known as purpose-based quantum simulators, would already be of significant importance for the understanding of quantum physics. The basic idea is to engineer the Hamiltonian of the quantum model of interest in a highly controllable quantum system and to retrieve all of the desired information with repeated measurements of its properties. Many research fields would eventually benefit from such devices: for example, two-dimensional and three-dimensional many-body physics, non-equilibrium dynamics or lattice gauge theories.

In recent years, the scientific community has been considering several quantum technologies such as cold atoms, trapped ions or superconducting circuits as examples of the most promising candidates for the realisation of a wide variety of dedicated quantum simulations. Indeed, these platforms are genuine quantum systems where the available experimental techniques offer an impressive degree of control together with high-fidelity measurements, thus combining two fundamental requirements for a quantum simulator. Among the most recent experimental achievements, just to mention a few, such as, the observation of Anderson localisation in disordered Bose-Einstein condensates (BECs), the research on itinerant ferromagnetism with cold fermions or the reconstruction of the equation of state of fermionic matter in extreme conditions, such as in neutron stars.

The advantages of quantum simulation are numerous: first, one can use it to study physical systems which are not experimentally accessible (systems of large or small scales, for example), or to observe the physical properties of unreal physical systems, which are not known to be found in nature, but can be mapped to the simulating systems. So far, a lot of quantum simulations were suggested, and some were even experimentally implemented. The simulated systems come from almost every area of physics: condensed matter and relativistic quantum physics, gravity and general relativity, and even particle physics and quantum field theory.  The last of these is the topic of this review, specifically gauge theories. While quantum simulations have been proposed (and even realised) for condensed matter models, gauge theories are a newer branch where quantum technologies might be employed. See also \cite{mwr2014,bazavov2015gauge,zou2014progress,garcia2015fermion,kasper2016schwinger,dehkharghani2017quantum,kasper2017implementing,kaplan2018gauss,51atom} and for a general review on quantum simulation \cite{trabesinger2012quantum}.

\section{Quantum information techniques}
\label{QItechiques}

\subsection{Tensor networks for Lattice Gauge Theories}

As mentioned before, classical numerical simulations are playing a leading role in the understanding of lattice gauge theories. In particular, in recent years, there has been a boost in the development of tensor network methods to simulate lattice gauge theories. There are different approaches, that range from the exploitation of mappings of some theories to spin models~\cite{banuls2013matrix,Banuls2013}, to the development of gauge invariant tensor networks in the quantum link formulation~\cite{silvi2014lattice,Dalmonte2016,montangero2018,funcke2019topological,PhysRevD.99.074501,banuls2019review}. This section reviews some of the studies that appeared in the last years, covering most of the available approaches for Abelian and non-Abelian lattice gauge theories~\cite{banuls2013matrix,Banuls2013,silvi2014lattice,tagliacozzo2014tensor,buyens2014matrix,rico2014tensor,haegeman2015gauging}. 

In the following subsections, a selection of works performed along these lines is described in some detail.

\subsubsection{Matrix Product States for Lattice Field Theories\cite{banuls2013matrix,Banuls2013}}

The Schwinger model \cite{schwinger62,Coleman1976}, i.e., QED in one spatial dimension, is arguably the simplest theory of gauge-matter interaction, and yet it exhibits features in common with more complex models (like QCD) such as confinement or a non-trivial vacuum. Therefore, it constitutes a fundamental benchmark to explore the performance of lattice gauge theory techniques. In particular it has been extensively used in the last years to probe the power of TN  as alternative methods to conventional Monte Carlo-based lattice techniques for solving quantum field theories  in the continuum.

The first such study was carried out by Byrnes and coworkers \cite{Byrnes2002} using the original DMRG formulation, and it already improved by orders of magnitude the precision of the ground state energy and vector particle mass gap, with respect to results obtained by other numerical techniques, although the precision decreased fast for higher excitations. The application of TN formulated algorithms, including extensions to excited states, time evolution and finite temperature has allowed a more systematic exploration of the model in recent years.

The discretised Hamiltonian of the model, in the Kogut-Susskind formulation with staggered fermions \cite{Kogut1975} reads
\begin{eqnarray}
H=&-\frac{i}{2 a}\sum_n \left (  \phi^{\dagger}_n e^{i\theta_n}  \phi_{n+1} - \mathrm{H.c.}\right )   
+m\sum_n(-1)^n\phi_n^{\dagger}\phi_n  \nonumber \\
&+\frac{a g^2}{2} \sum_n (L_n+\alpha)^2,
\label{eq:schwinger}
\end{eqnarray}
where $ \phi_{n}^{\dagger}$ represents the creation operator of a spin-less fermion on lattice site $n$, and $U_n=e^{i\theta_n}$ is the link operator between sites $n$ and $n+1$. $L_n$, canonical conjugate to $\theta_n$, corresponds to the electric field on the link, and $\alpha$ corresponds to a background field. Physical states need to satisfy Gauss' law as an additional constraint, $L_n-L_{n-1}=\phi_n^{\dagger}\phi_n-\frac{1}{2}\left [1-(-1)^n\right]$. In the continuum, the only dimensionless parameter of the model is the fermion mass $m/g$ (expressed in terms of the coupling). The discretisation introduces one more parameter, namely the lattice spacing $a g$. For convenience, the Hamiltonian is often rescaled and expressed in terms of the dimensionless parameters $x=1/(a g)^2$, $\mu=2\sqrt{x}m/g$, with the continuum limit corresponding to $x\to \infty$. The local Hilbert space basis for the fermionic sites can be labeled by the occupation of the mode, $\phi^{\dagger}_n \phi_{n} \in\{0,1\}$ (for site $n$), while the basis elements for the links can be labeled by the integer eigenvalues of $L_n$, $\ell_n$. Using this basis, an MPS ansatz can be optimised to approximate the ground state or the excitations.

\begin{figure}
\begin{minipage}[b]{.3\columnwidth}
\includegraphics[height=.9\columnwidth]{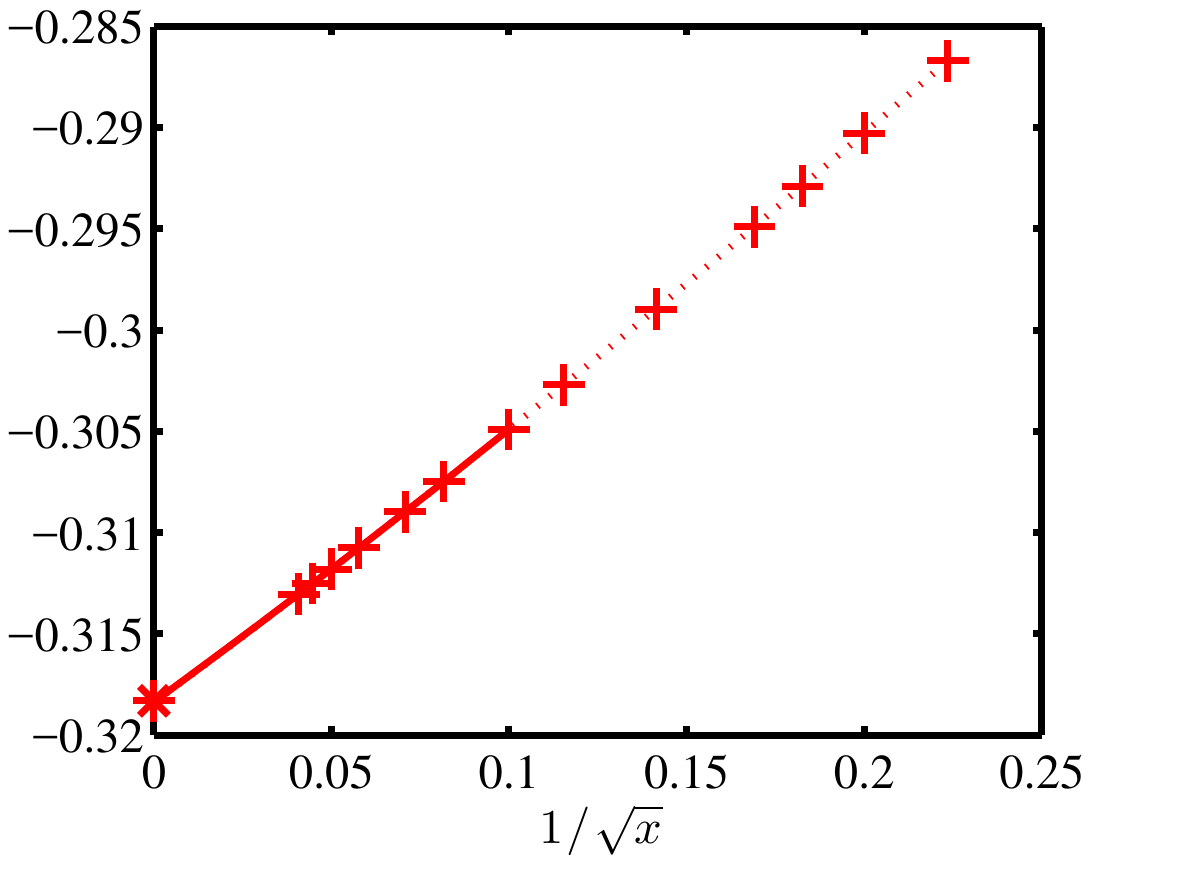}
\end{minipage}
\hspace{.01\columnwidth}
\begin{minipage}[b]{.3\columnwidth} 
\includegraphics[height=.9\columnwidth]{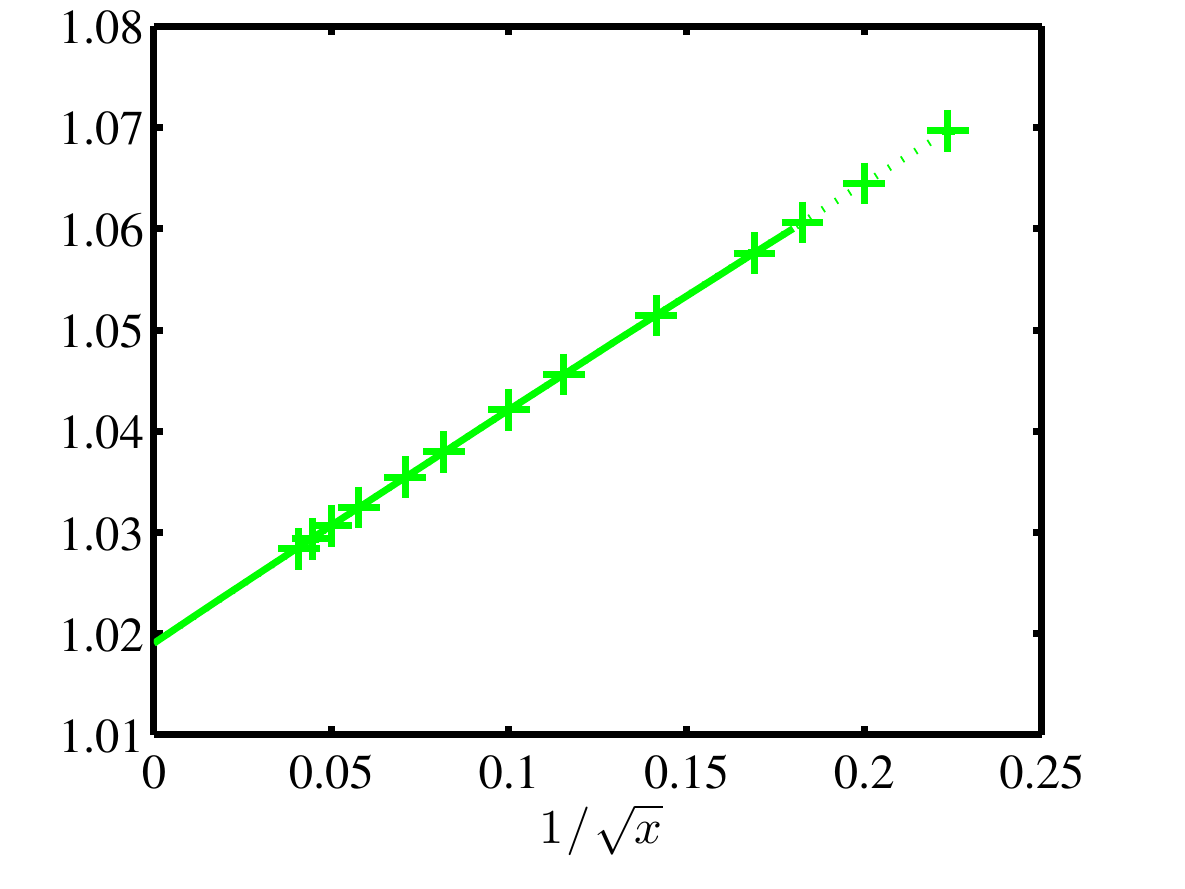}
\end{minipage}
\hspace{.01\columnwidth}
\begin{minipage}[b]{.3\columnwidth} 
\includegraphics[height=.9\columnwidth]{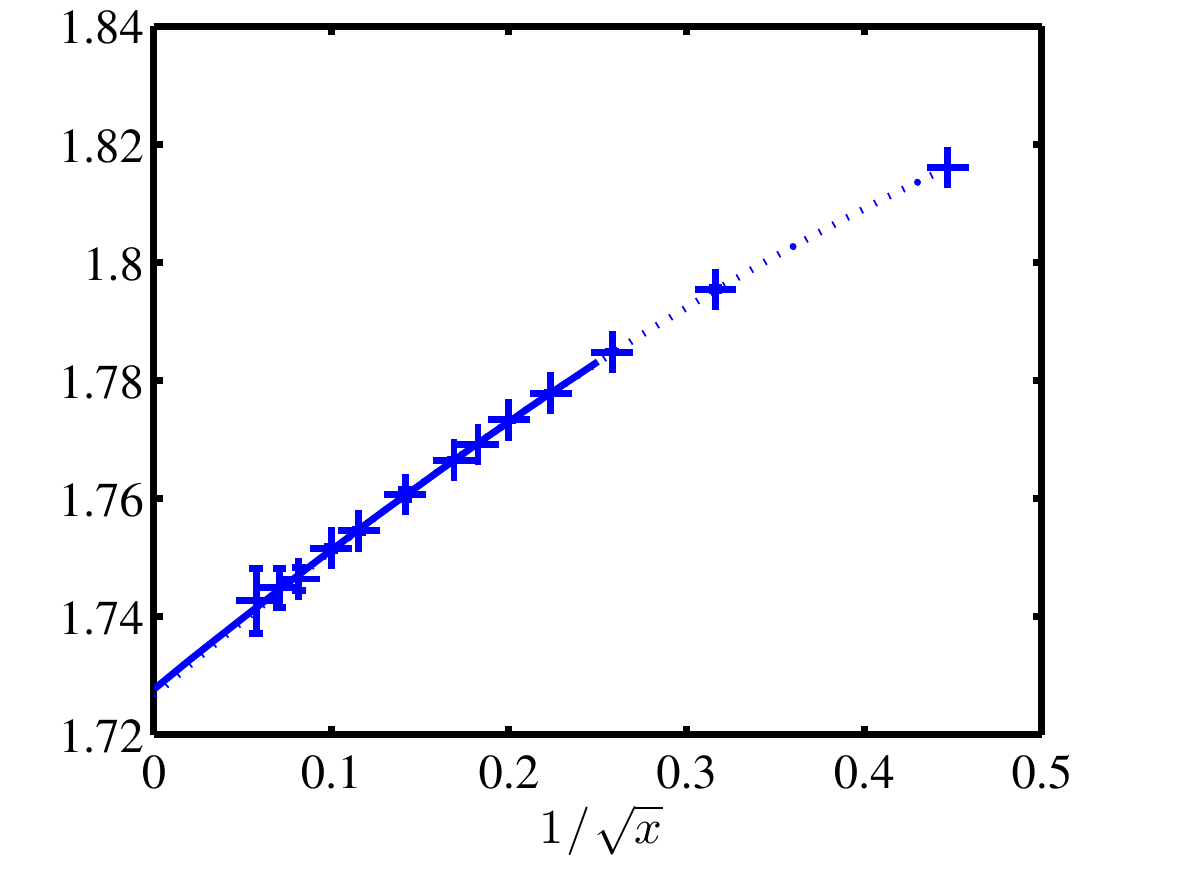}
\end{minipage}
\caption{(from \cite{Banuls2013}) Example of the extrapolations in lattice spacing for the energy density of the ground state (left), and the mass gaps of vector (center) and scalar (right) particles for fermion mass $m/g=0.25$. The solid lines show the fitted curves that produce the final value, and the dashed lines a different fit to estimate the error.}
\label{fig:massMPS}
\end{figure}

Instead of working with explicit gauge degrees of freedom, it is possible to integrate them out using Gauss' law, and to work directly in the physical subspace. This results in a Hamiltonian expressed only in terms of fermionic operators, but with non-local interactions among them. Additionally, a Jordan-Wigner transformation can be applied to map the model onto a more convenient spin Hamiltonian \cite{Banks:1975gq}. In \cite{Banuls2013} a systematic study of the mass spectrum in the continuum was performed using MPS with open boundary conditions, in the absence of a background field, for different values of the fermion mass. The ground state and excitations of the discrete model were approximated by MPS using a variational algorithm, and the results were successively extrapolated in bond dimension, system size (individual calculations were done on finite systems) and lattice spacing, in order to extract the continuum values of the ground state energy density and the mass gaps (Fig.~\ref{fig:massMPS} illustrates the continuum limit extrapolations). These steps resemble those of more usual lattice calculations, so that also standard error analysis techniques could be used to perform the limits and estimate errors, and thus gauge the accuracy of the method. Values of the lattice spacing much smaller than the usual ones in similar Monte Carlo calculations could be explored, and very precise results were obtained for the first and second particles in the spectrum (respectively vector and scalar), beyond the accuracy of earlier numerical studies (see table \ref{tab:specSchw}).

\begin{table}
\begin{center}
\begin{tabular}{|c|c|c|c|c|}
\hline
&\multicolumn{2}{c|}{$M_V/g$}
&\multicolumn{2}{c|}{$M_S/g$} 
\\ 
\cline{2-5}
$m/g$ &
OBC \cite{Banuls2013} &
uMPS \cite{buyens2014matrix} &
OBC \cite{Banuls2013} &
uMPS \cite{buyens2014matrix} \\
\hline
0	&
0.56421(9) &
0.56418(2)& 
1.1283(10) &
- \\
\hline
0.125 &
0.53953(5) &
0.539491(8)&
1.2155(28)  &  1.222(4) \\ \hline
0.25 &
0.51922(5) & 0.51917(2)
&
1.2239(22)  &  1.2282(4)\\ \hline
0.5 &
0.48749(3) & 0.487473(7)
&
1.1998(17)  &  1.2004(1)\\ \hline
\end{tabular}
\caption{Binding energies, $M/g:=\omega-2 m/g$, with errors of the vector and scalar particles. Both results obtained with  open boundary finite MPS with gauge degrees of freedom integrated out (left columns) or gauge invariant uniform MPS \cite{buyens2014matrix} (right column) simulations are shown. In the case of massless fermion, the analytical values are
$M_V/g=0.5641895$
and $M_S/g=1.12838$.}
\label{tab:specSchw}
\end{center}
\vspace{-.5cm}
\end{table}

\begin{figure}
\begin{minipage}[b]{.4\columnwidth}
\includegraphics[height=.9\columnwidth]{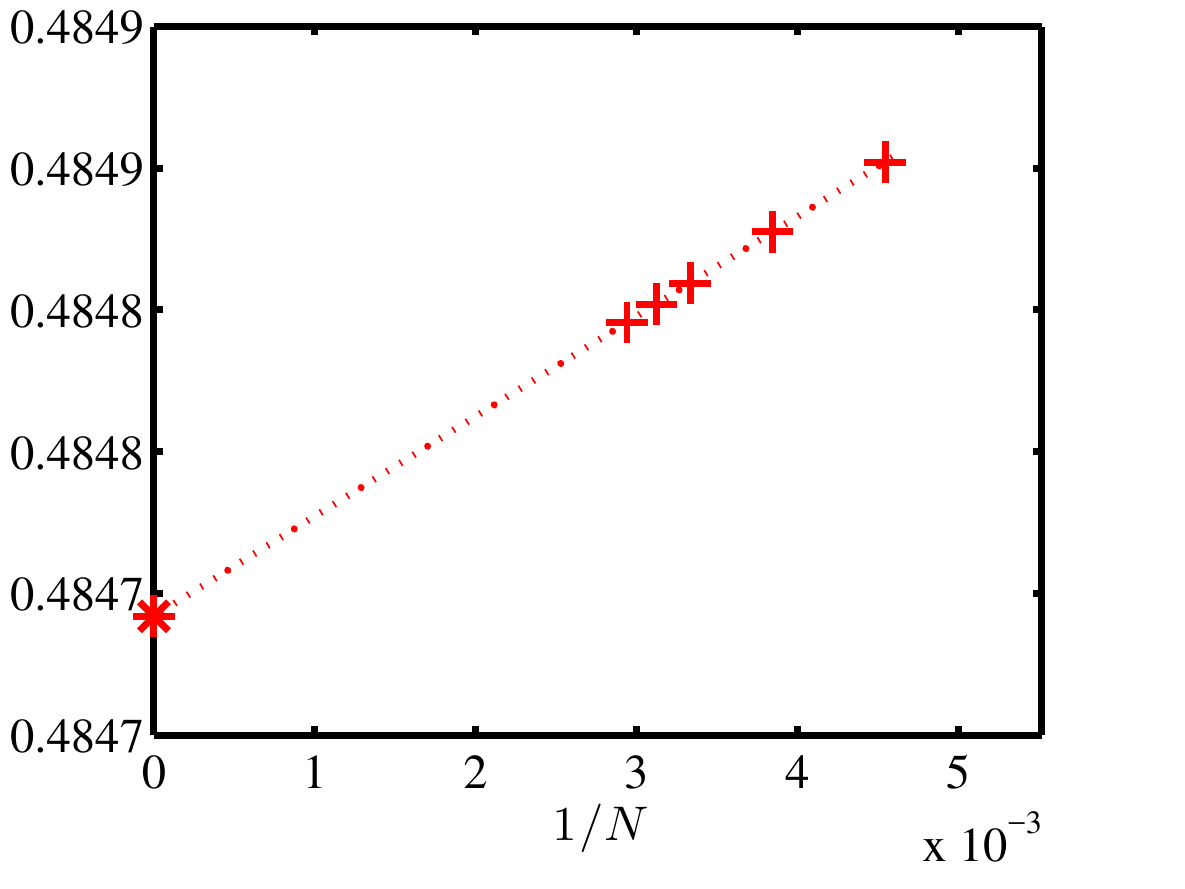}
\end{minipage}
\hspace{.1\columnwidth}
\begin{minipage}[b]{.4\columnwidth} 
\includegraphics[height=.9\columnwidth]{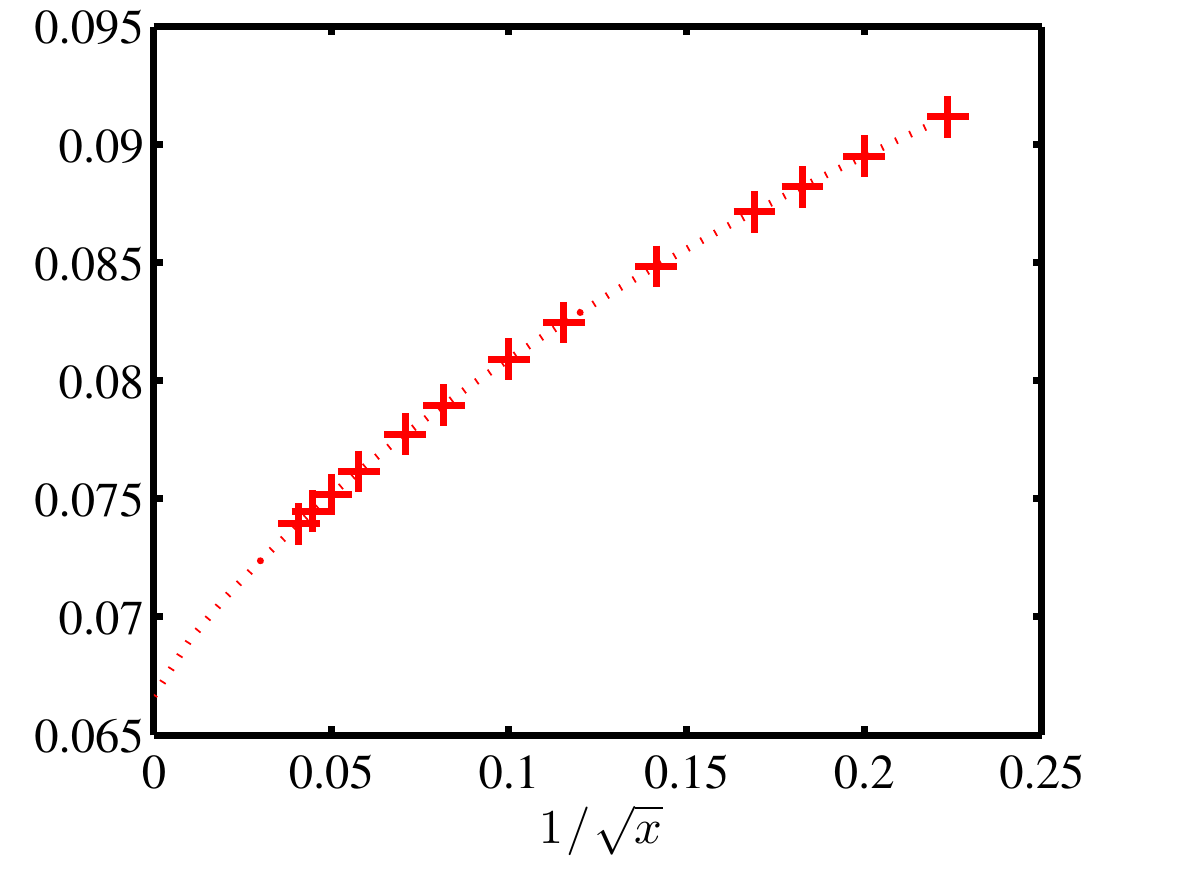}
\end{minipage}
\caption{Example of the condensate extrapolations in finite size (left) and lattice spacing (right) for fermion mass $m/g=0.25$  \cite{banuls2013matrix}. The left plot corresponds to fixed lattice parameter $x=100$. On the right, the divergent part corresponding to the non-interacting case has already been subtracted. The dashed lines show the fitted curves.}
\label{fig:condMPS}
\end{figure}

\begin{table}
\begin{center}
\begin{tabular}{|c|c|c|}
\hline
&
\multicolumn{2}{|c|}{Subtracted condensate} \\
\hline
$m/g$  & MPS with OBC & exact \\
\hline
0 & 0.159930(8) & 0.159929 \\
\hline
0.125 & 0.092023(4) & - \\
\hline
0.25 &  0.066660(11) & - \\
\hline
0.5 & 0.042383(22) & - \\
\hline
\end{tabular}
\end{center}
\caption{Values of the vacuum chiral condensate in the continuum for different fermion masses obtained with the MPS ansatz.}
\label{tab:condSchw}
\end{table}

Since the algorithms provide a complete ansatz for each excited state, other observables can be calculated. An interesting quantity is the chiral condensate, order parameter of the chiral symmetry breaking, and written in the continuum as $\Sigma=\langle \overline{\Psi}(x) \Psi(x) \rangle/g$. When computed on the lattice, the condensate has a UV divergence, which is already present in the free theory. Using the MPS approximations for the ground state, the continuum limit of the condensate was extracted in~\cite{banuls2013matrix} (some of these results were refined later in \cite{banuls2016thermalmass}). After subtracting the UV divergence, lattice effects were found to be dominated by corrections of the form $a \log a$. Systematic fitting and error analysis techniques were applied to obtain very precise estimations of the condensate for massless and massive fermions (table~\ref{tab:condSchw}, see also results with uniform MPS~\cite{buyens2014pos} and infinite DMRG \cite{zapp2017tensor}). In the former case the exact value can be computed analytically, but for the latter, very few numerical estimations existed in the literature.

These results demonstrate the feasibility of the MPS ansatz to efficiently find and describe the low-energy part of the spectrum of a LGT in a non-perturbative manner. Moreover they show explicitly how the errors can be systematically controlled and estimated, something fundamental for the predictive power of the method, if it is  to be used on theories for which no comparison to an exact limit is possible.

\subsubsection{Matrix product states for gauge field theories}

\begin{figure}[t]
\begin{subfigure}[b]{.24\textwidth}
\includegraphics[width=\textwidth]{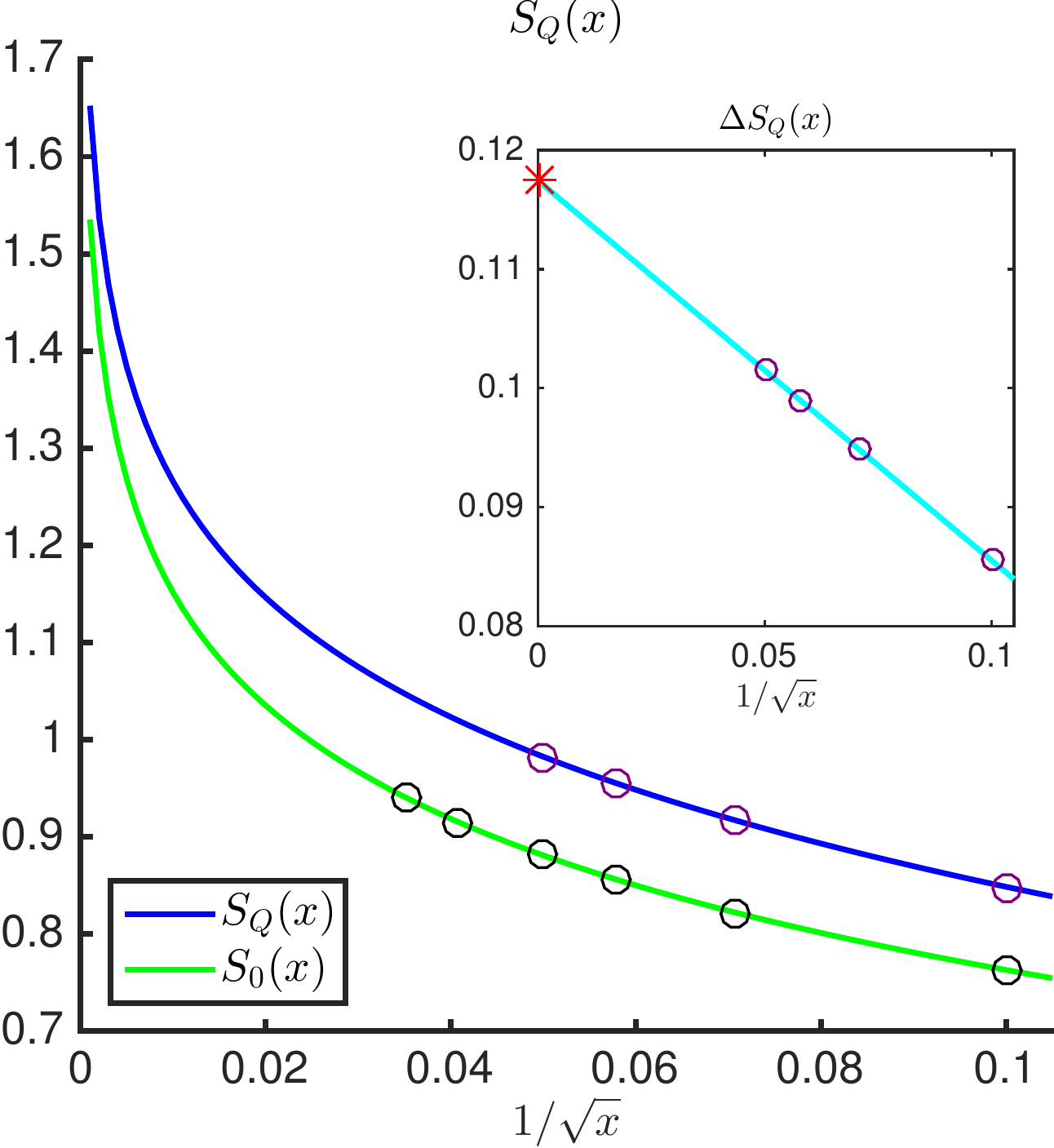}
\caption{\label{fig1VV}}
\end{subfigure}\hfill
\begin{subfigure}[b]{.24\textwidth}
\includegraphics[width=\textwidth]{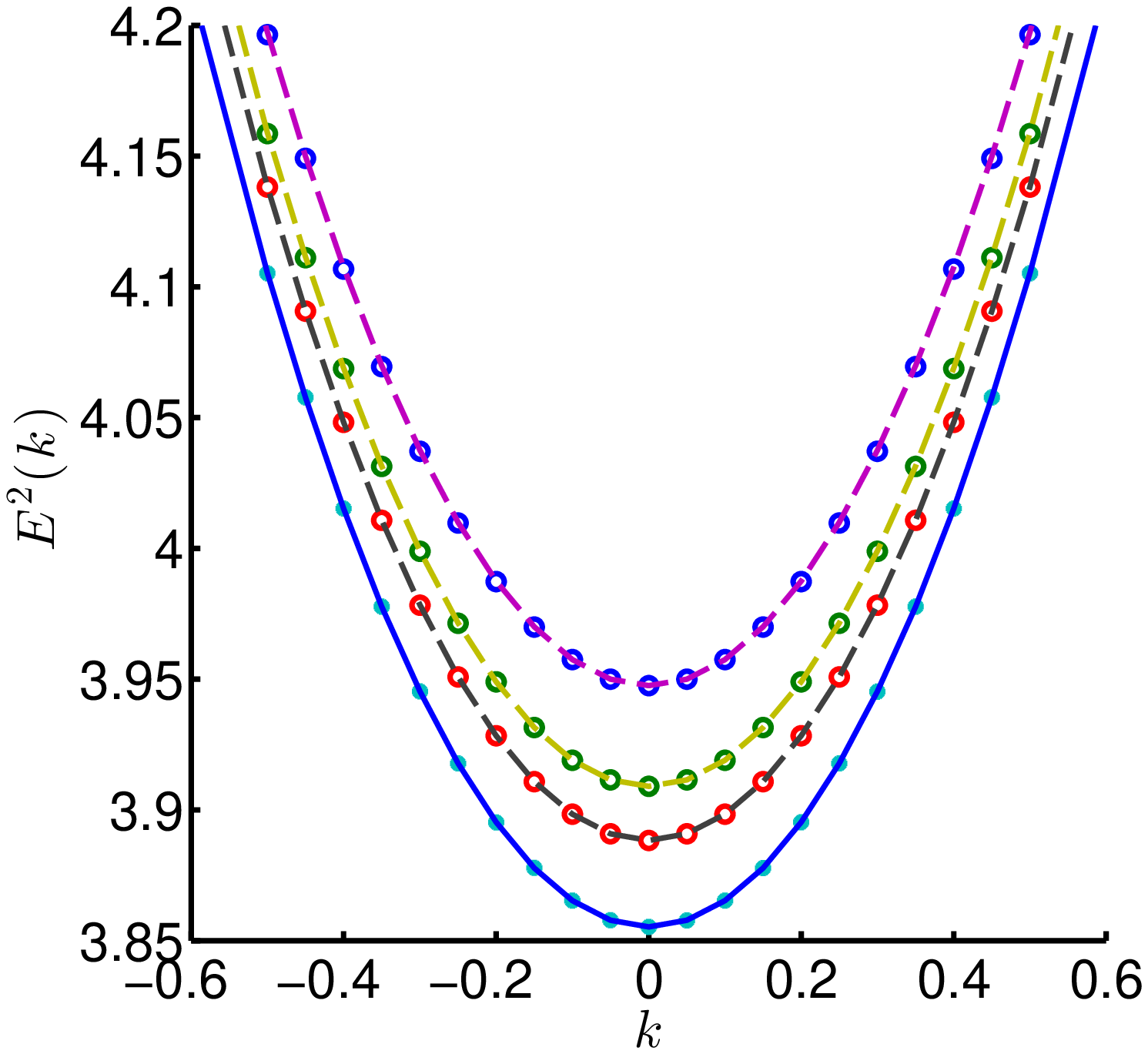}
\caption{\label{fig2VV} }
\end{subfigure}\vskip\baselineskip
\captionsetup{justification=raggedright}
\caption{(\ref{fig1VV}): Bipartite entanglement entropy for different ground state simulations at different lattice spacings $ga=1/\sqrt{x}$. Fits are of the form $-1/6 \log a + A + B a $   \cite{Buyens2015}. (\ref{fig2VV}): fit of the Einstein-dispersion relation $E^2=k^2 + m_v^2(a)$ to the numerical results for the lowest lying bosonic vector excitation, for $m/g = 0.75$, $ag = 1/10, 1/\sqrt{300}, 1/\sqrt{800}$. In this case the continuum extrapolation (full blue line) $a\rightarrow 0$ gives: $m_v(0)/g= 1.96347(3)$ \cite{buyens2014matrix}.}
\end{figure}

A different series of papers by Buyens et al.\cite{buyens2014matrix,buyens2014pos,Buyens2015,Buyens2015b,Buyens2016,Buyens2017,Buyens2017b} also thoroughly studied the aforementioned Schwinger model\cite{Schwinger1951} within the broad MPS framework. In this section, the general systematics of this approach is reviewed vis-\`{a}-vis the particularities that come with the simulation of gauge field theories in the continuum limit. An overview of the most important results that result from these simulations are also shown.

\emph{Continuum limit.} As in the approach of both Byrnes et al.\cite{Byrnes2002,Byrnes2002b} and Ba{\~n}uls et al. \cite{Banuls2013}, the simulations start from a discretisation of the QFT Hamiltonian with the Kogut-Susskind prescription \cite{Kogut1975} followed by a Jordan-Wigner transformation. But different from \cite{Byrnes2002,Byrnes2002b, Banuls2013}, the simulations \cite{buyens2014matrix,buyens2014pos,Buyens2015,Buyens2015b,Buyens2016,Buyens2017,Buyens2017b} are performed directly in the thermodynamic limit, avoiding the issue of finite-size scaling. From the lattice point of view, the QFT limit is then reached by simulating the model near (but not at) the continuum critical point \cite{Kogut1979}. Upon approaching this critical point the correlation length in lattice units diverges $\xi/a\rightarrow \infty$. Large scale correlations require more real-space entanglement, specifically for the Schwinger model the continuum critical point is the free Dirac-fermion $c=1$ CFT, implying that the bipartite entanglement entropy should have a UV-divergent scaling of $1/6\ln (\xi/a)$\cite{calabrese2004entanglement}. This was confirmed explicitly by the numerical MPS simulations of the ground state of the Schwinger model \cite{buyens2014matrix, Buyens2015}, as shown in figure \ref{fig1VV}. Notice the same UV scaling for ground states in the presence of an electric background field $Q$, leading to a UV finite subtracted entropy (see the inset), that can be used as a probe of the QFT IR physics \cite{Buyens2015}. 

\begin{figure}[t]
\begin{subfigure}[b]{.24\textwidth}
\includegraphics[width=\textwidth]{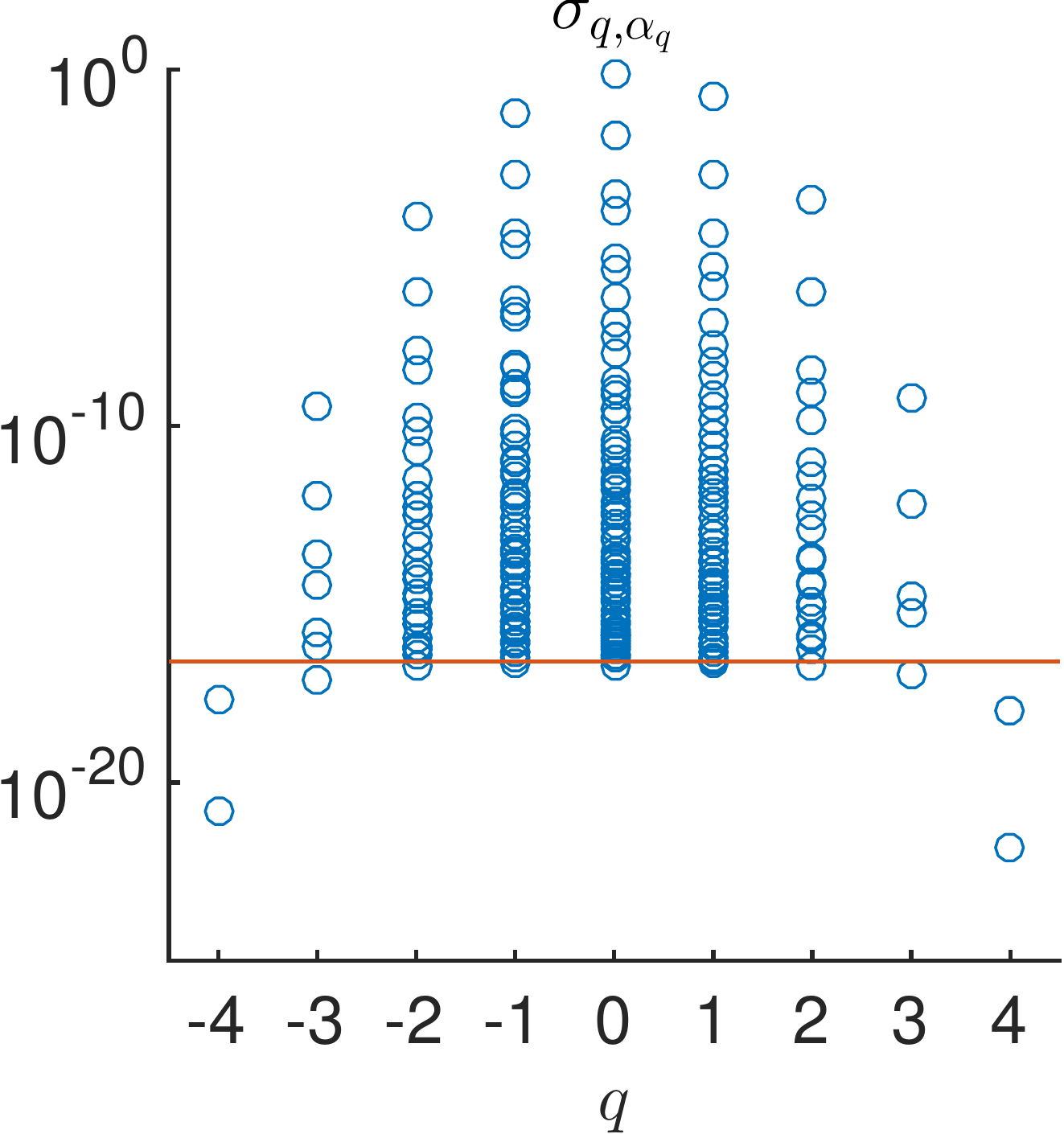}
\caption{\label{fig3VV}}
\end{subfigure}\hfill
\begin{subfigure}[b]{.24\textwidth}
\includegraphics[width=\textwidth]{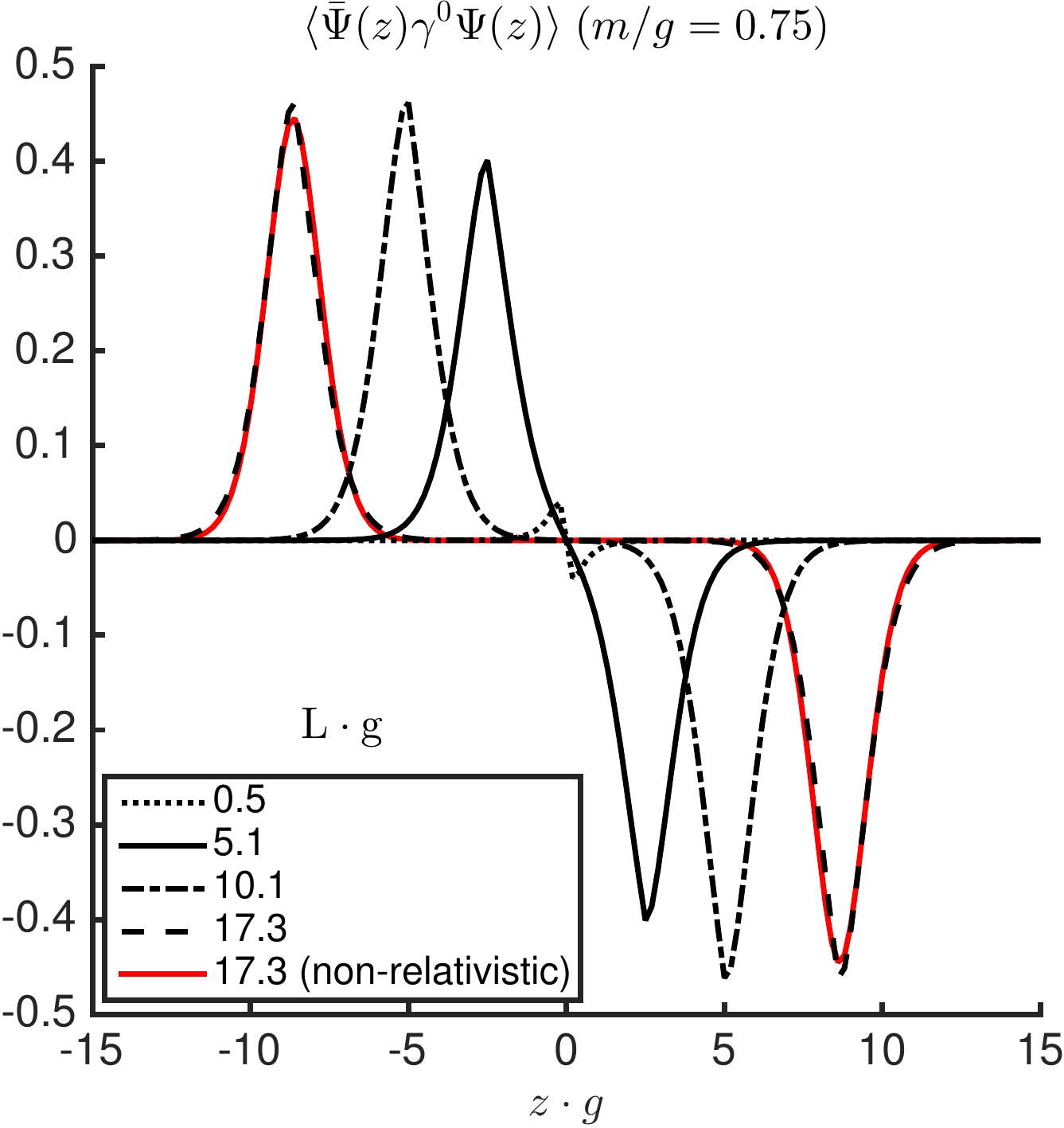}
\caption{\label{fig4VV}}
\end{subfigure}\vskip\baselineskip
\captionsetup{justification=raggedright}
\caption{(\ref{fig3VV}): A typical Schmidt spectrum (single cut on the infinite line) for a converged ground state simulation; justifying the truncation at electric field values $|q|=3$ for a Schmidt precision $\epsilon=2.5 \times10^{-17}$ (orange line) \cite{Buyens2017b}. (\ref{fig4VV}): Charge distribution of the light fermions around unit probe point charges at different inter-charge distances $Lg$, for $m/g=0.75$ \cite{Buyens2015}.}
\end{figure}

For the MPS simulations this UV divergence of the entanglement requires bond dimensions $D$ that grow polynomially with the inverse lattice spacing, $D\sim a^{-n}$. But it turns out that, despite this polynomial growth one can simulate the Schwinger model sufficiently close to its continuum critical point $a\rightarrow 0$, with a relatively low computational cost.  In the different papers simulations were performed up to $a\approx 1/(30 g)$, corresponding to a correlation length $\xi/a\approx 15-35$, depending on the particular ratio of the fermion mass and gauge coupling $m/g$ in the Hamiltonian. The simulations at different decreasing values of $a$ then allow for very precise continuum extrapolations, as illustrated in figure \ref{fig2VV} for the dispersion relation of the (lowest lying) excitation \cite{buyens2014matrix}. Notice that in contrast to e.g. $d=3+1$ QCD,  $d=1+1$ QED is a super-renormalizable theory, with a finite continuum extrapolation of the particle excitation masses $m_{phys}$ in terms of the bare parameters $(m,g)$ of the theory: $m_{phys}(a)=m_{phys}(0)+ \mathcal{O}(a)$.   A further study demonstrated that even simulations with lattice spacings $a\geq 1/(10 g)$ (implying a smaller computational cost) are already sufficient for continuum extrapolations with four digit precision \cite{Buyens2017b}. 

\emph{ Truncating the gauge field.} The numerical Hamiltonian MPS simulations require finite local Hilbert spaces, which is in apparent conflict with the bosonic gauge degrees of freedom that come with the continuous $U(1)$ group of the Schwinger model.  As became evident in the work of Buyens et al., these bosonic fields can be efficiently truncated in the electric field basis, leading to an effective finite local Hilbert space appropriate for the simulations. In figure \ref{fig3VV} the distribution of the Schmidt values \footnote{By the singular value decomposition, any matrix $M$ can be decomposed in a positive semidefinite diagonal matrix $D$ and two unitaries matrices $U$ and $V$ such that $M=UDV^{\dagger}$. The diagonal elements of the matrix $D$ are called the Schmidt values or Schmidt spectrum} is shown over the different electric field eigenvalues $q$ for a particular ground state simulation. Notice that the electric field values are discrete in the compact QED formulation. As one can see from the figure, the contribution from the higher electric field values decays rapidly, in fact exponentially, and it was shown that this decay remains stable towards the continuum limit \cite{Buyens2017b}. For a given Schmidt precision one can therefore indeed safely truncate in $q$. Most simulations used $q\in [-3,3]$.    

\emph{Gauge invariance.} As was discussed already in previous sections, the Kogut-Susskind set-up starts from the Hamiltonian QFT formulation in the time-like axial gauge $A_0=0$, with the physical states obeying the Gauss constraint ${\nabla}E=\rho$. This is indeed equivalent to requiring the physical states to be invariant under local gauge transformations. The resulting lattice Hamiltonian then operates on a Hilbert space of which only a subspace of gauge invariant states, obeying a discretised version of Gauss' law, is actually physical. The simulations of Buyens et al. exploited this gauge invariance by constructing general gauge invariant MPS states \cite{buyens2014matrix} and simulating directly on the corresponding gauge invariant manifold. As shown in \cite{buyens2014pos}, for ground state simulations, working with explicit gauge invariant states leads to a considerable reduction in the computation time. The reason lies in the sparseness of the matrices appearing in gauge invariant MPS states; but also in the fact that the full gauge variant Hilbert space contains pairwise excitations of non-dynamical point charges, separated by short electric field strings of length $L\sim a$. In the continuum limit this leads to a gapless spectrum for the full Hilbert space, whereas the spectrum on the gauge invariant subspace remains gapped. Such a nearly gapless spectrum requires many more time steps before convergence of the imaginary time evolution towards the proper ground state. As such, these test simulations on the full Hilbert space \cite{buyens2014pos} are consistent with Elitzur's theorem \cite{Elitzur1975}, which states that a local gauge symmetry cannot be spontaneously broken, ensuring the same gauge invariant ground state on the full gauge variant Hilbert space.  

\emph{Results.} Using the Schwinger model \cite{Coleman1976,Adam1996} as a very nice benchmark model for numerical QFT simulations, the results of the numerical simulations \cite{buyens2014matrix,buyens2014pos,Buyens2015,Buyens2015b,Buyens2016,Buyens2017,Buyens2017b} were verified successfully against these analytic QFT results in the appropriate regimes. In addition, where possible, the results were compared with the numerical work of \cite{Byrnes2002, Byrnes2002b, Banuls2013}, and found to be in perfect agreement within the numerical precision. Taken together, the tensor network simulations of Byrnes, Ba{\~n}uls, Buyens et al., form the current state of art of numerical results on the Schwinger model. Now, a selection of the results of Buyens et al. are discussed:

\emph{Ground state and particle excitations}. By simulating the ground state and constructing ansatz states on top of the ground state, MPS techniques allow for an explicit determination of the approximate states corresponding to the particle excitations of the theory \cite{Haegeman2012}. For the Schwinger model three particles were found \cite{buyens2014matrix}: two \emph{vector} particles (with a quantum number $C=-1$ under charge conjugation) and one \emph{scalar} particle ($C=+1$). For each of these particles, the obtained dispersion relation is perfectly consistent with an effective Lorentz symmetry at small momenta, as illustrated in Fig.~\ref{fig2VV}. The second vector excitation was uncovered for the first time, confirming prior expectations from strong coupling perturbation theory \cite{Coleman1976, Adam1996}. See the extrapolated mass values obtained for the scalar and first vector particle in absence of a background field in Table 3. Furthermore, in \cite{Buyens2015b} the excitations were studied in presence of a background electric field. By extrapolating towards a vanishing mass gap for a half-integer background field, this allowed for a precise determination of the critical point $(m/g)_c=0.3308$ in the phase diagram \cite{Buyens2017b}.

\begin{table}
\begin{tabular}{|c|c|c|c|c|}
\hline
$m/g$ & $- \omega_0$ & $M_{v,1}$ & $M_{s,1}$ & $M_{v,2}$   \\
\hline
0 &0.318320(4)         &  0.56418(2)&   &  \\
0.125&        0.318319(4) &0.789491(8)  &1.472(4)  &2.10 (2)  \\

0.25 &0.318316(3)         &  1.01917 (2)&  1.7282(4)&2.339(3)  \\
0.5 &        0.318305(2) &  1.487473(7)& 2.2004 (1) & 2.778 (2) \\
0.75 &        0.318285(9) &  1.96347(3)& 2.658943(6) &3.2043(2)  \\
1 &0.31826(2)&2.44441(1)         &3.1182 (1)  & 3.640(4)  \\
\hline
\end{tabular}
\caption{Energy density and masses of the one-particle excitations (in units $g=1$) for different $m/g$. The last column displays the result for the heavy vector boson \cite{buyens2014matrix}.}
\end{table}

\emph{String breaking.} By probing the vacuum of a confining theory with a heavy charge/anti-charge pair, one can investigate the detailed physics of string formation and breaking, going from small inter-charge distances to larger distances. In the latter case the heavy charges get screened by the light charged particles that are created out of the vacuum. This string breaking picture was studied in detail for the Schwinger model in \cite{Buyens2015}. Figure \ref{fig4VV} shows one of the results on the light particle charge density for different distances between the heavy charges. At small distances there is only a partial screening, whereas at large distances the screening is complete: for both fully integrated clouds, the total charge is exactly $\pm 1$. For large values of $Lg$, the string is completely broken and the ground state is described by two free particles, i.e. mesons. Notice the red line in the plot which depicts the corresponding analytic result of the ground state charge distribution for the non-relativistic hydrogen atom in $d=1+1$. Finally, also fractional charges were studied in \cite{Buyens2015}, explicitly showing for the first time the phenomenon of partial string breaking in the Schwinger model.           

\subsubsection{Tensor networks for Lattice Gauge Theories and Atomic Quantum Simulation\cite{rico2014tensor}}
\label{LGTN}
In \cite{rico2014tensor}, an exact representation of gauge invariance of quantum link models, Abelian and non-Abelian, was given in terms of a tensor network description. The starting point for the discussion are LGTs in the Hamiltonian formulation, where gauge degrees of freedom $U_{x,y}$ are defined on links of a lattice, and are coupled to the matter ones $\psi_{x}$, defined on the vertices. In the quantum link formulation, the gauge degrees of freedom are described by bilinear operators (Schwinger representation). This feature allows one to solve exactly, within the tensor network representation, the constraints imposed by the local symmetries of this model.

Quantum link models have two independent local symmetries, (i) one coming from the Gauss law and (ii) the second from fixing a representation for the local degree of freedom. (i) Gauge models are invariant under local symmetry transformations. The local generators of these symmetries, $G_{x}$, commute with the Hamiltonian, $\left[ H, G_{x} \right]=0$. Hence, $G_{x}$ are constants of motion or local conserved quantities, which constrain the physical Hilbert space of the theory, $G_{x}|\text{phys} \rangle =0~ \forall x$, and the total Hilbert space splits in a physical or gauge invariant subspace and a gauge variant or unphysical subspace: $\mathcal{H}_{\text{total}} = \mathcal{H}_{\text{phys}} \oplus \mathcal{H}_{\text{unphys}}$. This gauge condition is the usual Gauss' law. (ii) The quantum link formulation of the gauge degrees of freedom introduces an additional constraint at every link, that is, the conservation of the number of link particles, $N_{x,y}=c^{\dagger}_{y} c_{y} + c^{\dagger}_{x} c_{x} =N$.  Hence, $\left[H,N_{x,y}\right]=0$ which introduces a second and independent local constraint in the Hilbert space.

\begin{figure}
\includegraphics[width=0.5\textwidth]{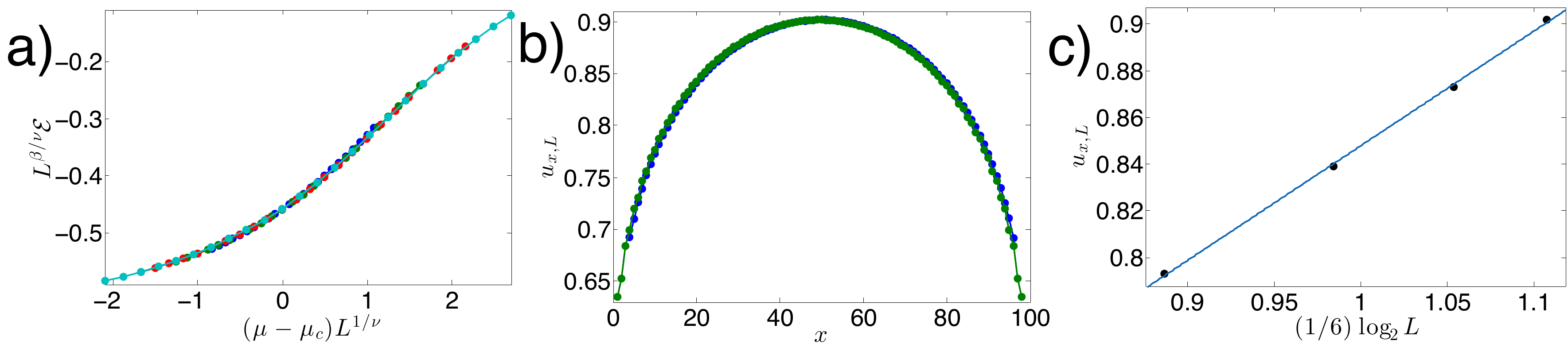}
\caption{Results  for  the  model  with  spin 1  on  the  links:  a)  Electric  flux $\mathcal{E}$ for $L=\{ 40,60,80,100 \}$, $g^{2}/2=1$ and $ \epsilon = 1/2$ with an estimate of the critical exponents $\nu \sim 1$ and $\beta \sim 1/8$ where the overlap among the different curves is maximal.  b) Uniform part of the entanglement entropy (green plot, first order approximation, i.e. $u_{x,L} =\frac{1}{2} \left( u_{x,L} + u_{x+1,L} \right)$, and blue plot third order approximation). c) Fit to $u_{x,L}=\frac{c}{6} \log{ \left[ 2 L \pi \sin{ \left( \pi x L \right)} \right] }+ a$, where $c= 0.49 \pm 0.04$.  Both, critical exponents and central charge are consistent with the Ising universality class at the phase transition taken from\cite{rico2014tensor}. }
\label{app5}
\end{figure}

More concretely, in a fermionic Schwinger representation of a non-Abelian $U(N)$ quantum link model, the gauge operators $U^{ij}_{x,y}$ that live on the links $\langle x, y \rangle$ of a $d$-dimensional lattice, with color indices $i,j$ are expressed as a bilinear of fermionic operators, $U^{ij}_{x,y}=c^{i}_{x}c^{j \dagger}_{y}$. In this link representation, the number of fermions per link is a constant of motion $N_{x,y}=\sum_{i} c^{i \dagger}_{y} c^{i}_{y} + c^{i \dagger}_{x} c^{i}_{x} =N$. In models with matter, at every vertex $x$ of the lattice, there is a set of fermionic modes $\psi^{i}_{x}$ with color index $i$.

The left and right generators of the $SU(N)$ symmetry are defined as $L^{a}_{x,y}=\sum_{i,j}  c^{i\dagger}_{x} \lambda^{a}_{i,j} c^{j}_{x}$ and $R^{a}_{x,y}=\sum_{i,j}  c^{i\dagger}_{y} \lambda^{a}_{i,j} c^{j}_{y}$, with $\lambda^{a}_{i,j}$ the group structure constants. Hence, the non-Abelian generators of the gauge symmetry are given by $G^{a}_{x} = \sum_{i,j} \psi^{i\dagger}_{x}  \lambda^{a}_{i,j} \psi^{j}_{x}+ \sum_{\hat{k}} \left[ L^{a}_{x,x+\hat{k}} + R^{a}_{x-\hat{k},x}  \right]$, with $\hat{k}$ the different directions in the  lattice. There  are  also  similar  expressions  for the Abelian part of the group $G_{x}$. 

The ``physical'' Hilbert subspace is defined as the one that is annihilated by every generator, i.e., $G_{x} |\text{phys}\rangle = G^{a}_{x} |\text{phys}\rangle= 0 ~ \forall x,a$. A particular feature of quantum link models is that, these operators being of bosonic nature (they are bilinear combinations of fermionic operators), the spatial overlap between operators at different vertices $x$ or $y$ is zero, i.e., $G^{a}_{x} G^{b}_{y}=0 , \forall a,b \text{ and } x\neq y$, even between nearest-neighbours.  In this way, (i) the gauge invariant Hilbert space (or Gauss' law) is fixed by a projection, which is defined locally $A\left[ s_{x} \right]$ on the ``physical'' subspace $\{ |s_{x} \rangle \}$ with $A\left[ s \right]_{ n_{c}, n_{\psi}}=\langle s | n^{i}_{c},n^{j}_{\psi} \rangle$, where $n^{i}_{c}$, $n^{j}_{\psi}$ is some configuration of occupations of fermionic modes $c^{i}$ and $\psi^{j}$.

Finally, (ii) the second gauge symmetry is controlled by the fermionic number on the link, which is ensured by the product of the nearest-neighbour projectors $A\left[ s_{x} \right]$ being non-zero only when $N= \sum_{i} n^{i}_{c,y}+ n^{i}_{c,x}$.

The $U(1)$ gauge invariant model in $(1+1)$ dimensions is defined by the Hamiltonian,
\begin{equation}
\begin{split}
H = &\frac{g^{2}}{2} \sum_{x} \left[ E_{x,x+1} - \left( -1 \right)^{x} E_{0} \right]^{2} + \mu \sum_{x} \left( -1 \right)^{x} \psi^{\dagger}_{x} \psi_{x} \\
&- \epsilon \sum_{x} \psi^{\dagger}_{x} U_{x, x+1} \psi_{x+1} + \text{H.c.},
\end{split}
\end{equation}
where $\psi_{x}$ are spin-less fermionic operators with staggered mass term $\mu$ living on the vertices of the one-dimensional lattice. The bosonic operators $E_{x,x+1}$ and $U_{x,x+1}$, the electric and gauge fields, live on the links of the one-dimensional lattice.

The Hamiltonian is invariant under local $U(1)$ symmetry transformations, and also it is invariant under the discrete parity transformation $P$ and charge conjugation $C$. The total electric flux, $\mathcal{E}= \sum_{x} \langle E_{x,x+1} \rangle / L$ is the order parameter and locates the transition. It is zero in the disordered phase, non-zero in the ordered phase, and changes sign under the $C$ or $P$ symmetry, i.e., $^{P}\mathcal{E}=^{C}\mathcal{E}=-\mathcal{E}$.

In this framework, in \cite{rico2014tensor} the phase diagram of $(1+1)$D quantum link version of the Schwinger model is characterised in an external classical background electric field: the quantum phase transition from a charge and parity ordered phase with non-zero electric flux to a disordered one with a net zero electric flux configuration is described by the Ising universality class (see Fig.~\ref{app5}). The thermodynamical properties and phase diagram of a one-dimensional $U(1)$ quantum link model are characterised, concluding that the model with half-integer link representation has the same physical properties as the model with integer link representation in a classical background electric field $E_{0}= \frac{1}{2}$. 

\subsubsection{Tensor Networks for Lattice Gauge Theories with continuous groups\cite{tagliacozzo2014tensor}}

\begin{figure}
\begin{center}
\includegraphics[width=0.3\textwidth]{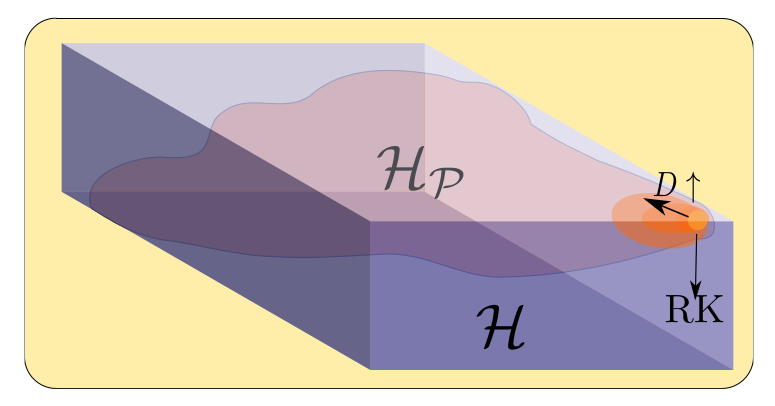}
\caption{The Hilbert space ${\cal H}$ of a quantum many body system (represented here by a 3D box) is exponentially large, since it is the tensor product of the Hilbert spaces of the constituents. Gauge symmetry allows to identify a smaller space, called the physical Hilbert space ${\cal H}_P$. This is the subspace spanned by those states that fulfill all the local constraints imposed by the gauge symmetry and is represented by a membrane inside $\cal{H}$.}
\label{pHs}
\end{center}
\end{figure}

The main difference between lattice gauge theories and generic many-body theories is that they require to work on an artificially enlarged Hilbert space, where the action of the group that generates the local invariance can be defined. The physical Hilbert space \cite{buividovich_entanglement_2008} is then embedded into the tensor product Hilbert space of the constituents by restricting it to those states that fulfill the Gauss law, that is to those states that are gauge invariant (see Fig.~\ref{pHs} for a graphical description). A generic gauge transformation is built out of local operators $A_s(g)$ that represent the local rotation at site $s$ corresponding to a certain element of the group $g$. The \emph{physical Hilbert space} (or gauge invariant Hilbert space) ${\cal H}_p$ is defined as the space spanned by all those states that are invariant under all  $A_s(g)$,   
\begin{align}
{\cal H}_p \equiv & \left\{ \ket{\phi} \in \mathbb{C}(G)^L, \right. \nonumber\\ 
&\left. A_s(g)\ket{\phi} = \ket{\phi} \quad \forall s \in \Lambda, g\in G \right\}, \label{eq:gauge}
\end{align}
where $s$ are the sites of the lattice $\Lambda$, $L$ is the number of links, and $g$ is an arbitrary group element.

In \cite{Tagliacozzo2011} the group algebra $\mathbb{C}(G)$ is considered as the local Hilbert space, as suggested in the original Hamiltonian description of lattice gauge theories \cite{Kogut1975,creutz_gauge_1977}. In \cite{Tagliacozzo2011} by exploiting the locality of  the operators $A_s(g)$ and the fact that they mutually commute, it is shown that the projection onto ${\cal H}_p$ is compatible with a tensor network structure. In particular, the projector is built as hierarchical tensor networks such as the MERA \cite{vidal_entanglement_2007} and the Tree Tensor Network \cite{tagliacozzo_simulation_2009}. While the MERA is computationally very demanding, a hybrid version of it has been built, that allows to construct the physical Hilbert space by using a MERA and then use a Tree Tensor network on the physical Hilbert space as a variational ansatz. In the same paper, it is also highlighted how the construction of a physical Hilbert space can be understood as a specific case of a duality such as the well known duality between the $Z(2)$ gauge theory and the Ising model \cite{savit_duality_1980}. 

\begin{figure}
\begin{center}
\includegraphics[width=0.3\textwidth]{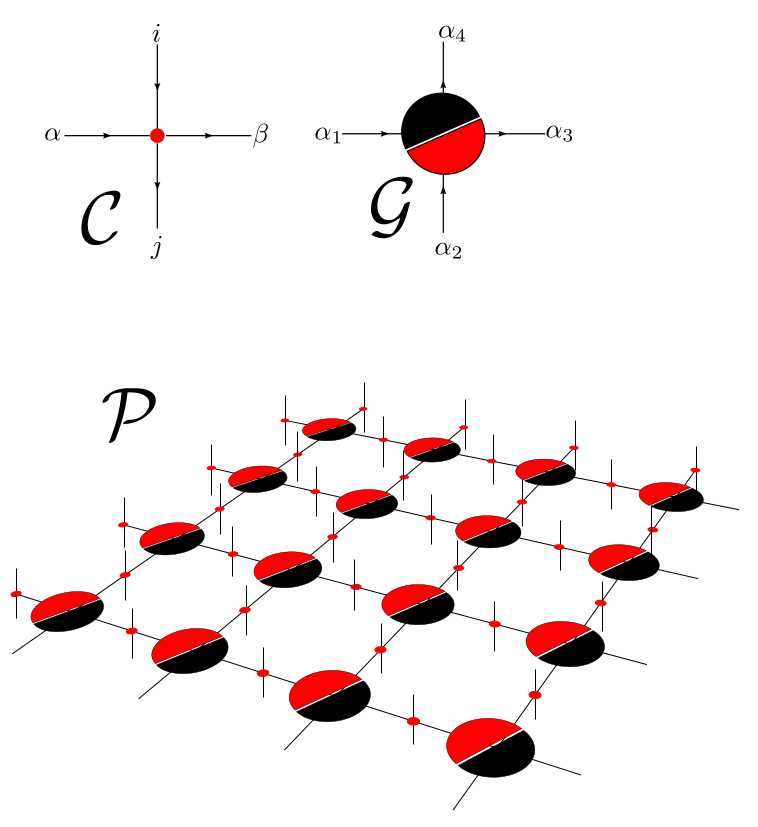}
\caption{The projector on the gauge invariant states defined through the contraction of the two tensors ${\cal C}$ that copy the physical Hilbert space onto the auxiliary Hilbert space and ${\cal G}$ that selects only configurations fulfilling the gauge invariance condition. The case of a $4\times 4$ square lattice with PBC is presented.}
\label{poh}
\end{center}
\end{figure}

The idea \cite{Tagliacozzo2011} is very flexible and general but  strongly relies on using $\mathbb{C}(G)$ as the local Hilbert space for every constituent. Since the  group algebra contains an orthogonal state for every distinct group element, $g$, the local Hilbert space  becomes infinite dimensional in the case of continuous groups such as e.g. $U(1)$ and $SU(N)$.

\begin{figure}
\begin{center}
\includegraphics[width=0.3\textwidth]{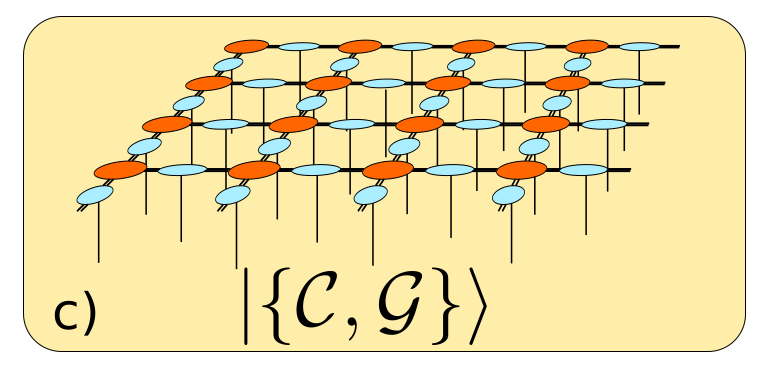}
\caption{Variational ansatz for gauge invariant states on a lattice of $4\times 4$ sites and periodic boundary conditions. The network contains one {\cal C} per link of the lattice, and one {\cal G} every site. The double lines connecting the tensors are used to remind that each of the elementary tensors has a double structure, one part dictated by the symmetry and the other one containing the actual variational parameters.}
\label{gip}
\end{center}
\end{figure}

Furthermore, the numerical results with iPEPS in the context of strongly correlated fermions in two dimensions were very promising \cite{corboz_competing_2014}, and thus it was decided to generalise the construction to PEPS tensor networks in \cite{tagliacozzo2014tensor}. There, it was understood that there is a unifying framework for all the Hamiltonian  formulations of lattice gauge theories that can be  based on a celebrated theorem in group theory, stating that the group algebra can be decomposed as the sum of all possible irreducible representations  $\mathbb{C}(G) = {\oplus_r}( r\otimes \bar{r} )$, where $r$ is an irreducible representation and $\bar{r}$ is its conjugate (see Figs.~\ref{poh} and \ref{gip}).  If the group is compact, the irreducible representations are finite dimensional. 

By decomposing $\mathbb{C}(G)$ into the direct sum of all the irreducible representations and truncating the sum to only a finite number of them, a formulation of LGT is obtained on finite dimensional Hilbert spaces. For Abelian gauge theories furthermore this procedure \cite{tagliacozzo2013optical} leads to the already known gauge magnets or link models \cite{horn_finite_1981,Orland1990,Chandrasekharan:1996ih}.

With this group theoretical picture in mind, it is very easy to directly construct both the projector onto the physical Hilbert space as a tensor network, and tensor network ansatz for states defined on it. The general recipe is given in \cite{tagliacozzo2014tensor}. Here for concreteness, the construction is shown for a two-dimensional square lattice. The tensor network is composed of two elementary tensors. The first one, ${\cal C}^{\alpha, j}_{i, \beta}$, a four-index tensor that has all elements zero except for those corresponding to $\alpha =  i =   \beta = j$. ${\cal C}$ is applied to each of the lattice sites and acts as a copy tensor that transfers the physical state of the links (encoded in the leg $i$) to the auxiliary legs $\alpha, \beta$.

The two auxiliary legs are introduced to bring the information to the two sites of the lattices that the link connects. Thus, the copy tensor ${\cal C}$ allows the decoupling of the gauge constraint at the two sites and to impose the Gauss law individually.

This operation is performed at each site by the second type of tensor, ${\cal G}^{\alpha_1 \alpha_2}_{ \alpha_3 \alpha_4}$, onto the trivial irreducible representation contained in the tensor product Hilbert space $\mathcal{H}_{\alpha_1}\otimes\mathcal{H}_{\alpha_2}\otimes\mathcal{H}_{\alpha_3}\otimes \mathcal{H}_{\alpha_4}$. The contraction of one $\cal{C}$ for every link with one $\cal{G}$ for every site gives rise to the desired projector onto ${\cal H}_p$ with the structure of a PEPS.

Alternatively, the projector onto ${\cal H}_p$ can be incorporated into a variational iPEPS ansatz for gauge invariant states, by promoting each of its tensor elements to a degeneracy tensor along the lines used to build symmetric tensor network states first introduced in \cite{singh_tensor_2009}. The gauge invariant tensor network can thus be interpreted as an iPEPS with a fixed tensor structure dictated by the gauge symmetry, where each element is again a tensor. These last tensors collect the variational parameters of the ansatz.

\subsection{Phase diagram and dynamical evolution of Lattice Gauge Theories with tensor networks}

Despite their impressive success, the standard LGT numerical calculations based on Monte Carlo sampling are of limited use for scenarios that involve a sign problem, as is the case when including a chemical potential. This constitutes a fundamental limitation for LQCD regarding the exploration of the QCD phase diagram at non-zero baryon density. In contrast, TNS methods do not suffer from the sign problem, which makes them a suitable alternative tool for exploring such problems, although, it is challenging to simulate high-dimensional systems.

In this section, it is shown how tensor network techniques could go beyond Monte Carlo calculations, in the sense, of being able to perform real-time calculations and phase diagrams with finite density of fermions. Examples of these achievements appear in \cite{banuls2016thermalmass,Banuls2015,pichler2016real,silvi2017finite,banuls2017density,banuls2017efficient,sala2018pos,tschirsich2019phase}.

\subsubsection{Real-time Dynamics in U(1) Lattice Gauge Theories with Tensor Networks\cite{pichler2016real}}

One of the main applications of tensor network methods is  real-time dynamics. Motivated by experimental proposals to realise quantum link model dynamics in optical lattice experiments, Ref.~\cite{pichler2016real} studied the quench dynamics taking place in quantum link models (QLMs) when starting from an initial product state (which is typically one of the simplest experimental protocols). In particular, the model under investigation was the U(1) QLM with $S=1$ variables as quantum links, whose dynamics is defined by the Hamiltonian
\begin{eqnarray}
 H&=&-t\sum_x \left[ \psi^{\dagger}_x U_{x,x+1}^{\dagger} \psi_{x+1} + \psi^{\dagger}_{x+1} U_{x,x+1} \psi_{x} \right] \nonumber\\
  & &+m\sum_x (-1)^x\psi^{\dagger}_x\psi_x
  +\frac{g^2}{2}\sum_xE^2_{x,x+1}
   \label{eq:Hamiltonian}
\end{eqnarray}
where $\psi_x$ defines staggered fermionic fields, $U_{x,x+1}=S^+_{x,x+1}$ and $E_{x,x+1}=S^z_{x,x+1}$ are quantum link spin variables; while the three Hamiltonian terms describe minimal coupling, mass, and electric field potential energy, respectively. 

\begin{figure}
\begin{center}
\resizebox{0.6\columnwidth}{!}{\includegraphics{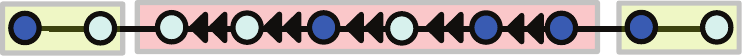}}
\resizebox{0.85\columnwidth}{!}{\includegraphics{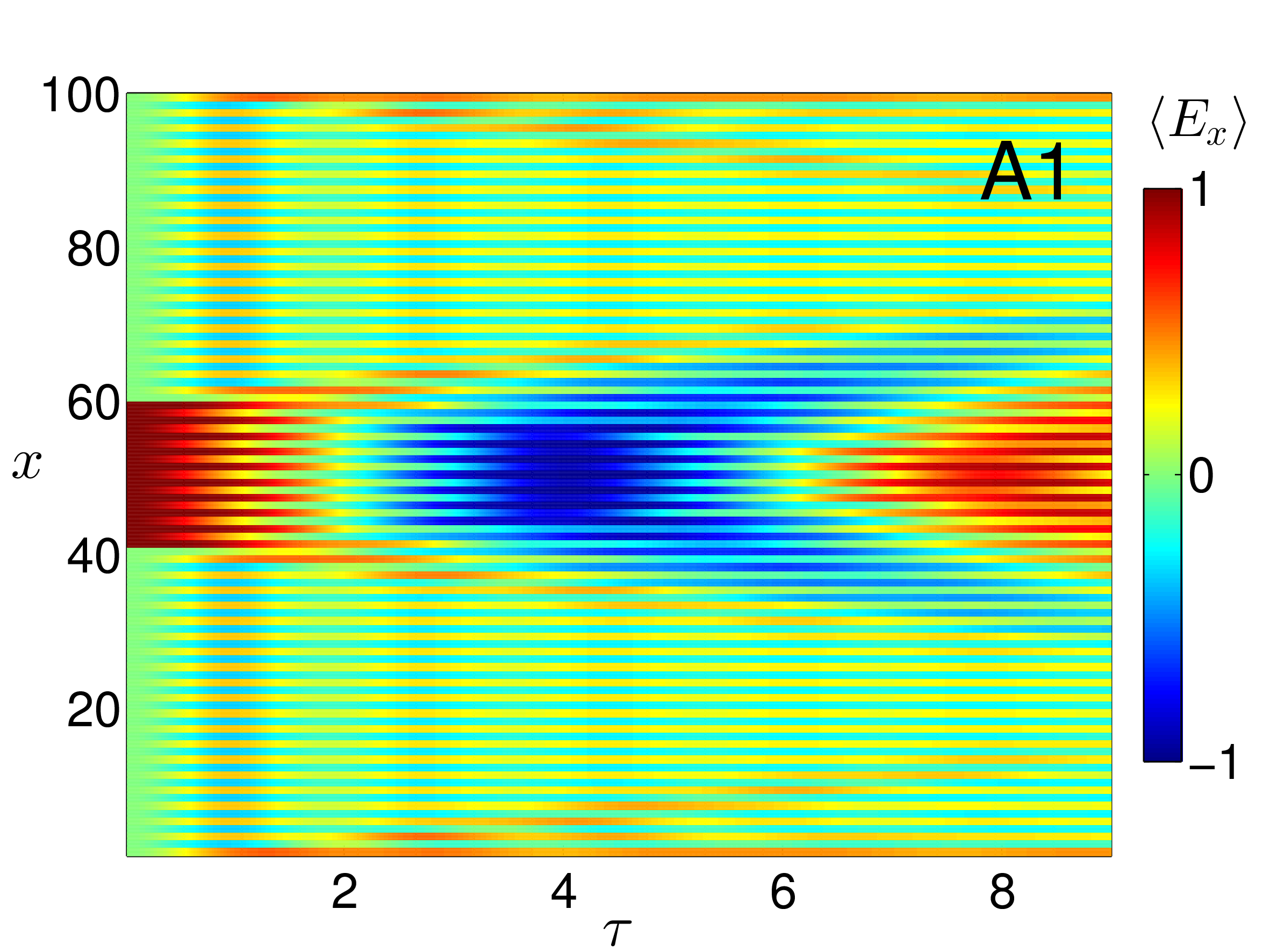}}
\end{center}

\caption{Real-time evolution of an electric flux string of length $L=20$ embedded in a larger lattice (of length $N=100$) in the vacuum state (the initial cartoon state is sketched on the top). The electric flux real-time evolution is shown for $m=0, g=0$, and the time is in units of $t=1$. Figure adapted from Ref.~\cite{pichler2016real}.}
\label{fig:QLM_SB}
\end{figure}

Several types of time evolutions were investigated. Fig.~\ref{fig:QLM_SB} presents the time evolution corresponding to string breaking dynamics: the initial state, schematically depicted on the top of the main panel, consists of a charge and anti-charge separated by a string of electric field (red region), and surrounded by the bare vacuum (light yellow). After quenching the Hamiltonian dynamics (in this specific instance, with $m=g=0$), the string between the two dynamical charges breaks (as indicated by a mean value of the electric field around 0 after a time $\tau t\simeq2$), and the charges spread in the vacuum region. For this specific parameter range, an anti-string is created at intermediate time-scales $\tau t\simeq 4$. Such string dynamics has also a rather clear signature in the entanglement pattern of the evolving state: in particular, it was shown how the speed of propagation of the particle wave-front extracted from the local value of the electric field was in very good agreement with the one extracted from the bipartite entanglement entropy.

With the same algorithm, it is possible to simulate the time evolution of a rather rich class of initial states up to intermediate times. As another example, Ref.~\cite{pichler2016real} also investigated the scattering taking place between cartoon meson states at strong coupling $g^2\gg 1$ (at smaller coupling, investigating scattering requires a careful initial state preparation, where a finite-momentum eigenstate is inserted ad hoc onto the MPS describing the dressed vacuum). Rather surprisingly, even these simplified scattering processes were found to generate a single bit of entanglement in a very precise manner. 

Closer to an atomic physics implementation with Rydberg atoms is the work \cite{notarnicola2019real} where different string dynamics are explored to infer information about the Schwinger model.

\subsubsection{Finite-density phase diagram of a (1+1)-d non-Abelian lattice gauge theory with tensor networks\cite{silvi2017finite}}

By means of the tools introduced in Sec.~\ref{LGTN}, in~\cite{silvi2017finite,Silvi2019} the authors have studied the finite-density phase diagram of a non-Abelian $SU(2)$ and $SU(3)$ lattice gauge theory in (1+1)-dimensions. 

In particular, they introduced a quantum link formulation of an $SU(2)$ gauge invariant model by means of the Hamiltonian
\begin{equation} 
\label{eq:model}
H =  H_{\mathrm{coupl}} + H_{\mathrm{free}} + H_{\mathrm{break}}
\end{equation}
where the first term introduces the coupling between gauge fields and matter, as
\begin{equation}
 H_{\mathrm{coupl}} = t \sum_{j = 1}^{L-1} \sum_{s,s' = \uparrow, \downarrow} c^{[M] \dagger}_{j,s}
 U_{j,j+1;s,s'} c^{[M]}_{j+1,s'} + \mbox{H.c.},
\end{equation}
where $c$ and $U$ are the matter field and parallel transport operators, $j \in \{1, \cdots ,L-1\}$ numbers the lattice sites, and $s \in \{\uparrow, \downarrow\}$. The second gauge term accounts for the energy of the free field
\begin{equation}
\begin{split}
 H_{\mathrm{free}} &= \frac{g_0^2}{2} \sum_{j = 1}^{L}  \left[ \vec{J}^{[R]}_{j-1,j} \right]^2 + \left[ \vec{J}^{[L]}_{j,j+1} \right]^2 \\
 & = 2 g_1^2
 \sum_{j = 1}^{L} \left( 1 - n^{[L]}_{j,\uparrow} n^{[L]}_{j,\downarrow} - n^{[R]}_{j+1,\uparrow} n^{[R]}_{j+1,\downarrow} \right),
\end{split}
\end{equation}
written in terms of the fermion occupation $n^{[\tau]}_{j,s} = c^{[\tau] \dagger}_{j,s} c^{[\tau]}_{j,s}$, where $g_1 = g_0 \sqrt{3/8}$. The last term $H_{\mathrm{break}}$ has to be introduced to resolve the undesired accidental local conservation of the number of fermions $\sum_{s = \uparrow, \downarrow} \left( n^{[R]}_{j,s} + n^{[M]}_{j,s} + n^{[L]}_{j,s} \right)$ around every site $j$, that results in a $U(2)$ theory. This last term breaks this invariance and thus, the final theory is an $SU(2)$ gauge invariant one~\cite{silvi2017finite}. 

Finally, by means of a finite-size scaling analysis of correlation functions, the study of the entanglement  entropy and fitting of the central charge of the corresponding conformal field theory, the authors present the rich finite-density phase diagram of the Hamiltonian~\eqref{eq:model}, as reported in Fig.~\ref{fig:phasedia}. In particular, they identify different phases, some of them appearing only at finite densities and supported also by some perturbative analysis for small couplings. At unit filling the system undergoes a phase transition from a meson superfluid, or meson BCS, state to a charge density wave via spontaneous chiral symmetry breaking. At filling two-thirds, a charge density wave of mesons spreading over neighbouring sites appears, while for all other fillings explored, the chiral symmetry is restored almost everywhere, and the meson superfluid becomes a simple liquid at strong couplings.

\begin{figure}
   \begin{overpic}[width = \columnwidth, unit=1pt]{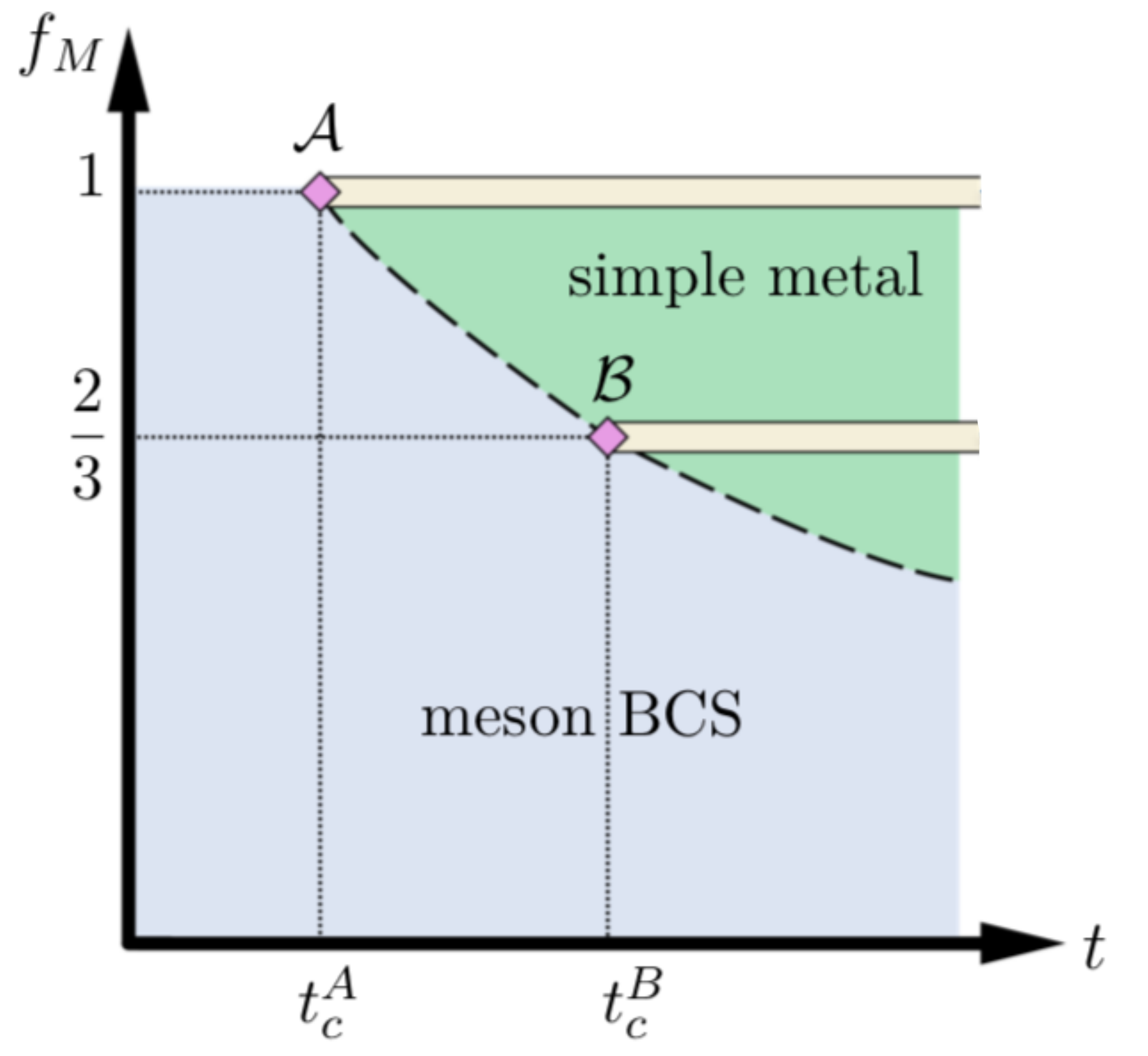}

   \end{overpic}
   \caption[width = \columnwidth]{  \label{fig:phasedia}
Phase diagram of the $SU(2)$ lattice gauge model in the quantum link formalism as a function of the matter-field coupling and the matter filling $( t, f_M)$. Two insulating phases appear at large coupling $t$ and $f_M = 1, 2/3$ embedded in a meson BCS and a simple liquid phase. Figure from~\cite{silvi2017finite}.}
\end{figure}

Very recently,  also a one-dimensional $SU(3)$ gauge theory has been studied with the same approach.  In~\cite{Silvi2019},  working on and extending the results reviewed in the previous paragraph, the authors present an $SU(3)$ gauge invariant model in the quantum link formulation, and perform an extended numerical analysis on the different phases of the model. For space reasons, the model Hamiltonian is not displayed here and the interested reader is referred to the original publication. However, the main results are: at filling $\nu = 3/2$, the Kogut-Susskind vacuum, a competition between a chiral and a dimer phase separated by a small gapless window has been reported. Elsewhere, only a baryonic liquid is found. The authors also studied the binding energies between excess quarks on top of the vacuum,  finding that a single baryon state (three excess quarks) is a strongly bound one, while two baryons (six quarks) weakly repel each other. The authors concluded that the studied theory -- differently from three dimensional QCD -- disfavours baryon aggregates, such as atomic nuclei.

\begin{figure}
   \begin{overpic}[width = \columnwidth, unit=1pt]{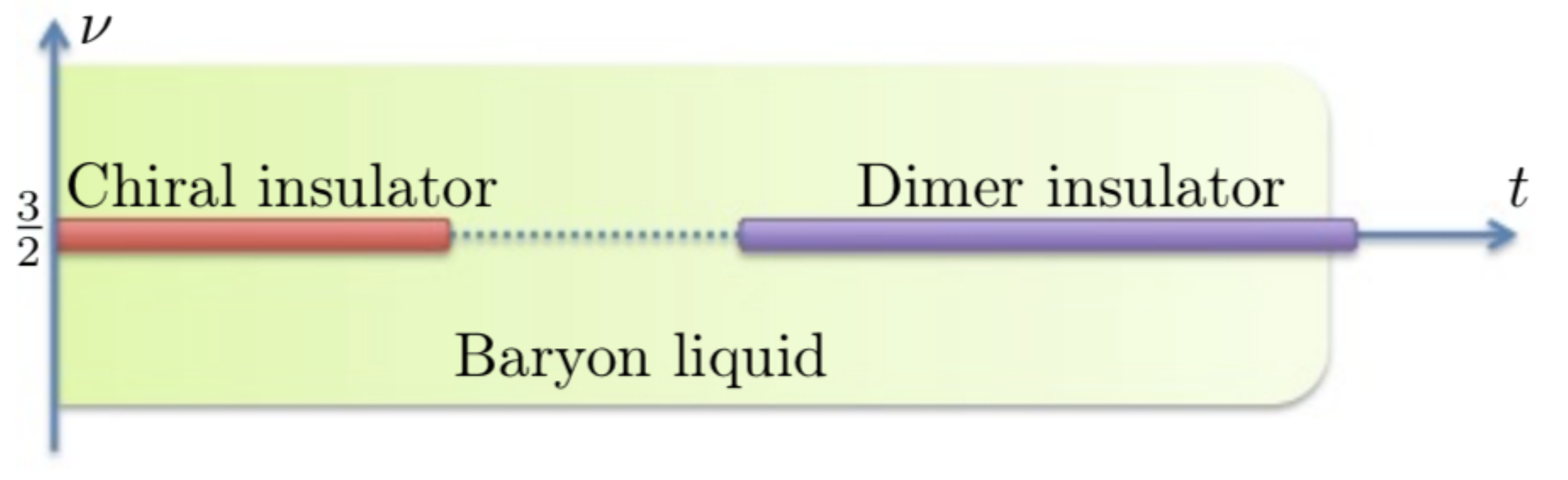}

   \end{overpic}
   \caption[width = \columnwidth]{  \label{fig:phasedia}
Phase diagram of the $SU(3)$ lattice gauge model in quantum link formalism as a function of the  space matter-field coupling $t$ and the matter filling $\nu$, for bare mass $m=0$. Figure from~\cite{Silvi2019}.}
\end{figure}

\subsubsection{Density Induced Phase Transitions in the Schwinger Model: A Study with Matrix Product States\cite{banuls2017density}}

\begin{figure}
\begin{minipage}[b]{.4\columnwidth}
\includegraphics[height=.9\columnwidth]{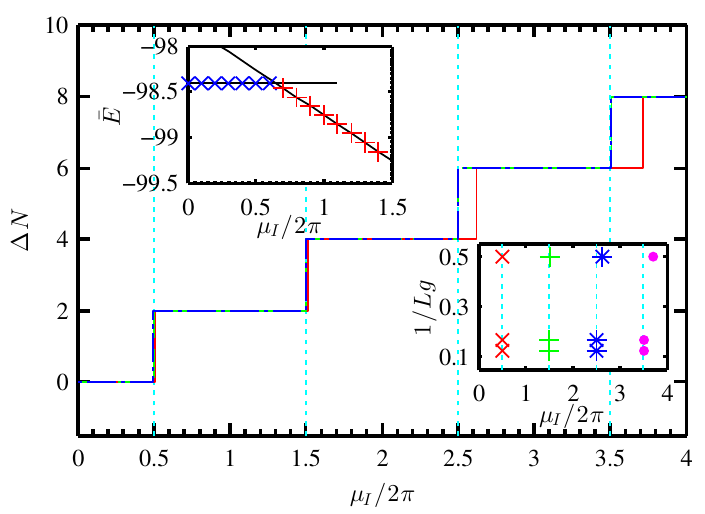}
\end{minipage}
\hspace{.1\columnwidth}
\begin{minipage}[b]{.4\columnwidth} 
\includegraphics[height=.9\columnwidth]{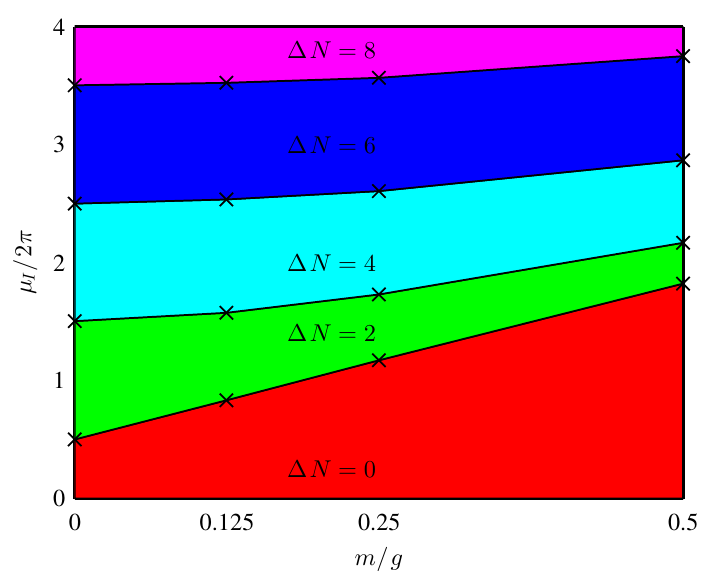}
\end{minipage}
\caption{(from \cite{banuls2017density})
Left: Continuum value, for massless fermions, of the imbalance $\Delta N$ in the ground state as a function of the rescaled isospin chemical potential, for increasing volumes 2 (red solid), 6 (green dashed) and 8 (blue dash-dotted line). The vertical lines correspond to the analytical prediction for the phase transitions. The lower inset shows explicitly the volume dependence of the successive transition points. Right: Phase diagram in the mass vs. isospin chemical potential plane for  volume $Lg=8$. The black crosses mark the computed data points, the different colours indicate the different phases.}
\label{fig:multiflavor}
\end{figure}

The potential of TNS methods to deal with scenarios where standard Monte Carlo techniques are plagued by the sign problem was explicitly demonstrated in \cite{banuls2017density} by studying the multi-flavour Schwinger model, in a regime where conventional Monte Carlo suffers from the sign problem. The Hamiltonian of the Schwinger model (\ref{eq:schwinger}) can be modified to include several fermionic flavours, with independent masses and chemical potential, that do not interact directly with each other but only through the gauge field. In the case of two-flavours with equal masses studied in \cite{banuls2017density} the model has an $SU(2)$ isospin symmetry between the flavours. For vanishing fermion mass and systems of fixed volume, the analytical results~\cite{Narayanan2012,lohmayer2013phase} demonstrate the existence of an infinite number of particle number sectors, characterised by the imbalance between the number of fermions of both flavours. The different phases are separated by first order phase transitions that occur at fixed and equally separated values of the (rescaled) isospin chemical potential, independent of the volume, so that the isospin number of the ground state varies in steps as a function of the chemical potential (see left panel of Fig. ~\ref{fig:multiflavor}). 

The numerical calculations in \cite{banuls2017density} used MPS with open boundary conditions, and followed the procedure described in \cite{Banuls2013}, with the exception that the lattices considered had constant volume, $Lg=N/\sqrt{x}$. Thus there was no need for a finite-size extrapolation of the lattice results ($N\to\infty$), but a sufficiently large physical volume $Lg$ had to be considered. The results reproduced with great accuracy the analytical predictions at zero mass, as shown in the left panel of Fig.~\ref{fig:multiflavor}. While the analytical results can only cope with massless fermions, the MPS calculation can be immediately extended to the massive case, for which no exact results exist. For varying fermion masses, the phase structure was observed to vary significantly. The location of the transitions for massive fermions depends on the volume, and the size of the steps is no longer constant. The right panel of Fig. ~\ref{fig:multiflavor} shows the mass vs. isospin phase diagram for volume $Lg=8$.

The MPS obtained for the ground state allows further investigation of its properties in the different phases. In particular, the spatial structure of the chiral condensate, which was studied in \cite{banuls2017density} and \cite{banuls2016multif}. While for $\Delta N=0$, the condensate is homogeneous, for non-vanishing isospin number it presents oscillations, with an amplitude close to the zero density condensate value and a wave-length that, for a given volume, decreases with the isospin number or imbalance, but decreases with $Lg$.

\subsubsection{Efficient Basis Formulation for (1+1)-Dimensional $SU(2)$ Lattice Gauge Theory: Spectral calculations with matrix product states\cite{banuls2017efficient}}

A non-Abelian gauge symmetry introduces one further step in complexity with respect to the Schwinger model, even in 1+1 dimensions. The simplest case, a continuum $SU(2)$ gauge theory involving two fermion colours, was studied numerically with MPS in \cite{banuls2017efficient}, using a lattice formulation and numerical analysis in the spirit of \cite{Banuls2013}.

The discrete Hamiltonian in the staggered fermion formulation reads  \cite{Kogut1975} 
\begin{eqnarray}
H=&\frac{1}{2a} \sum_{n=1}^{N-1}\sum_{\ell,\ell'=1}^2 \left({\phi_{n}^{\ell}}^\dagger U_{n}^{\ell\ell'}\phi_{n+1}^{\ell'}+\mathrm{H.c.}\right)
\nonumber \\
&+ m\sum_{n=1}^N\sum_{\ell=1}^2 (-1)^n{\phi_{n}^{\ell}}^\dagger\phi_{n}^{\ell} +\frac{ag^2}{2}\sum_{n=1}^{N-1} \mathbf{J}_{n}^2.
\label{eq:su2}
\end{eqnarray}
The link operators $U^{\ell\ell'}_n$ are $SU(2)$ matrices in the fundamental representation, and can be interpreted as rotation matrices. The Gauss law constraint is now non-Abelian, $G^\tau_m|\Psi\rangle=0$, $\forall m, \tau$, with generators $G^\tau_m =L^\tau_m-R^\tau_{m-1}-Q^\tau_m$, where $Q_m^\tau=\sum_{\ell=1}^2\frac{1}{2}{\phi_m^\ell}^\dagger\sigma^\tau_{\ell \ell'}\phi_m^{\ell'}$ are the components of the non-Abelian charge at site $m$ (if there are external charges, they should be added to $Q_m$). $L^\tau$ and $R^\tau$, $\tau\in\{x,y,z\}$, generate left and right gauge transformations on the link, and the colour-electric flux energy is $\mathbf{J}_{m}^2 = \sum_\tau L^\tau_mL^\tau_m = \sum_\tau R^\tau_mR^\tau_m$. The Hilbert space of each link is analogous to that of a quantum rotor, and its basis elements, for the $m$-th link, can be labeled  by the eigenvalues of $\mathbf{J}_{m}$, $L^{z}_m$ and $R^{z}_m$, as $ |j_m \ell_m \ell'_m\rangle$.

As in the case of the Schwinger model, it is possible to truncate the gauge degrees of freedom. This can be achieved in a gauge invariant manner~\cite{zohar2015quantum} and was applied to study the string breaking phenomenon in the discrete theory in~\cite{Kuehn2015}. However, in order to attain precise results that permit the extraction of a continuum limit, it is convenient to work in a more efficient basis, in which gauge degrees of freedom are integrated out. A first step to reduce the number of spurious variables is the color neutral basis introduced in ~\cite{Hamer1977,Hamer1982a}. In \cite{banuls2017efficient}, building on that construction, a new formulation of the model on the physical subspace is introduced in which the gauge degrees of freedom are completely integrated out. Nevertheless, it is still possible to truncate the maximum colour-electric flux at a finite value $j_{\mathrm{max}}$ in a gauge invariant manner, and analyse the effect of this truncation on the physics of the model. This is relevant, for instance, to understand how to extract continuum quantities from a potential quantum simulation of the truncated theory. To this end, different quantities were computed and extrapolated to the continuum, including the ground state energy density, the entanglement entropy in the ground state, the vector mass gap and its critical exponent for values of the maximum colour-electric flux $j_{\mathrm{max}}=1/2,1,3/2,2$. 

\begin{figure}
\begin{minipage}[c]{.48\columnwidth}
\includegraphics[width=.9\columnwidth]{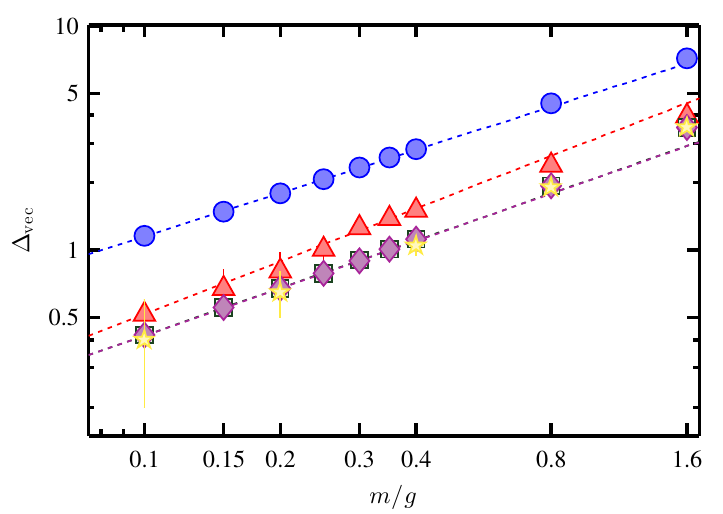} 
\end{minipage}
\begin{minipage}[c]{.48\columnwidth}
\includegraphics[width=.9\columnwidth]{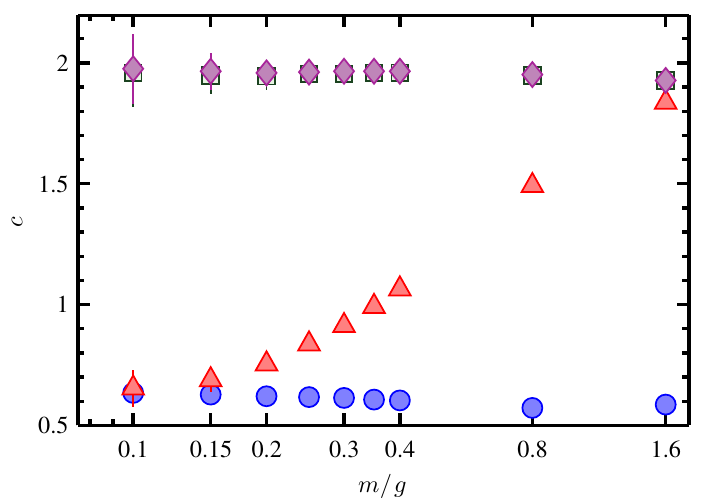}
\end{minipage}
\caption{(from \cite{banuls2017efficient})
Left: Final value of the vector mass gap (after continuum extrapolation) as a function of the fermion mass using truncations $j_{\mathrm{max}}=1$ (red triangles), $3/2$ (green squares) and $2$ (magenta diamonds), with the yellow stars showing results from a strong coupling expansion \cite{Hamer1982a}. The blue circles correspond to $j_{\mathrm{max}}=1/2$, although the continuum estimation is not reliable in that case. The dotted lines represent the best fit of the form $\gamma (m/g)^\nu$. Right: Central charges extracted from the scaling of the entanglement entropy (see panel d in Fig. \ref{fig:su2_entropy}) for different fermion masses and $j_{\mathrm{max}}=1/2$ (blue circles), $1$ (red triangles), $3/2$ (green squares), and $2$(magenta diamonds). }
\label{fig:su2_cont}
\end{figure}

The results demonstrated that, while a small truncation is enough to obtain the correct continuum extrapolated  ground state energy density, the situation varies for the mass gap. In particular (see left panel of  Fig.\ref{fig:su2_cont}), if the truncation is too drastic, it fails to produce a reliable extrapolation, and only $j_{\mathrm{max}}>1$ allowed for precise estimations of the vector mass, and the extraction of a critical exponent as the gap closes for massless fermions.

\begin{figure}
\begin{minipage}[c]{\columnwidth} 
\includegraphics[width=.95\columnwidth]{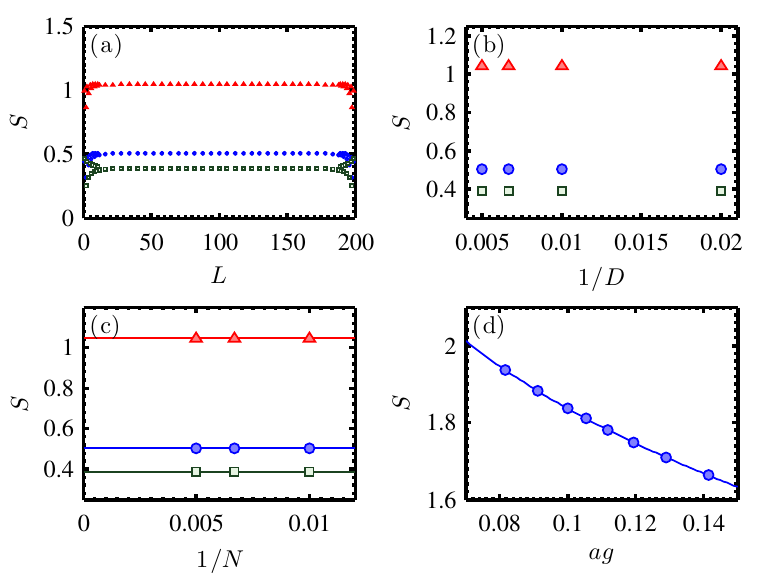}\\
\end{minipage}
\caption{(from \cite{banuls2017efficient}) Ground state entropy in the $SU(2)$ LGT. In (a) the different contributions (distillable in blue circles) are shown for the entanglement entropy of a chain of $200$ sites, for fermion mass $m/g=0.8$ and using $j_{\mathrm{max}}=2$. Panels (b) and (c) show the extrapolations in bond dimension and system size, and panel (d) shows the continuum limit for the total entropy,  exhibiting the divergent $\log ag$ term.}
\label{fig:su2_entropy}
\end{figure}

Particularly interesting is the study of the entanglement entropy of the vacuum, which can be easily computed from the MPS ansatz, as already demonstrated in \cite{Buyens2015} for the Schwinger model. The gauge constraints are not local with respect to a straightforward bipartition of the Hilbert space~\cite{ghosh2015entanglement,van2016entanglement}, and different contributions to the entropy can be identified, of which only one is distillable, while the others respond only to the gauge invariant structure of the state. In \cite{banuls2017efficient}, these different contributions were computed and their scaling analysed (see Fig. \ref{fig:su2_entropy} a). For a massive relativistic QFT, as is the case here for non-vanishing fermion mass, the total entanglement entropy is predicted to diverge as $S=(c/6)\log_2(\xi/a)$ \cite{calabrese2004entanglement}, where $c$  is the central charge of the conformal field theory describing the system at the critical point, in the $SU(2)$ case, $c=2$. This effect could also be studied from the MPS data (Fig. \ref{fig:su2_entropy} d), and it was also found to be sensitive to the truncation, leading to the conclusion that truncations of $j_{\mathrm{max}}\leq 1$  would not recover the continuum theory in the limit of vanishing lattice spacing, as shown by Fig. \ref{fig:su2_cont}.

\subsubsection{Gaussian states for the variational study of (1+1)-dimensional lattice gauge models \cite{sala2018pos}}

Gaussian states~\cite{Bravyi2005,Kraus2009,Peschel2009}, whose density matrix can be expressed as the exponential of a quadratic function in the creation and annihilation operators, are widely used to describe fermionic as well as bosonic quantum many-body systems. They fulfill Wick's theorem and thus can be completely described in terms of a covariance matrix, with a dimension that scales only linearly in the system size. This provides a very efficient representation of the quantum many-body state, which can be used as a variational ansatz. But in systems with interacting bosons and fermions, as is the case for lattice gauge theories with gauge and matter degrees of freedom, Gaussian states present a severe limitation, since they cannot describe any correlations between the two types of fields. 

However, as was recently shown \cite{Shi2018}, generalised ans\"atze that combine non-Gaussian unitary transformations with a Gaussian ansatz in the suitable basis, can be successfully used to approximate static and dynamic properties of systems containing fermions and bosons, also in higher dimensions. In \cite{sala2018pos} this approach was shown to work for (1+1)-dimensional lattice gauge theories. More explicitly, a set of unitary transformations was introduced that completely disentangle the gauge and matter degrees of freedom for any gauge symmetry given by a compact Lie group and a unitary representation. This allows for new ways of studying these lattice gauge theories.

The particular cases of U(1) and $SU(2)$ were explicitly studied in  \cite{sala2018pos}. For U(1), the resulting Hamiltonian is the one proposed by Hamer, Weihong, and Oitmaa in \cite{Hamer1997}, which has been used for numerical calculations \cite{Hamer1997,Banuls2013,Banuls2015}, and has been experimentally implemented with trapped ions in a pioneering quantum simulation \cite{martinez2016real}. The general character of the decoupling transformations thus provides alternative formulations of other lattice gauge theories which can be suitable for experimental implementation with the advantage of being directly defined in the physical space and not requiring the explicit realisation of any gauge degrees of freedom.
 
\begin{figure}
\begin{minipage}[c]{.45\columnwidth}
\includegraphics[width=.9\columnwidth]{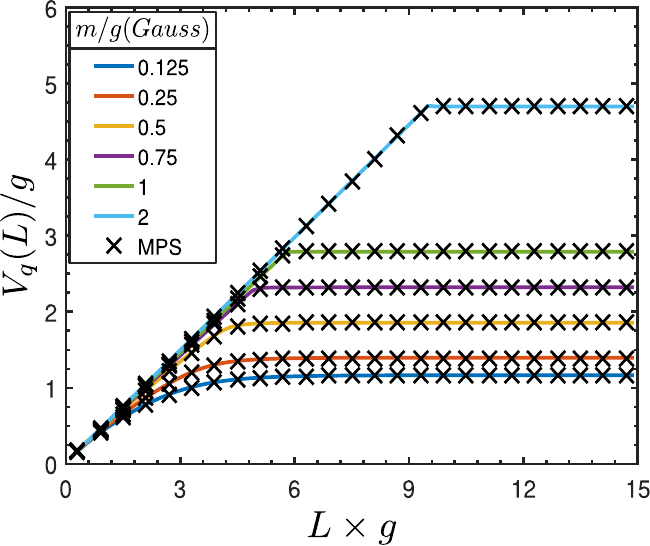}
\end{minipage}
\begin{minipage}[c]{.45\columnwidth} 
\includegraphics[width=.95\columnwidth]{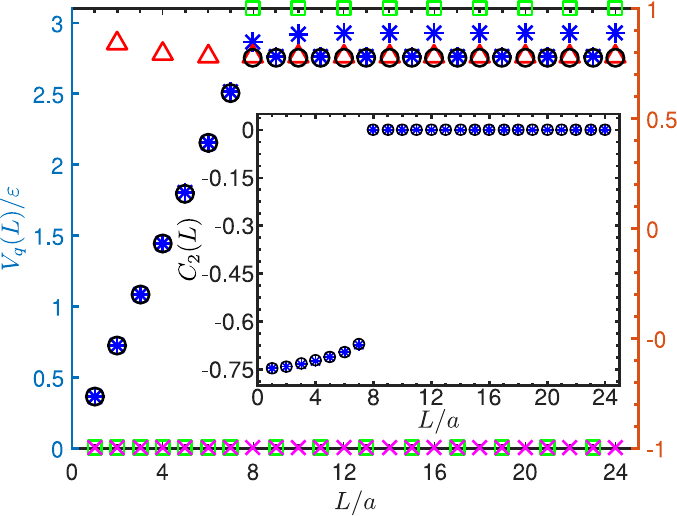}
\end{minipage}\\
\begin{minipage}[c]{.45\columnwidth}
\includegraphics[width=.9\columnwidth]{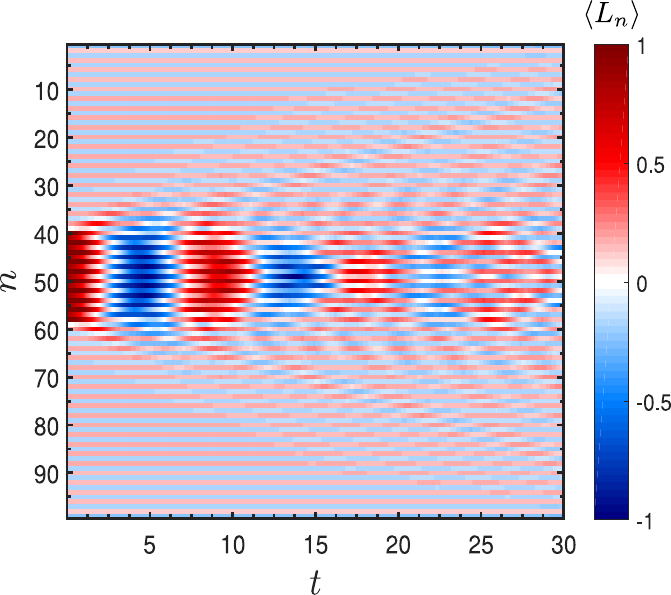}
\end{minipage}
\begin{minipage}[c]{.45\columnwidth} 
\hfill
\includegraphics[width=.95\columnwidth]{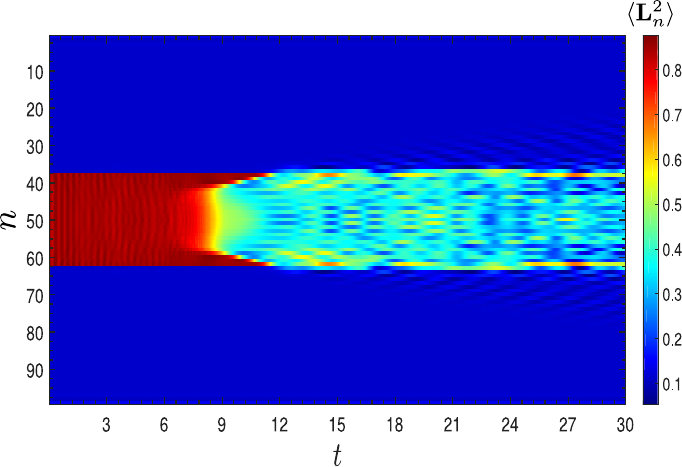}
\end{minipage}
\caption{ String breaking in U(1) (left column) and $SU(2)$ (right column) LGT (from \cite{sala2018pos}). The upper row of plots shows the static potential between two external charges as a function of the distance, at fixed lattice spacing, for different values of the fermion mass, computed with MPS (crosses on the left, circles on the right) and a Gaussian ansatz (solid lines on the left, triangles and asterisks on the right). The left plot corresponds to two unit charges in the U(1) LGT, and several fermion masses, while the right plot shows the corresponding calculation in the $SU(2)$ case for a pair of external static charges carrying $s=1/2$. The lower row shows the real-time evolution of a string created on top of the interacting vacuum. On the left, for the U(1) case, the edges are dynamical charges and can propagate, while on the right, for SU(2), they are static. In both cases, the color indicates the chromo-electric flux on each link as a function of time.}
\label{fig:gaussian}
\end{figure}

With a numerical perspective, \cite{sala2018pos} addressed the decoupled formulation using a Gaussian variational ansatz, and used it to investigate static and dynamical aspects of string breaking in the Abelian U(1) and non-Abelian $SU(2)$ gauge models. In the U(1) case, the formulation directly allows the study of the real-time string breaking phenomenon in the presence of static external or dynamic charge. In the $SU(2)$ case, only the case of static external charges was studied, using another unitary transformation that decouples them from the dynamical fermions. The Gaussian approach was capable of capturing the essential features of the phenomenon, both the static properties and the out-of-equilibrium dynamics (see Fig. \ref{fig:gaussian}). The results showed excellent agreement with previous TNS simulations over a broad range of the parameter space, despite the number of variational parameters in the Gaussian ansatz being much smaller. The approach could be extended and used for further non-equilibrium simulations of other LGTs.
 
\subsubsection{Thermal evolution of the Schwinger model \cite{banuls2016thermalmass,Banuls2015}}

\begin{figure}
\begin{minipage}[b]{.4\columnwidth}
\includegraphics[height=.9\columnwidth]{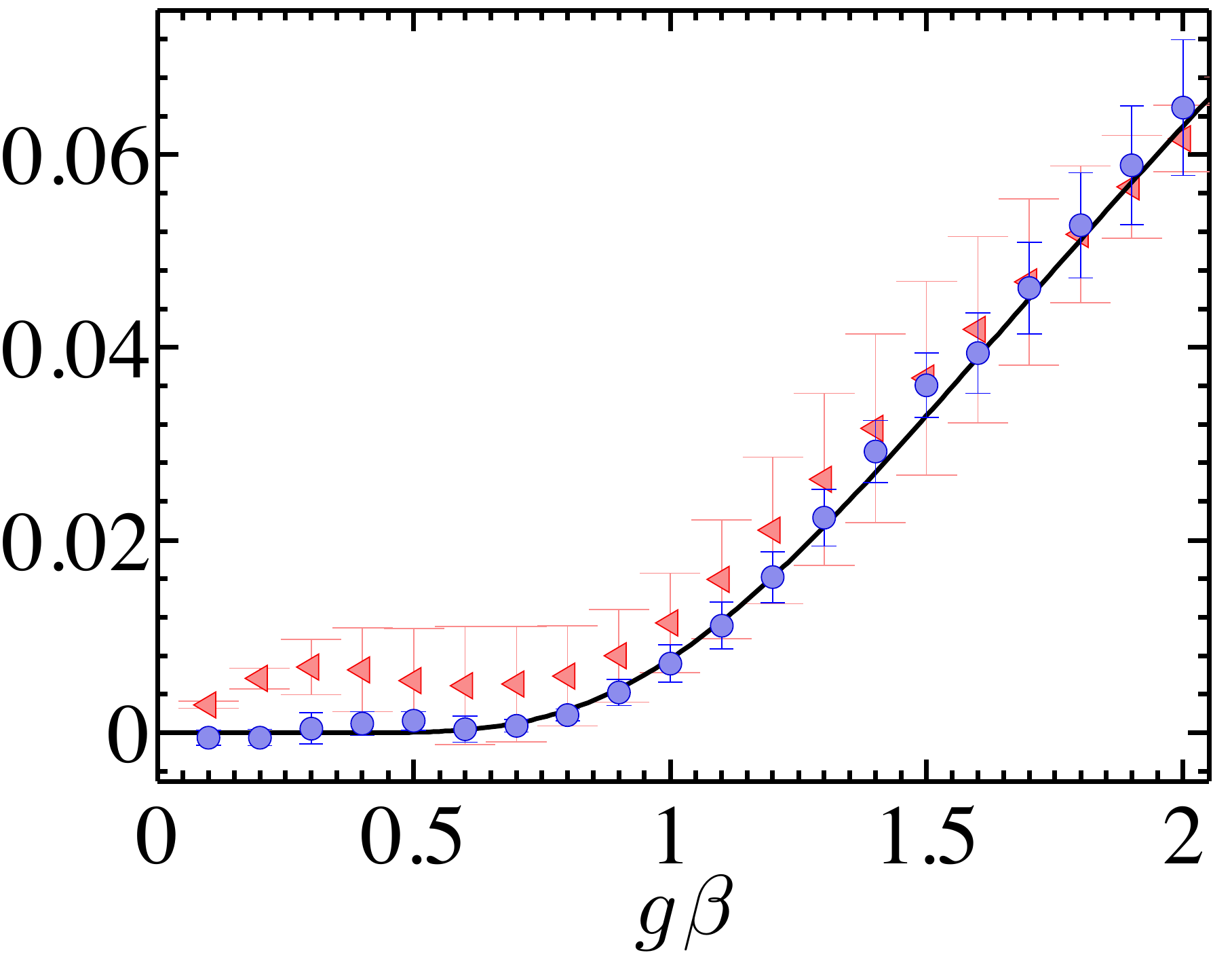}
\end{minipage}
\hspace{.02\columnwidth}
\begin{minipage}[b]{.4\columnwidth} 
\includegraphics[height=1.05\columnwidth]{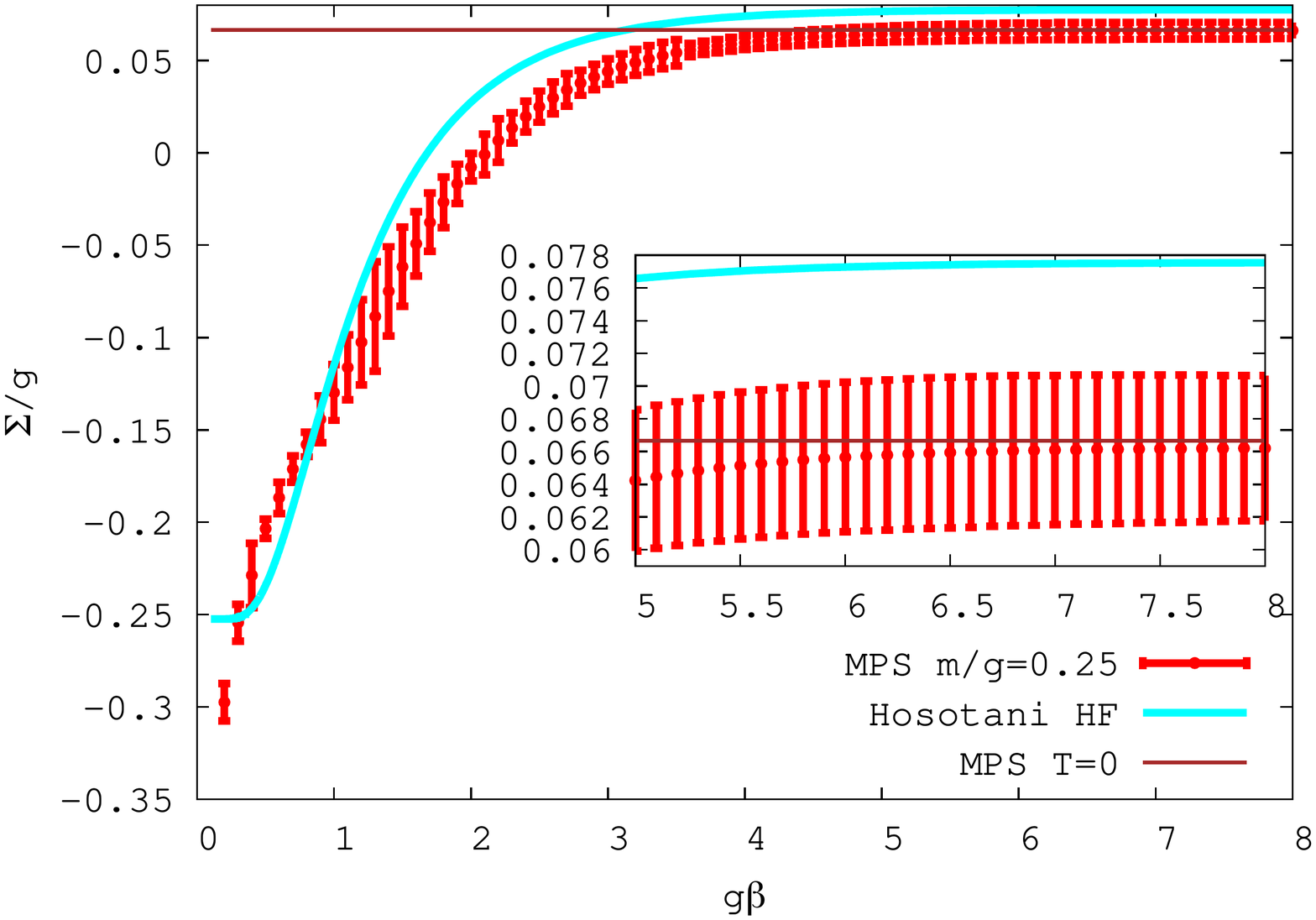}
\end{minipage}
\caption{
Chiral condensate in thermal equilibrium in the continuum as a function of temperature for massless (left, from \cite{Banuls2015}) and massive fermions (right, from \cite{banuls2016thermalmass}). The solid line on the left shows the analytical prediction for the restoration of the chiral symmetry, while the data points show results obtained with MPO using two different approximations for the evolution operators. For the massive case on the right the horizontal line shows the value at zero temperature (computed numerically with MPS) and the solid blue line the approximation by Hosotani and Rodriguez \cite{Hosotani1998} (which is exact only at very small masses).}
\label{fig:thermal}
\end{figure}

TNS ans\"atze can also describe density operators, in particular thermal equilibrium states, and can therefore be used to study the behaviour of a LGT at finite temperature. This approach was followed in  \cite{Banuls2015,banuls2016thermalmass}, which employed a purification ansatz \cite{Verstraete2004a}  to represent the thermal equilibrium state as a matrix product operator (MPO). At infinite temperature, $g\beta=0$, the thermal equilibrium state is maximally mixed, and has an exact representation as a simple MPO. By applying imaginary time evolution on this MPO \cite{Verstraete2004a,Zwolak2004}, a whole range of temperatures can be studied. A relevant observable to analyse in the case of the Schwinger model is again the chiral condensate (see Fig.\ref{fig:thermal}). In the massless case, the chiral symmetry is broken (due to an anomaly), and the condensate has a non-zero value in the ground state. The symmetry is smoothly restored at infinite temperature, as demonstrated analytically in \cite{Sachs:1991en}.

In  \cite{Banuls2015,banuls2016thermalmass} a finite size MPO ansatz was used in the physical subspace, i.e., after integrating out the gauge degrees of freedom. The imaginary time (thermal) evolution was applied in discrete steps, making use of a Suzuki-Trotter approximation, and after each step a variational optimisation was used to truncate the bond dimension of the ansatz. In the physical subspace, the long-range interactions among charges pose a problem for standard TN approximations of the exponential evolution operators. Two alternatives were considered to apply the discrete steps as MPO. In one of them, the long-range exponential was approximated by a Taylor expansion. In the other, an exact MPO expression of the exponential of the long-range term was used, taking advantage of the fact that it is diagonal in the occupation basis. For the application to be efficient, a truncation was introduced in this MPO, which was equivalent to a cut-off in the electric flux that any link in the lattice can carry. The second approach was found to be more efficient, and a small cut-off was sufficient for convergence over the whole range of parameters explored. With this method, the chiral condensate in the continuum was evaluated from inverse temperature $g\beta=0$ to $g\beta \sim \mathcal{O}(10)$ both for massless and massive fermions. For non-vanishing fermion masses, the condensate diverges in the continuum, and a renormalisation scheme has to be adopted that subtracts the divergence. In \cite{banuls2016thermalmass} this was achieved by subtracting the value of the condensate at zero temperature in the non-interacting case, after the finite-size extrapolation.

The continuum limit was performed for each value of the temperature in a manner similar to \cite{Banuls2013}. The width of the time step in the Trotter approximation introduced an additional source of error, and required an additional extrapolation. However, the form of the step width extrapolation is given by the order of the Suzuki-Trotter approximation, and this step did not affect the final precision, which again turned out to be controlled by the continuum limit.

All in all, the technique allowed for reliable extrapolations in bond dimension, step width, system size and lattice spacing, with a systematic estimation and control of all error sources involved in the calculation. Notably, although the large temperature regime of the lattice model is easier to describe by a MPO, the lattice effects are also more important, which resulted in larger errors after the continuum extrapolation. As the temperature decreases, the errors from the lattice effects become less relevant, but the truncation errors from the MPO approximation accumulate, so that they dominate the low-temperature regime. In conclusion, these results further validate the TNS approach as a tool to investigate the phase diagram of a quantum gauge theory.

\subsubsection{Finite temperature and real-time simulation of the Schwinger model \cite{buyens2014matrix,Buyens2016,Buyens2017}} 

\emph{Finite temperature.} In \cite{Buyens2016} different aspects of the finite temperature physics of the Schwinger model were studied. For different temperatures the appropriate gauge invariant Gibbs states were obtained from imaginary time evolution on a purification of the identity operator. Among the different results here the computation of the temperature dependent renormalised chiral condensate is quoted, in agreement with the analytical result for $m/g=0$ and the numerical results of \cite{Banuls2015} for $m/g\neq 0$. Furthermore, the study of the temperature-dependence of the energy density in an electric background field allowed for the study of an effective deconfinement transition. For half-integer background fields the expected restoration of the $C$ symmetry at non-zero temperature was also verified.     

\begin{figure}[t]
\begin{subfigure}[b]{.24\textwidth}
\includegraphics[width=\textwidth]{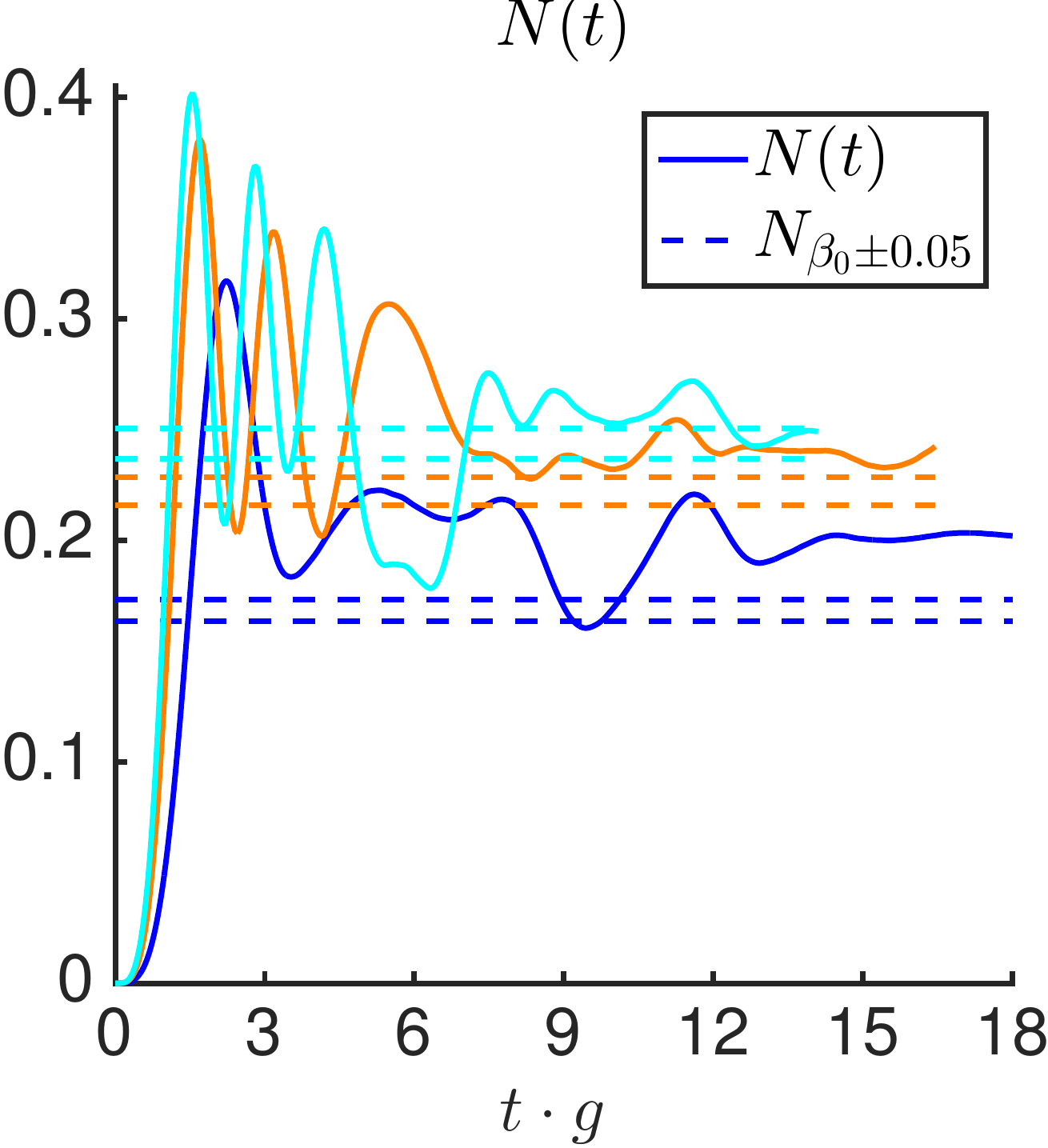}
\caption{}
\end{subfigure}\hfill
\begin{subfigure}[b]{.24\textwidth}
\includegraphics[width=\textwidth]{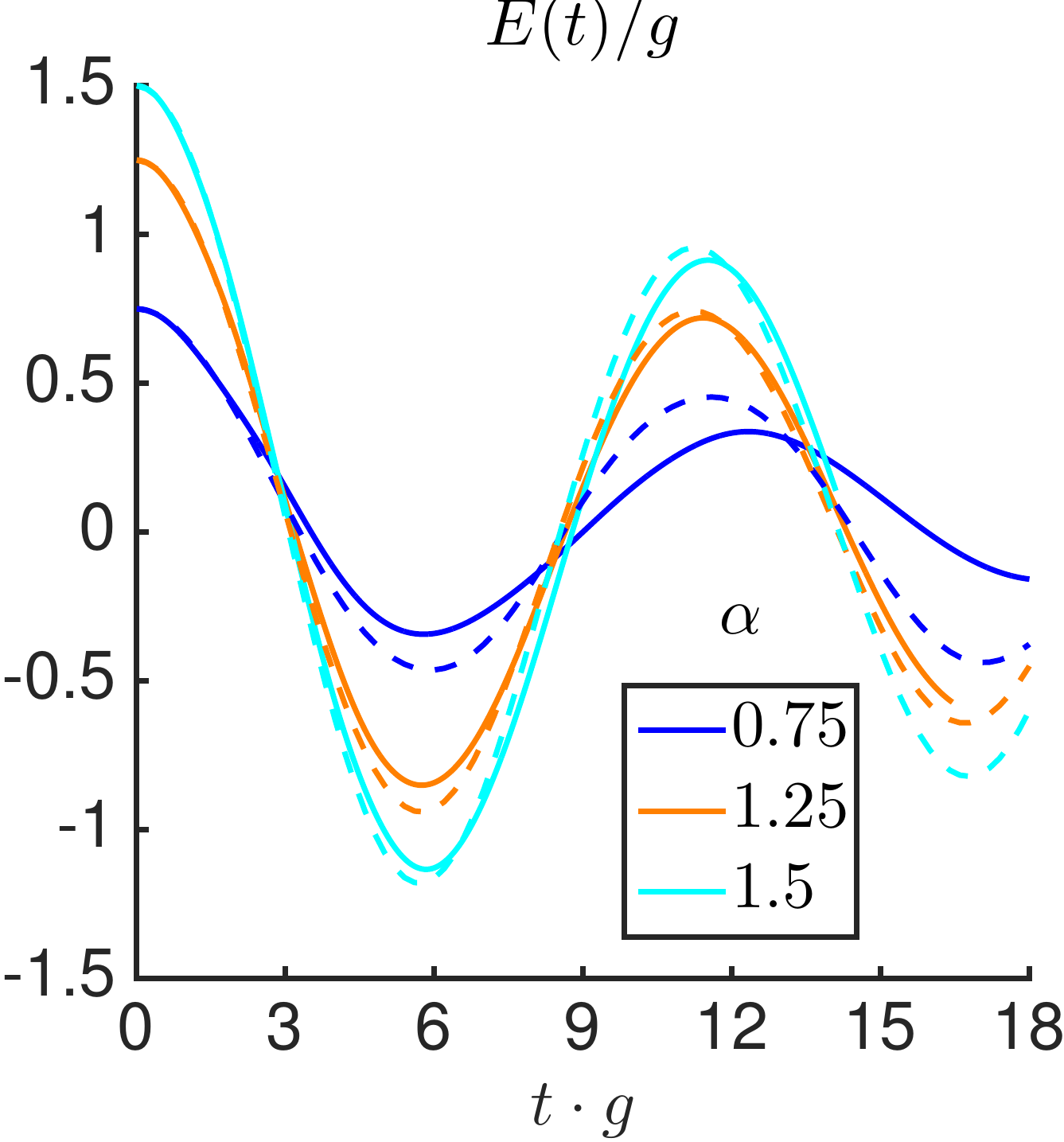}
\caption{}
\end{subfigure}\vskip\baselineskip
\captionsetup{justification=raggedright}
\caption{\label{figrealtimeVV}: Real-time evolutions for electric field quenches, $E(0)/g=0.75,1,25, 1.5$, for $m/g=0.25$ . $N(t)$ is the (vacuum subtracted) particle density, $E(t)$ is the electric field. In (a) the dotted lines show the corresponding thermal values (within a temperature interval $\Delta \beta=0.1$). In (b) the dotted lines show the result for the corresponding semi-classical simulations \cite{Buyens2017}.}\end{figure}

\emph{Real-time simulations.} Finally, some of the most relevant results on the real-time simulations of \cite{buyens2014matrix, Buyens2017} are: an intriguing effect in the Schwinger model concerns the non-equilibrium dynamics after a quench that is induced by the application of a uniform electric field onto the ground state at time $t=0$. Physically, this process corresponds to the so-called Schwinger pair creation mechanism \cite{Schwinger1951} in which an external electric field separates virtual electron-positron dipoles to become real electrons and positrons. The original derivation \cite{Schwinger1951} involved a classical background field, neglecting any back-reaction of the created particle pairs on the electric field. In \cite{Kluger1992,Hebenstreit2013} this back-reaction was taken into account at the semi-classical level. The real-time MPS simulations of \cite{buyens2014matrix, Buyens2017} provide the first full quantum simulation of this non-equilibrium process. In Fig.~\ref{figrealtimeVV} a sample result is shown, for the case of large initial electric fields. After a brief initial time interval of particle production, a regime can be observed with damped oscillations of the electric field and particle densities going to a constant value. This is in line with the semi-classical results, but as can be seen in figure \ref{figrealtimeVV}(b)  there is a quantitative difference, with a stronger damping, especially for the smaller fields. Finite temperature MPS simulations also allow a comparison with the purported equilibrium thermal values (figure \ref{figrealtimeVV} (a)).    

\subsubsection{Phase Diagram and Conformal String Excitations of Square Ice using Gauge Invariant Matrix Product States~\cite{tschirsich2019phase}}

The examples discussed above widely demonstrate the computational capabilities of tensor network methods in dealing with (1+1)-d lattice gauge theories. Ref.~\cite{tschirsich2019phase} reports instead results on a two-dimensional  $U(1)$ gauge theory, the (2+1)-d quantum link model, also known as square ice (for tensor network results on a theory with discrete gauge group, see Ref.~\cite{Tagliacozzo2011}). 

The main difference between square ice and a conventional $U(1)$ LGT is that the gauge fields now span a two-dimensional Hilbert space, and parallel transporters are replaced by spin operators. The system Hamiltonian reads:
\begin{equation}
H = \sum_{\square} {\left( -f_\square + \lambda f_\square^2 \right)}
\label{eq:hamiltonian}
\end{equation}
where the summation goes over all plaquettes of a square lattice, and the plaquette operator $f_\square = \sigma_{\mu_1}^+ \sigma_{\mu_2}^+ \sigma_{\mu_3}^- \sigma_{\mu_4}^- + \mathrm{H.c.}$ flips the spins on the links  $\mu_1,...,\mu_4$ of oriented plaquettes. The first term corresponds to the magnetic field interaction energy, while the second term is a potential energy for flippable plaquettes. There is no direct electric field energy since the spin representation is $S=1/2$.

The phase diagram of the model has been determined using a variety of methods, including exact diagonalisation~\cite{Shannon2004CyclicExchangeXXZ} and quantum Monte Carlo~\cite{Banerjee2013QuantumLinkDeconfined}. There are two critical points: one is the so-called Rokshar-Kivelson point at $\lambda=1$, where the ground state wave function is factorised into an equal weight superposition of closed loops~\cite{Rokhsar1988Dimers}. This points separates a columnar phase at $\lambda>1$ from a resonating valence-bond solid (RVBS). The latter is separated from a N\'eel phase by a weak first order transition point at around $\lambda\simeq0.36$~\cite{Shannon2004CyclicExchangeXXZ,Banerjee2013QuantumLinkDeconfined}. All of these phases are confining. The richness of its phase diagram and the possibility of carrying out precise MC simulations make this an ideal model for testing tensor network techniques for (2+1)-d lattice gauge theories.

Ref.~\cite{tschirsich2019phase} presents an analysis based on several observables computed in $L_x\times L_y$ cylinder geometries to mitigate entanglement growth as a function of the system size. The method of choice was an iTEBD algorithm applied on an MPS ansatz. Beyond simplicity and numerical stability of the algorithm, the main technical advantage of this approach is that re-arranging the MPS in columns allows the integration of the Gauss law in a relatively simple manner.

In the first part, conventional LGT diagnostics, such as the scaling of order parameters for the ordered phase, and the decay of Wilson loops, were analysed. The main conclusion is that TN methods can reach system sizes well beyond ED with the necessary accuracy for determining order parameters and correlation functions. However, the system sizes achieved (up to 600 spins) were smaller when compared to the ones accessible with QMC: this prevented, for instance, a systematic study of Wilson loops, that were found to be particularly sensitive to finite volumes and open boundary conditions.

\begin{figure}
\resizebox{0.99\columnwidth}{!}{\includegraphics{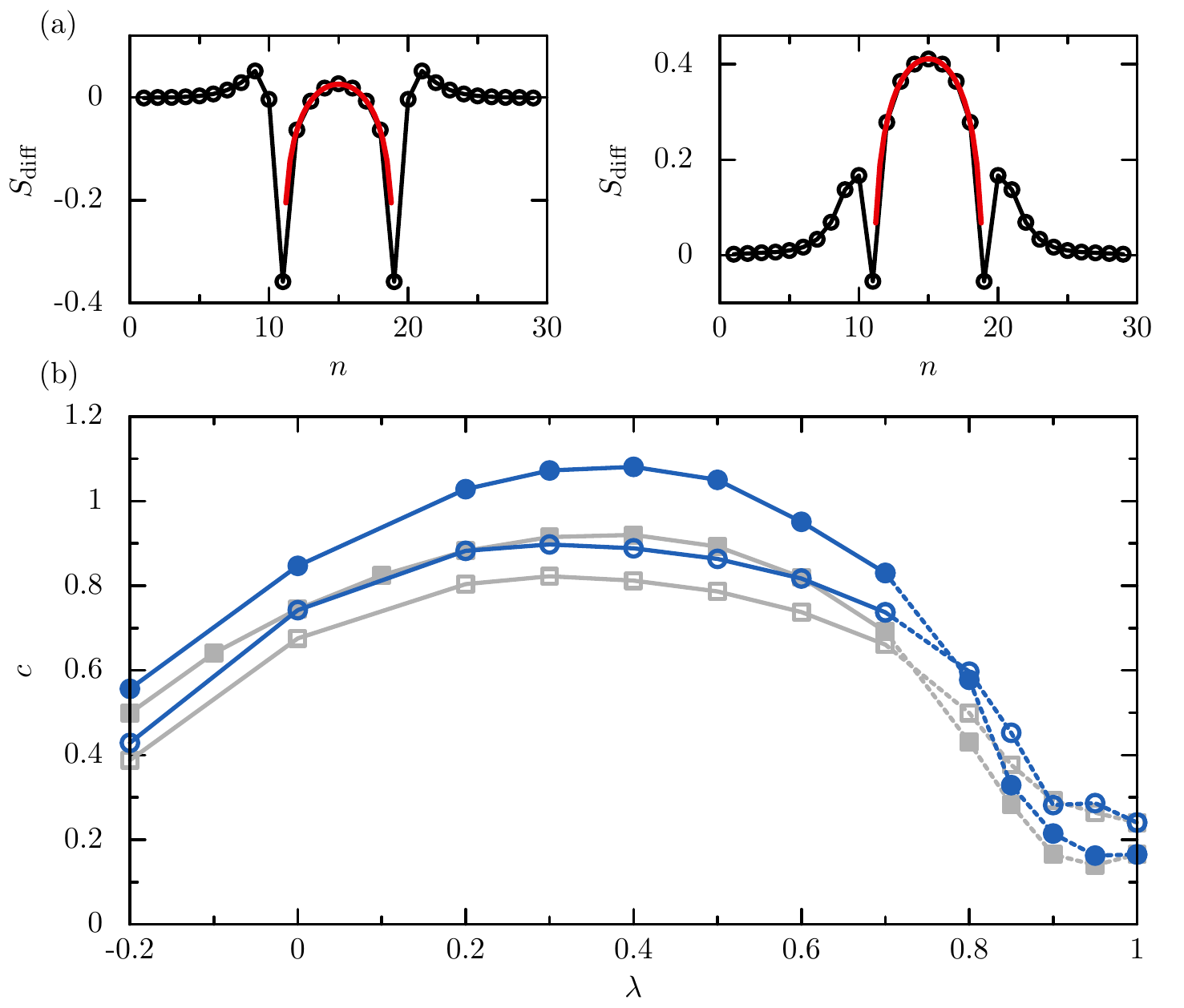}}
\caption{(a) Increase in von Neumann entropy after insertion of two static charges $\pm1$ at distance $\ell=9$ on a $30\times10$ lattice. Entropies taken at vertical cuts of the cylinder between sites $n,n+1$. Between the charges, a Calabrese--Cardy fit (red) yields a central charge $c$ as shown for $\lambda=-0.2$ (left) and $\lambda=0.7$ (right). (b) Central charges $c$ over coupling parameter $\lambda$ for system sizes $24\times8$ (grey boxes) and $30\times10$ (blue points), $\ell=5$ (open) and $\ell=9$ (filled symbols). For $\lambda\in\mathopen{[}0.7,1\mathclose{)}$ (dashed lines), resonant plaquettes on both sub-lattices where found in superposition when charges where present, increasing the entanglement entropy compared to the system without charges. For $\lambda=1$, exact ground state results are shown. Bare fit-errors are smaller than point sizes. Figure taken from Ref.~\cite{tschirsich2019phase}. }
\label{fig:Ent_Ice}
\end{figure}

The second part of the work instead focused on entanglement properties of string states, which are generated by introducing two static charges in the system at given distance $\ell$. Some results on the entropy difference between the string and ground states are depicted in Fig.~\ref{fig:Ent_Ice}a-b: in the region of space between the two charges  ($n\in[11,20]$ in the panels), this entropy difference is compatible with a conformal field theory scaling, with a central charge that is compatible with 1 in a large parameter regime within the RVBS phase (Fig.~\ref{fig:Ent_Ice}c), where finite volume effects are moderate. These results represent an entanglement proof of the conformal behaviour of string excitations in a confining theory, which is a well established fact in non-Abelian (3+1)-d cases as determined from the string energetics~\cite{L_scher_2002}.

\section{Quantum computation and digital quantum simulation}
\label{QCDQS}

There are two avenues towards quantum simulations - analog and digital. In analog simulations, the degrees of freedom of the original system and the dynamical evolutions, are mapped to the simulating system. In digital simulations, the simulating system is evolving forward in time stroboscopically, by applying a sequence of short quantum operations. In this section, the digital quantum simulation approach to high-energy models is reviewed, while the analogue quantum simulation is described in the following one. 

\subsection{Quantum and Hybrid Algorithms for Quantum Field Theories}

Quantum information, in general, and quantum computation, in particular, have brought new tools and perspectives for the calculation and computation of strongly correlated quantum systems. Understanding a dynamical process as a quantum circuit or the action of a measurement as a projection in a Hilbert space are just two instances of this quantum framework. In this section, two relevant articles \cite{jordan2012quantum,lu2018simulations} are described where these new approaches are used.

\subsubsection{Quantum Algorithms for Quantum Field Theories\cite{jordan2012quantum}}

Quantum computers can efficiently calculate scattering probabilities in $\phi^{4}$ theory to arbitrary precision at both weak and strong coupling with real-time dynamics, contrary to what is achieved in LGT where the scattering data can also be computed in Euclidean simulations \cite{luscher1986volume,wiese1989identification,luscher1991signatures}. In \cite{jordan2012quantum}, Jordan et al. developed a (constructive) quantum algorithm that could compute relativistic scattering probabilities in a massive quantum field theory with quartic self-interactions ($\phi^{4}$ theory) in space-time of four and fewer dimensions and solve the equations of QFT efficiently that can be compared with the data from particle accelerators. The proposed algorithm is polynomial in the number of particles, their energy, and the desired precision. In the limit of the so-called strong coupling of QFT, the algorithm actually provides an exponential acceleration with respect to the best known classical algorithms.

This work is based on three important technical achievements. First, continuous fields can be accurately represented by a finite number of qubits whose coordinates form a lattice. This achievement is highly non-trivial, because QFTs are contaminated by infinite values of various quantities that must be cured by using renormalisation and regularisation methods, both of which feature naturally in the discussed algorithm. Second, one bottleneck for an efficient implementation of the simulation, the preparation of the initial state, is achieved by a preparation of particles in the form of wave packets. Third, the time evolution is split into the action of local quantum gates. This procedure works well for local theories (field theories discretised on a finite space-time mesh) whose accuracy and convergence must be controlled. Hence, quantum computation for continuous fields can be achieved in a controlled way and with an exponential quantum speedup.

More concretely, the scalar field Hamiltonian in $D-1$ spatial dimensions reads $H= H_{\pi} + H_{\phi}$ with
\begin{equation}
\begin{split}
H_{\pi} =& \sum_{x} \frac{a^{D-1}}{2}  \pi \left( x \right)^{2} \\
H_{\phi} =& \sum_{x} \frac{a^{D-1}}{2} \left( \nabla \phi \left( x \right)^{2} + m_{0}^{2} \phi \left( x \right)^{2} + \frac{\lambda_{0}}{12} \phi \left( x \right)^{4} \right)
\end{split}
\end{equation}
The conjugate variables $\phi (x)$ and $\pi (x)$ obey the canonical commutation relations $\left[ \phi \left( x \right) , \phi \left( y \right) \right] = \left[ \pi \left( x \right) , \pi \left( y \right) \right]  = 0$, $\left[ \phi \left( x \right) , \pi \left( y \right) \right] = \frac{i}{a^{D-1}} \delta \left(x ,y\right)$.

If the coefficient $\lambda_{0}$ vanishes, then the Hamiltonian is quadratic in the variables $\phi$ and $\pi$. In that case, the theory is Gaussian, describes a massive non-interacting particle and it can be solved exactly. The complete Hamiltonian is the sum of two terms, one is diagonal in the $\pi$ basis, while the other is diagonal in the $\phi$ basis. Choosing a small time step $\epsilon$, then $\exp{ \left( -i \epsilon H \right)} \sim \exp{ \left( -i \epsilon H_{\pi} \right)} \exp{ \left( -i \epsilon H_{\phi} \right)} + O \left( \epsilon^{2} \right)$. It is easy to simulate time evolution governed by the diagonal Hamiltonian $H_{\pi}$ or $H_{\phi}$, evolving the system using the field Fourier transform to alternate back and forth between the $\pi$ and $\phi$ bases, and applying a diagonal evolution operator in each small time step. 

\subsubsection{Simulations of Subatomic Many-Body Physics on a Quantum Frequency Processor\cite{lu2018simulations}}

The emerging paradigm for solving optimisation problems using near-term quantum technology is a kind of hybrid quantum-classical algorithm.  In this scheme, a quantum processor prepares an $n$-qubit state, then all the qubits are measured and the measurement outcomes are processed using a classical optimiser; this classical optimiser instructs the quantum processor to alter slightly how the $n$-qubit state is prepared. This cycle is repeated many times until it converges to a quantum state from which the approximate solution can be extracted. When applied to classical combinatorial optimisation problems, this procedure goes by the name Quantum Approximate Optimisation Algorithm (QAOA) \cite{farhi2014quantum}. But it can also be applied to quantum problems, like finding low-energy states of many-particle quantum systems (large molecules, for example). When applied to quantum problems this hybrid quantum-classical procedure goes by the name Variational Quantum Eigensolver (VQE).

Hence, VQE algorithms provide a scalable path to solve grand challenge problems in subatomic physics on quantum devices in the near future. One way to implement VQE optically is using a quantum frequency processor (QFP). A variety of basic quantum functionalities have recently been demonstrated experimentally in this approach. A QFP is a photonic device that processes quantum information encoded in a comb of equi-spaced narrow band frequency bins. Mathematically, the QFP is described by a unitary mode transformation matrix $V$ that connects input and output modes. 

In \cite{lu2018simulations} it was demonstrated how augmenting classical  calculations with their quantum counterparts offers a roadmap for quantum-enabled subatomic physics simulations. More concretely, a subatomic system can be simulated as a collection of nucleons with effective field theory (EFT) parameters input from experimental data or ab-initio calculations. Using a photonic QFP, the ground state energies of several light nuclei using experimentally determined EFT parameters were computed in \cite{lu2018simulations}.

The VQE algorithm calculates the binding energies of the atomic nuclei $^{3}H$, $^{3}He$, and $^{4}He$. Further, for the first time, a VQE was employed to determine the effective interaction potential between composite particles directly from an underlying lattice quantum gauge field theory, the Schwinger model. This serves as an important demonstration of how EFTs themselves can be both implemented and determined from first principles by means of quantum simulations.

\subsection{Digital quantum simulation with trapped ions}

Due to the high degree of quantum control of trapped ions platforms, they can be seen as prototypes of universal quantum simulators. In the following sections, two applications are described that realise experimentally the idea of a quantum simulator for high-energy processes. 

\subsubsection{Real-time dynamics of lattice gauge theories with a few-qubit quantum computer\cite{martinez2016real,muschik2017u}}

\begin{figure*}[t]
\centering
\includegraphics[width=\textwidth, angle=0]{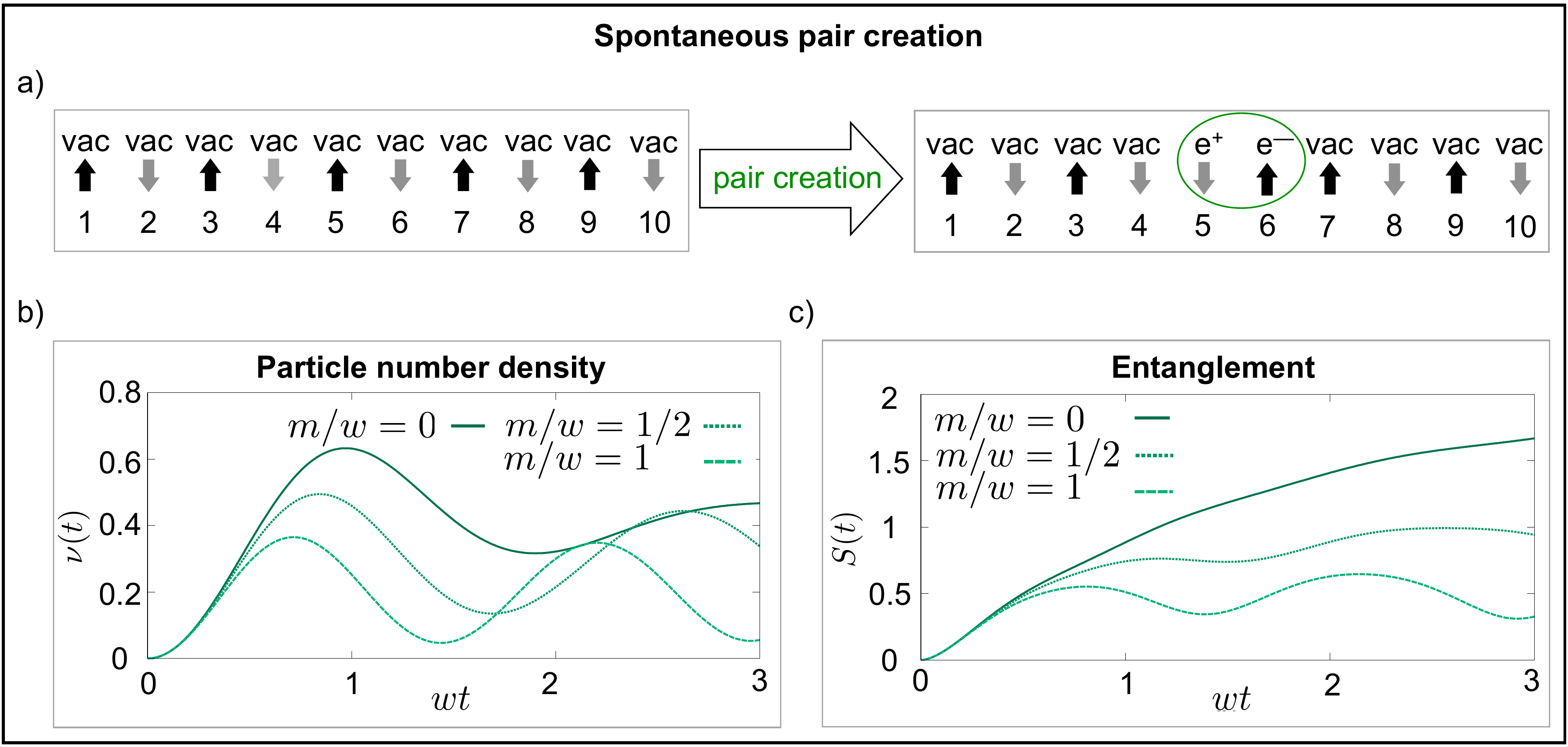}
\caption{Simulation of particle production out of the bare vacuum. (a) Pair creation in the encoded Schwinger model. The left spin configuration corresponds to the bare vacuum state. The right configuration displays a state with one particle-anti-particle pair. (b,c) Instability of the bare vacuum: (b) Particle number density $\nu (t)$ and (c) entanglement entropy $S (t)$ for $J/w=1$ and different values of $m/w$, where $J$ and $w$ quantify the electric field energy and the rate at which particle-anti-particle pairs are produced, and $m$ is the fermion mass, for the lattice Hamiltonian $\hat{H}_{\text{lat}}=w \sum_{n} \left[ \hat{\sigma}^{+}_{n} \hat{\sigma}^{-}_{n+1} + \text{H.C.} \right] + \frac{m}{2} \sum_{n} \left( -1 \right)^{n}  \hat{\sigma}^{z}_{n} + J \sum_{n} \hat{L}^{2}_{n}$. (b) After a fast transient pair creation regime, the increased particle density favours particle-anti-particle recombination inducing a decrease of $\nu(t)$. This non-equilibrium interplay of regimes with either dominating production or recombination continues over time and leads to an oscillatory behaviour of  $\nu(t)$ with a slowly decaying envelope. (c) The entanglement entropy $S(t)$ quantifies the entanglement between the left and the right half of the system, generated by the creation of particle-anti-particle pairs that are distributed across the two halves. An increasing particle mass $m$ suppresses the generation of entanglement. From \cite{muschik2017u}
}
\end{figure*}

The article \cite{martinez2016real} reports on the first digital quantum simulation of a gauge theory from high-energy physics and entails a theoretical proposal along with its realisation on a trapped ion quantum computer~\cite{Blatt2012,Schindler2013}. This simulation addresses quantum electrodynamics in one spatial and one temporal dimension, the so-called Schwinger model~\cite{Schwinger1,Schwinger2}. 

In~\cite{martinez2016real,muschik2017u}, the coherent real-time dynamics of spontaneous particle-anti-particle pair creation has been studied. Such dynamical processes cannot be addressed with conventional Markov Chain Monte Carlo methods due to the sign problem~\cite{calzetta_book} preventing the simulation of time evolutions. The experiment performed in~\cite{martinez2016real,muschik2017u} realises a Trotter time evolution~\cite{Lloyd96} based on the Kogut-Susskind Hamiltonian formulation of the Schwinger model~\cite{KogutSusskindFormulation}. The simulation protocol used in this demonstration is custom-tailored to the experimental platform and based on eliminating the gauge degrees of freedom. The Schwinger model entails dynamical matter and gauge fields (electromagnetic fields). The gauge degrees of freedom are analytically integrated out~\cite{Encoding}, which leads to a pure matter model that can be cast in the form of an exotic spin model that features two-body terms and long-range interactions. The gauge bosons do not appear explicitly in the description but are included implicitly in the form of long-range interactions. The resulting encoded model is gauge invariant at all energy scales and allows one to simulate the full infinite-dimensional Hamiltonian. This approach is well matched to simulators based on trapped ions~\cite{Blatt2012,Schindler2013}, which naturally feature a long-range interaction and hence allow for a very efficient implementation of the encoded Schwinger model. The Trotter protocol that has been devised in~\cite{martinez2016real,muschik2017u} can be realised in a scalable and resource-optimised fashion. The number of Trotter steps is ideal for the required ion-ion coupling matrix and scales only linearly in the number of lattice sites $N$. Moreover, the protocol is designed such that Trotter errors do not lead to gauge variant contributions.

The experiment~\cite{martinez2016real,muschik2017u} has been carried out for $N=4$ lattice sites (i.e., using four qubits) and a gate sequence comprising more than $200$ gate operations. The used resources are high-fidelity local gate operations and the so-called M{\o}lmer-Sorensen gates~\cite{MSgates} with all-to-all connectivity between the individual ions. The qubit states are encoded in electronic sub-levels of the ions. In the experiment, a quench has been performed in which the bare vacuum (i.e., the ground state for infinite fermion mass) has been prepared, followed by a Trotterised time evolution under the encoded Schwinger Hamiltonian, which leads to the generation of particle-anti-particle pairs. This type of experiment can also be performed starting from the dressed vacuum (i.e., an eigenstate of the full Hamiltonian for finite fermion mass), which is a highly entangled state that can be prepared on a trapped ion quantum simulator using the method demonstrated in~\cite{kokail2018self} (see Sec.~\ref{SVVQS} below). In this case, a quench to generate pair creation events would involve the time evolution under the Schwinger Hamiltonian including background electric fields. Including electric background fields to the encoded Schwinger Hamiltonian leads to additional local terms and therefore requires only minor modifications in the quantum simulation protocol~\cite{muschik2017u}. Using trapped ions, high-fidelity measurements can be made using fluoresce detection~\cite{Blatt2012,Schindler2013}. In \cite{martinez2016real,muschik2017u} local measurements in the $z$-basis have been used to study the site-resolved particle number density and electric field distribution in real-time as a function of the fermion mass. This type of analysis can be directly scaled up to larger system sizes. The experiment~\cite{martinez2016real,muschik2017u} probed also the entanglement generated during pair creation, which can be done for small system sizes and involved the measurement of the density matrix of the spin system. As shown in~\cite{martinez2016real,muschik2017u}, the entanglement of the encoded model corresponds to the entanglement in the original model involving both gauge fields and fermions.

\subsubsection{Self-Verifying Variational Quantum Simulation of the Lattice Schwinger Model\cite{kokail2018self}}
\label{SVVQS}

In this article, a quantum co-processor successfully simulated particle physics phenomena on 20 qubits for the first time. The experiment uses new methods with a programmable ion trap quantum computer with 20 quantum bits as a quantum co-processor, in which quantum mechanical calculations that reach the limits of classical computers are outsourced. For this, a sophisticated optimisation algorithm has been developed that, after about 100,000 uses of the quantum co-processor by the classical computer, leads to the result. In this way, the programmable variational quantum simulator has simulated the spontaneous creation and destruction of pairs of elementary particles from a vacuum state on 20 quantum bits.

An analog quantum processor prepares trial states, quantum states that are used to evaluate physical quantities. The classical computer analyses the results of these evaluations, with the aim of optimising certain adjustable (variational) parameters on which the trial states depend. This computer then suggests improved parameters to its quantum co-worker in a feedback loop. In the study, the quantum device contains a line of atomic ions that each represent a qubit. This set-up is used to carry out quantum simulations of the ground state of electrons coupled to light, a system that is described by the theory of quantum electrodynamics in one spatial dimension. 

\subsection{Digital quantum simulation with superconducting circuits}

This section reviews the possibility to perform digital quantum simulation of lattice gauge theories with superconducting circuits\cite{mezzacapo2015non}.

\subsubsection{Non-Abelian $SU(2)$ Lattice Gauge Theories in Superconducting Circuits\cite{mezzacapo2015non}}

Superconducting circuits have proven to be reliable devices that can host quantum information and simulation processes. The possibility to perform quantum gates with high fidelities, together with high coherence times, makes them ideal devices for the realisation of digital quantum simulations. In \cite{mezzacapo2015non}, a digital quantum simulation of a non-Abelian dynamical $SU(2)$ gauge theory is proposed in a superconducting device. The proposal starts from a minimal setup, based on a triangular lattice, that can encode pure-gauge dynamics. The degrees of freedom of a single triangular plaquette of this lattice are encoded into qubits. Two implementations of this quantum simulator are described, using two different superconducting circuit architectures. A setup in which six tunable-coupling transmon qubits are coupled to a single microwave resonator is considered, and a device where six capacitively coupled Xmon qubits stand on a triangular geometry, coupled to a central auxiliary one. The experimental requirements necessary to perform the simulation on one plaquette are characterised and arguments for scaling to large lattices are also given in \cite{mezzacapo2015non}.

A minimal implementation of a pure $SU(2)$ invariant model in a triangular lattice is considered by using triangular plaquettes. In this case, the pure-gauge Hamiltonian on a single plaquette reads 
\begin{equation}
H_T=-J~\text{Tr}\left[U(\vec{x},\hat{\mu})U(\vec{x}+\hat{\mu},\hat{\nu})U(\vec{x}+\hat{\mu}+\hat{\nu},-\hat{\mu}-\hat{\nu})\right].
\end{equation}
This interaction corresponds to the magnetic term of a gauge invariant dynamics, which acts on closed loops.

Due to gauge invariance, the local Hilbert space of a link is four dimensional, and it can be faithfully spanned by two qubits, called ``position'' $\sigma^{a}_{\textrm{pos}}$ and ``spin'' qubit $\sigma^{a}_{m}$. In this subspace, it is useful to define the operators $\Gamma^{0} = \sigma^{x}_{\textrm{pos}} \sigma^{0}_{m}$, $\Gamma^{a} =\sigma^{y}_{\textrm{pos}} \sigma^{a}_{m}$, such that the total Hamiltonian is written as
\begin{eqnarray}
\label{HG}
H_{T}=&&-J\Big\{ \Gamma^0_{12}\Gamma^0_{23}\Gamma^0_{31}+ \sum_{abc}\epsilon_{abc}\Gamma^a_{12}\Gamma^b_{23}\Gamma^c_{31} \\
&&-\sum_{a}\Big[\Gamma^0_{12}\Gamma^a_{23}\Gamma^a_{31} + \Gamma^a_{12}\Gamma^0_{23}\Gamma^a_{31}+\Gamma^a_{12}\Gamma^a_{23}\Gamma^0_{31}\Big] \Big\} \nonumber.
\end{eqnarray}

\begin{figure}
\includegraphics[scale=0.283]{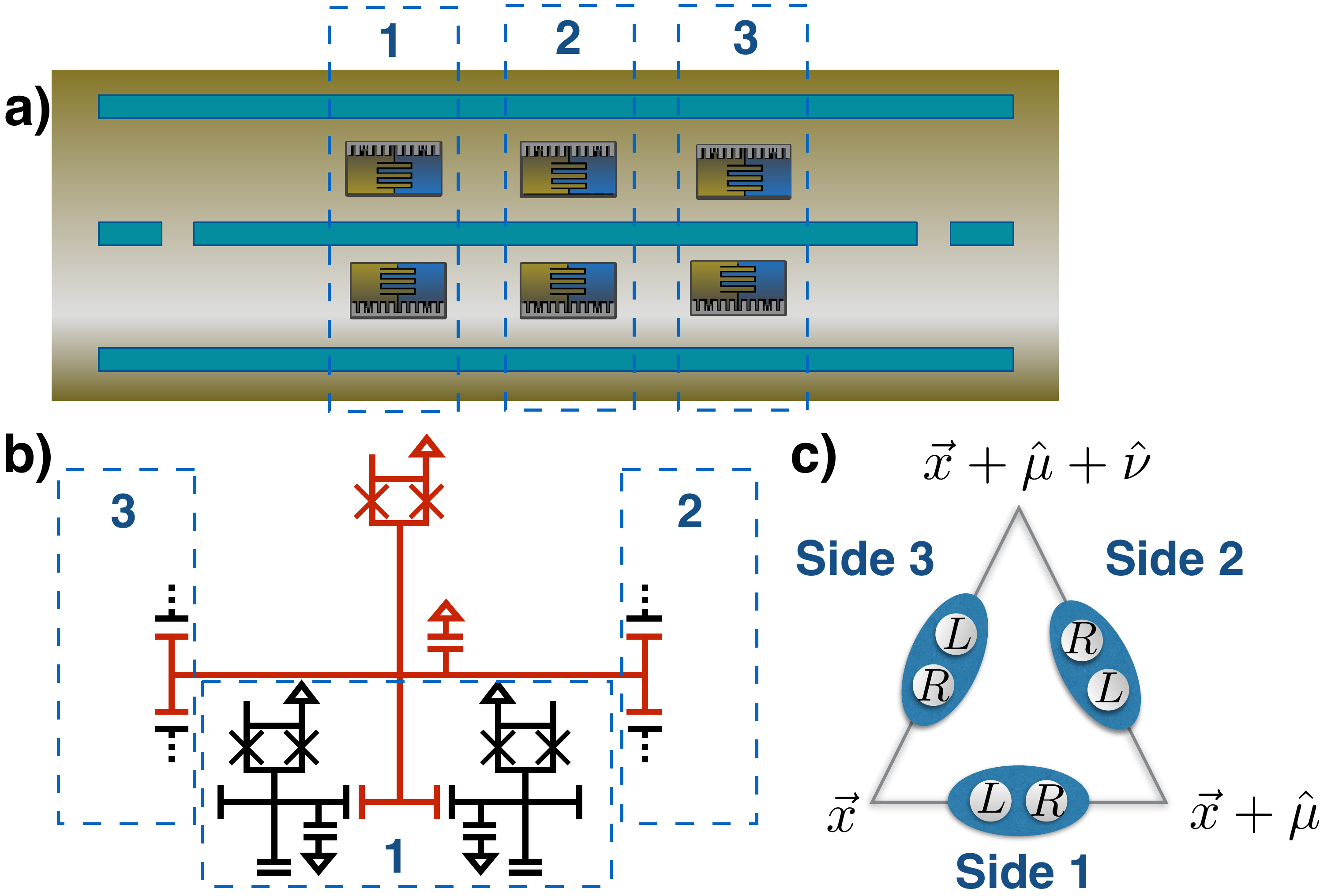}
\caption{(Color online) a) Six tunable-coupling transmon qubits coupled to a single microwave resonator. b) Six Xmon qubits on a triangular geometry, coupled to a central one. The box 1 in the scheme is implicitly repeated for the sides~2~and~3. Both setups can encode the dynamics of the $SU(2)$ triangular plaquette model schematised in c), where the left and right gauge degrees of freedom are explicitly depicted.\label{scheme} From \cite{mezzacapo2015non}.}
\end{figure}

In order to simulate the interaction of Eq.~(\ref{HG}), one can decompose its dynamics in terms of many-body monomials, and implement them sequentially with a digitised approximation. In a digital approach, one decomposes the dynamics of a Hamiltonian $H=\sum_{k=1}^mh_k$ by implementing its components stepwise, $e^{-iHt}\approx\left(\prod_{k=1}^m e^{-ih_kt/N}\right)^N$ (here and in the following $\hbar=1$), for a total of $m~\times~N$~gates, with an approximation error that goes to zero as the number of repetitions $N$ grows. In a practical experiment, each quantum gate $e^{-ih_kt}$ will be affected by a given error $\epsilon_k$. By piling up sequences of such gates, for small gate errors $\epsilon_k \ll 1$, the total protocol will be affected by a global error, which is approximately the sum $\epsilon\approx\sum_k\epsilon_k$. 

To simulate the pure-gauge interaction in a single triangular plaquette, first, a setup with six tunable-coupling transmon qubits coupled to a single microwave resonator is considered. Each tunable-coupling qubit is built using three superconducting islands, connected by two SQUID loops. Acting on these loops with magnetic  fluxes, one can modify the coupling of the qubits with the resonator, without changing their transition frequencies. By threading with magnetic fluxes at high frequencies, one can drive simultaneous red and blue detuned sidebands, and perform collective gates. Each many-body operator can be realised as a sequence of collective and single-qubit gates.

A second architecture is considered where six Xmon qubits in a triangular geometry are capacitively coupled with an additional central ancillary qubit. In this case, the collective interactions can be decomposed and performed with pairwise $C$-phase gates, using the central ancillary qubit to mediate non-nearest-neighbour interactions. In this way, the quantum simulation of one digital step of the Hamiltonian in Eq.~(\ref{HG}) will amount to realise \mbox{$168$~$C$-phase~gates} and a number of single-qubit rotations which is upper bounded by $520$. 

\subsection{Digital quantum simulation with ultra-cold atoms}

Digital quantum simulators show the possibility to achieve universal quantum computation. Among the most promising platforms are the ones built with ultra-cold atoms. In this section, several instances are shown using optical lattices and Rydberg platforms where even a completely gauge invariant simulation could be achieved \cite{weimer2010rydberg,tagliacozzo2013optical,tagliacozzo2013simulation,zohar2017digital,zohar2017digitalA}. 

\subsubsection{A Rydberg Quantum Simulator\cite{weimer2010rydberg}}

\begin{figure}[h!]
\includegraphics[width=0.95\columnwidth]{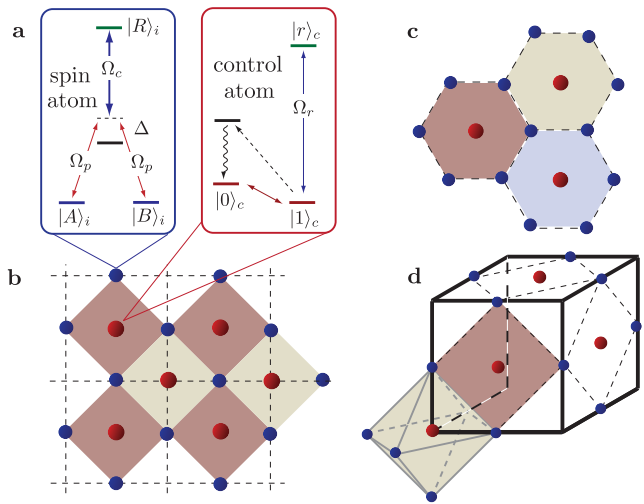} 
\caption{Setup of the system: a) Two internal states $|A\rangle_{i}$ and $|B\rangle_{i}$ give rise to an effective spin degree of freedom. These states are coupled to a Rydberg state $|R\rangle_{i}$ in two-photon resonance, establishing an electromagnetically induced transparency (EIT) condition. On the other hand, the control atom has two internal states $|0\rangle_{c}$ and $|1\rangle_{c}$. The state $|1\rangle_{c}$ can be coherently excited to a Rydberg state $|r\rangle_{c}$ with Rabi frequency $\Omega_{r}$, and can be optically pumped into the state $|0\rangle_{c}$ for initialising the control qubit. b) For the toric code, the system atoms are located on the links of a two-dimensional square lattice, with the control qubits in the centre of each plaquette for the interaction $A_{p}$ and on the sites of the lattice for the interaction $B_{s}$. Setup required for the implementation of the color code (c), and the $U(1)$ lattice gauge theory (d). From \cite{weimer2010rydberg}. }
\label{fig1} 
\end{figure}

A universal quantum simulator is a controlled quantum device that faithfully reproduces the dynamics of any other many-particle quantum system with short-range interactions. This dynamics can refer to both coherent Hamiltonian and dissipative open-system time evolution. Cold atoms in optical lattices, which are formed by counter-propagating laser beams, represent a many-particle quantum system, where the atomic interactions and dynamics of the particles can be controlled at a microscopic level by external fields. This high level of control and flexibility offers the possibility to use these systems as quantum simulators, i.e., as devices which can mimic the behaviour of other complex many body quantum systems and allow the study of their properties, dynamics and phases.

Stored cold atoms in deep lattices, in which atoms do not hop between the lattice sites, can be used to encode quantum bits in different electronic states of the atoms. Interestingly, although the atoms sit at different sites and do not collide, it is possible to induce very strong interactions between atoms separated by distances of several micrometers. This can be achieved by exciting them to electronically high-lying Rydberg states. These Rydberg interactions offer the possibility to realise fast quantum gates between remote atoms. Motivated by and building on these achievements, a digital Rydberg simulator architecture based on sequences of fast and efficient quantum gates between Rydberg atoms is developed in \cite{weimer2010rydberg}. This ``digital'' simulator offers promising perspectives for the simulation of complex spin models, which are of great interest both in quantum information science, condensed matter, and high-energy physics.

The proposed simulation architecture allows one to realise a coherent Hamiltonian as well as dissipative open-system time evolution of spin models involving $n$-body interactions, such as, e.g., the Kitaev toric code, colour code and lattice gauge theories with spin-liquid phases. The simulator relies on a combination of multi-atom Rydberg gates and optical pumping to implement coherent operations and dissipative processes. Highly excited Rydberg atoms interact very strongly, and it is possible to switch these interactions on and off in a controlled way by applying laser pulses. By choosing on which atoms to shine light, the properties of the quantum simulator can be precisely tuned.

As a key ingredient of the setup, extra auxiliary qubit atoms are introduced in the lattice, which play a two-fold role. First, they control and mediate effective $n$-body spin interactions among a subset of $n$ system spins residing in their neighbourhood of the lattice. This is achieved efficiently, making use of single-site addressability and a parallelised multi-qubit gate, which is based on a combination of strong and long-range Rydberg interactions and electromagnetically induced transparency (EIT). Second, the auxiliary atoms can be optically pumped, thereby providing a dissipative element, which in combination with Rydberg interactions results in effective collective dissipative dynamics of a set of spins located in the vicinity of the auxiliary particle, which itself eventually factors out from the system spin dynamics. The resulting coherent and dissipative dynamics on the lattice can be represented by, and thus simulates a master equation, where the Hamiltonian is the sum of $n$-body interaction terms, involving a quasi-local collection of spins in the lattice. The Liouvillian term in the Lindblad form governs the dissipative time evolution, where the many-particle quantum jump operators involve products of spin operators in a given neighbourhood.

\subsubsection{Optical Abelian Lattice Gauge Theories\cite{tagliacozzo2013optical}}

In \cite{tagliacozzo2013optical}, it is described how to perform a digital quantum simulation of the gauge-magnet/quantum link version of a pure U(1) lattice gauge theory with ultra-cold atoms, for a recent proposal of an analogue Rydberg simulator for the same theory see \cite{celi2019emerging}. Its phase diagram has been recently characterised by numerical investigations \cite{Banerjee2013QuantumLinkDeconfined}. The experiment aims at mapping the phase diagram of the spin 1/2 $U(1)$  quantum link model by measuring the string tension of the electric flux tube  between two static charges and its dependence on the distance. In the confined phase, the string tension is finite, and thus the energy of the system increases linearly with the inter-charge separation. Charges are thus bound together. In the deconfined phase the string tension vanishes and thus the charges can be arbitrarily far away with only a finite energy cost. 

In the proposed quantum simulation the gauge bosons are encoded in the hyper-fine levels of Rydberg atoms. The atoms are in a Mott-insulating phase with one atom per site. Extra atoms are needed in order to  collectively and coherently address several atoms at the same time. The simulation requires imposing the Gauss law and engineer the dynamics. The latter is obtained digitally decomposing unitary time evolution in elementary Trotter steps that can be performed by Rydberg gate operations. The former can be imposed by dissipation or by engineering digitally an energy penalty for the forbidden configurations. This is achieved by  using the Rydberg blockade  as first proposed in \cite{Mueller2009}. The key ingredient is the mesoscopic Rydberg gate in which one control atom is excited and de-excited from its Rydberg state and as a result of the blockade this affects several atoms inside its blockade radius.  The setup thus requires two set of atoms, atoms encoding the gauge boson degrees of freedom (one per link of the lattice) that are called \emph{ensemble} atoms. These atoms are controlled by addressing another set of atoms, the \emph{control} atoms. In this setup the control atoms are used in order to imprint the desired dynamics on the ensemble atoms.

In order to simulate the $U(1)$  quantum link model, one control atoms located at the center of  every plaquette and one control atom located at every site are used. The ensemble atoms are located at the center of the links of the lattice. The lattice spacing should also be engineered in such a way that only four atoms encoding the gauge boson degrees of freedom should be contained inside the blockade radius of the control atoms. Individually addressing and manipulating the control atoms via, e.g., a quantum-gas microscope is also needed. 

With this setup, an arbitrary Hamiltonian can be implemented on the atoms encoding the gauge boson degrees of freedom digitally, by decomposing it into a sequence of elementary operations, involving single-site rotations combined with the use of the mesoscopic Rydberg gate. As a result of the lattice geometry, the gate involves  one control atom  (either at one site or in the center of one plaquette) and the four ensemble atoms surrounding it.  This architecture is indeed sufficient to perform a universal quantum simulation of  Abelian lattice gauge theories \cite{weimer2010rydberg}.

The simulation requires two stages. During the first stage one starts from some trivial state and prepares the state to be studied such as, e.g., the ground state of the quantum link Hamiltonian. In a second stage, the mesoscopic Rydberg gates are reversed and the state of the system is transferred to the state of the control atoms, that if appropriately read out (through, e.g., a quantum-gas microscope), allow the measurement of the physical state of the system and its properties, such as, e.g., the string tension between two static charges.

The simulations are digital, in the sense that they require applying a discrete sequence of pulses to the atoms, whose nature and  duration can be found by using optimal control techniques. 

\subsubsection{Simulations of non-Abelian gauge theories with optical lattices\cite{tagliacozzo2013simulation}}

An important and necessary step towards the quantum simulation of QCD is the simulation of simpler non-Abelian gauge theories in two dimensions to study the interplay of electric and magnetic interactions with non-Abelian local symmetry. The minimal relevant example is given by $SU(2)$ gauge magnets or quantum link models \cite{Orland1990,Chandrasekharan:1996ih} with static charges considered in \cite{tagliacozzo2013simulation}. There it is shown how to characterise confinement in the model and determine its phase diagram by simulating it digitally with Rydberg atoms.

\begin{figure}[h!]
\begin{center}
\resizebox{0.99\columnwidth}{!}{\includegraphics{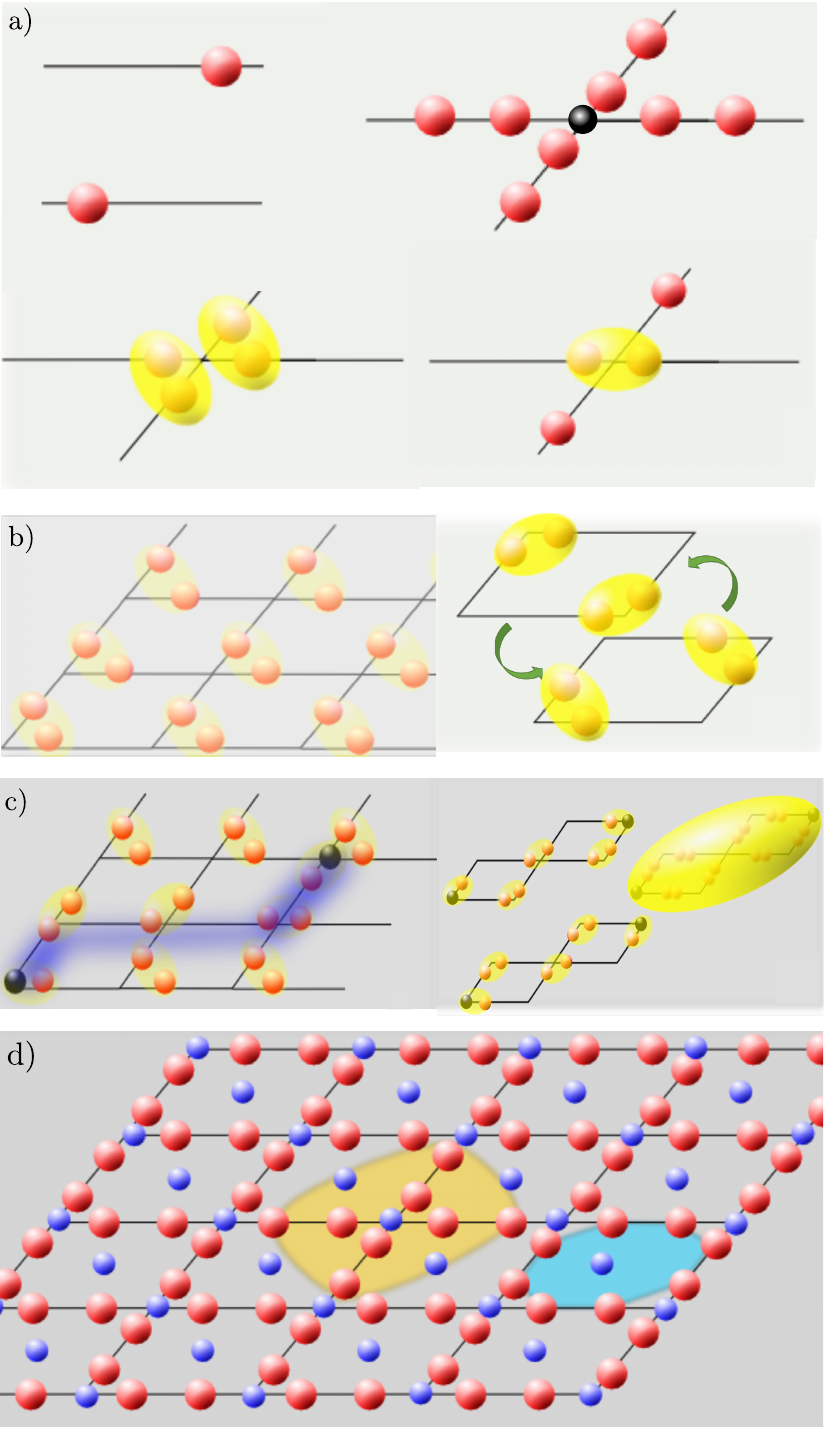}}
\caption{(color online) Non-Abelian gauge theories with Rydberg atoms.
a) $SU(2)$ gauge magnets: the gauge quanta on the links are a direct sum of spin $\frac 12$ at the links ends (red dots) while charges are spins on the sites (dark dots). Gauge invariance translates to singlet formation, some examples (without charges) with singlets in yellow. b) Electric interactions favour singlets in the left/down ends of the links. Magnetic interactions exchange parallel singlets on plaquettes and annihilate the other configurations. c) Linear confinement induced by a pair of opposite charges at strong and weak couplings, where the electric and magnetic terms dominate, respectively. d) Implementation scheme without charges: the Gauss law and the plaquette interactions are decomposed in elementary C-not gates that involve all physical qubits/atoms (in red)  within the yellow and blue blockade areas, respectively, of the auxiliary Rydberg atoms (in blue). For the full scheme see  \cite{tagliacozzo2013simulation}.} 
\label{fig:tagliacozzo2013simulation}
\end{center}
\end{figure}

For $SU(2)$, the quantum link is written as the direct sum of two spins $\frac 12$  sitting at each end of the link, see Fig. \ref{fig:tagliacozzo2013simulation} a). As in \cite{Kogut1975}, physical states, i.e., configurations allowed by gauge invariance, are determined through the (non-Abelian version of the) Gauss' law, and the dynamics comes from competition of electric (on each link) and magnetic (plaquette) interactions. In $SU(2)$ gauge magnets, the charges occupying the sites of the lattice are also represented as spins and the Gauss law demams that the total spin at each site, i.e., spins $\frac 12$ at the link ends coupled to the static charge residing at the site, is zero. Thus, for spin $\frac 12$ charges, physical states are singlet coverings. The electric term weights them depending on the position of the singlets while the plaquette interchanges singlet coverings (or annihilates them) as shown in Fig. \ref{fig:tagliacozzo2013simulation} b).     

The main features that $SU(2)$ gauge magnets share with QCD (and other non-Abelian gauge theories) are: the nature of confinement phases at weak (plaquettes dominate) and at strong coupling (electric terms dominate) and long-range entanglement between charges. To satisfy the Gauss law, the charges must form singlets with the nearby link spins, thus many singlets must be rearranged, and the allowed singlet coverings are different with respect to the ones of the vacuum, at least along a string between the charges. Such rearrangement generates long-range entanglement and costs an energy that increases linearly with the charge separation, i.e., linear confinement, see in Fig. \ref{fig:tagliacozzo2013simulation} c).    
To target such phenomena in a quantum simulator, it is enough to consider static spin $\frac 12$ charges \cite{tagliacozzo2013simulation}. Both spin $\frac 12$ or qubits on the links and on the sites are represented by ground and Rydberg states of atoms. The non-Abelian Gauss' law is converted into an energy penalty and added to the Hamiltonian. The dynamics of the generalised Hamiltonian is decomposed in a sequence of simultaneous Rabi transfers controlled by ancillary qubits and realised by Rydberg gates \cite{Mueller2009} see Fig. \ref{fig:tagliacozzo2013simulation} d), in a similar fashion as done for Abelian gauge theories \cite{weimer2010rydberg,tagliacozzo2013optical}.
In such a simulator, the ground state is prepared with a pair of opposite static charges at distance $L$ adiabatically (or super-adiabatically). By measuring the final state of the control qubits the energy of such a ground state can be computed with respect to the vacuum as a function of $L$, $E(L)$, and thus determines the string tension $\sigma=E(L)/L$. If $\sigma$ is finite for large $L$, there is a linearly confined phase.  The proposed Rydberg simulator can probe confinement at any coupling. By inspecting quantum correlations in the prepared ground state, it is also possible to experimentally access the long-range entanglement due to confinement in non-Abelian gauge theories. 

\subsubsection{Digital quantum simulation of $Z(2)$ lattice gauge theories with dynamical fermionic matter\cite{zohar2017digital,zohar2017digitalA}}

\begin{figure}[h!]
\begin{center}
\includegraphics[width=0.5\columnwidth]{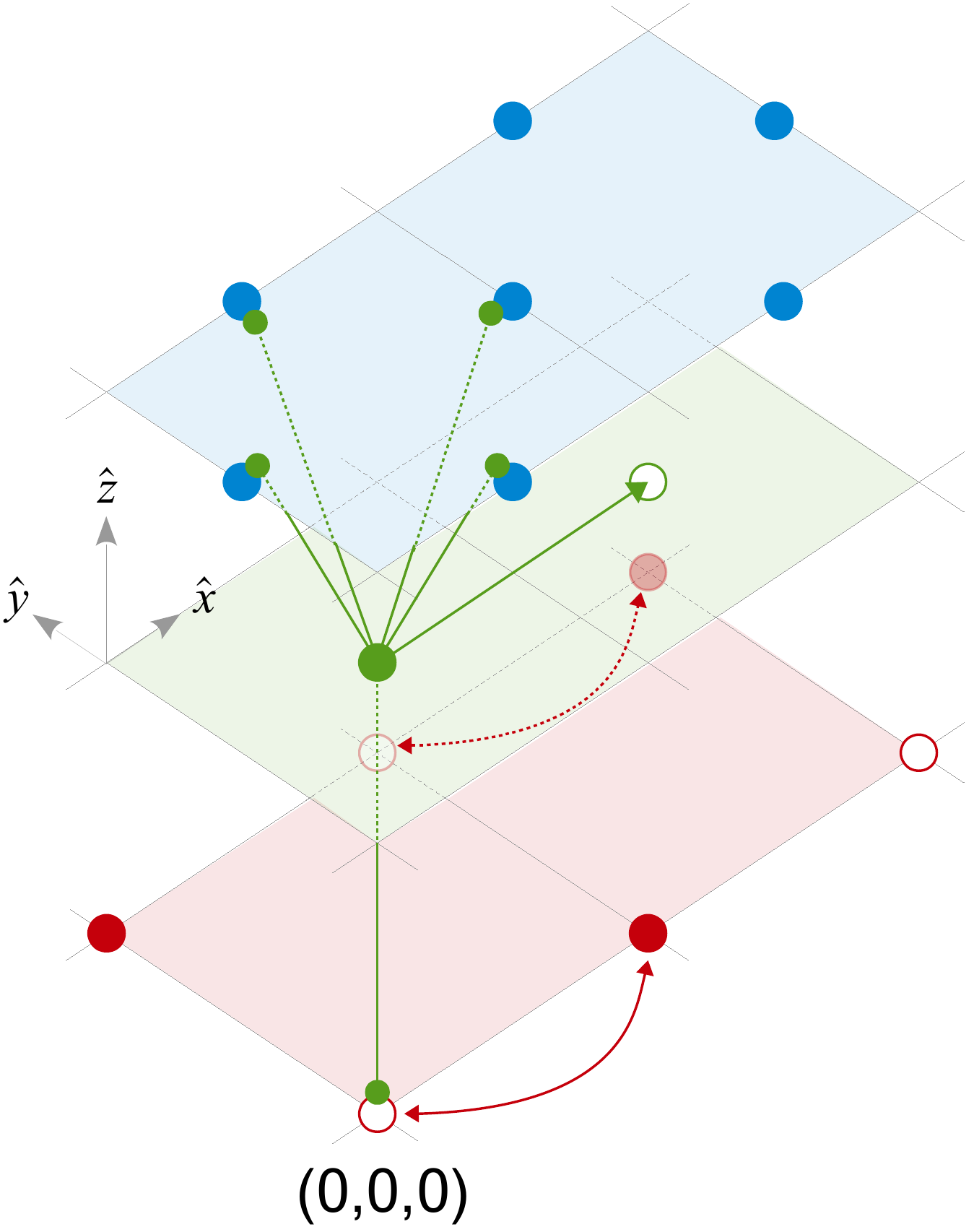} 
\end{center}
\caption{Different atomic species reside on different vertical layers. Green straight lines show how the auxiliary atoms have to move in order to realise interactions with the link atoms and the fermions, or to enter odd plaquettes. Red arrows show selective tunnelling of fermions across even horizontal links. From \cite{zohar2017digital}. }
\end{figure}

In a recent work \cite{zohar2017digital,zohar2017digitalA}, a digital scheme was introduced and its implementation with cold atoms was studied, for $Z(2)$ and $Z(3)$ lattice gauge theories. The scheme includes, in addition to the gauge and matter degrees of freedom, auxiliary particles that mediate the interactions and give rise to the desired gauge theory dynamics, by constructing stroboscopically the evolution from small time steps. The individual time steps respect local gauge invariance, so errors due to the digitisation will not break local gauge symmetry. Moreover, it is shown that the required three- and four-body interactions, can be obtained by a sequence of two-body interactions between the physical degrees of freedom and the  ancillary particles. The construction is general in form, and valid for any gauge group. Its generality and simplicity follows from the use of a mathematical quantum mechanical object called stator \cite{reznik2002remote,zohar2017half}.

\section{Analog Quantum simulations}
\label{QS}

\subsection{Analog quantum simulation of classical gauge potential}

The complete challenge of quantum simulating a lattice gauge theory has many interesting side products such as the study of classical gauge potential and the related topological insulators. In these cases, the gauge potential appears just as a classical configuration of the vector potential that is described by the minimal Wilson line in the lattice or the Peierls substitution of the hopping term. In the following, several theoretical proposals and an experimental realisation are reviewed \cite{bermudez2010wilson,hauke2012non,mancini2015observation,alaeian2019creating}.

\subsubsection{Wilson Fermions and Axion Electrodynamics in Optical Lattices\cite{bermudez2010wilson}}

Ultra-cold Fermi gases in optical superlattices can be used as quantum simulators of relativistic lattice fermions in $3+1$ dimensions. By exploiting laser-assisted tunnelling, an analogue of the so-called naive Dirac fermions is characterised in \cite{bermudez2010wilson}. An implementation of Wilson fermions is shown, and it is discussed how their mass can be inverted by tuning the laser intensities. In this regime of the quantum simulator, Maxwell electrodynamics is replaced by axion electrodynamics: a three dimensional topological insulator.

In particular, a concrete proposal for the realisation of laser-assisted tunnelling in a spin-independent optical lattice trapping a multi-spin atomic gas is presented. The setup consists of a spin-independent optical lattice that traps a collection of hyperfine states of the same alkaline atom, to which the different degrees of freedom of the field theory to be simulated are then mapped. Remarkably enough, it is possible to tailor a wide range of spin-flipping hopping operators, which opens an interesting route to push the experiments beyond the standard superfluid-Mott insulator transition. The presented scheme combines bi-chromatic lattices and Raman transfers, to adiabatically eliminate the states trapped in the middle of each lattice link. These states act as simple spectators that assist the tunnelling of atoms between the main minima of the optical lattice. This mechanism is clearly supported by numerical simulations of the time evolution of the atomic population between the different optical-lattice sites. 

\begin{figure}[h!]
\begin{center}
\includegraphics[width=0.7\columnwidth]{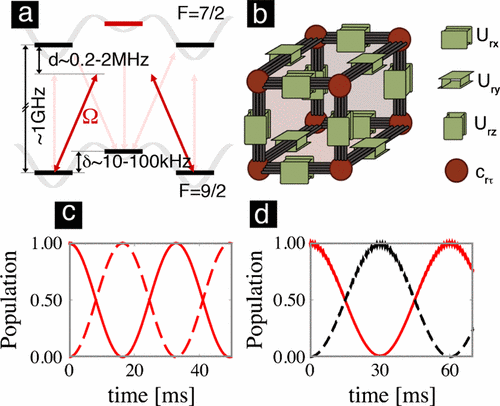} 
\end{center}
\caption{(a) Superlattice potential (grey lines). The hopping between $F=9/2$ levels is laser-assisted via an intermediate $F=7/2$ state. The coupling is induced by an off-resonant Raman transition with Rabi frequency. (b) Scheme of the four states of the $F=9/2$ manifold [(red) vertices], connected by laser-induced hopping [(green) boxes]. (c) Time-evolution of the populations of the neighbouring hyperfine levels. The solid (dashed) line is used for site $i$ ($i+1$); the red (black) line is used for $m_{F}=9/2$ ($m_{F}=7/2$). A clear spin-preserving Rabi oscillation between neighbouring sites is shown. (d) The same as before for a spin-flipping hopping. Notice the need for a superlattice staggering (10-20 kHz) in order to avoid on-site spin-flipping. From \cite{bermudez2010wilson}.}
\end{figure}

Such a device could have important applications in the quantum simulation of non-interacting lattice field theories, which are characterised, in their discrete version, by on-site and nearest-neighbour hopping Hamiltonians. Once the fields of the theory to be simulated are mapped into the atomic hyperfine states, the desired operators correspond to population transfers between such levels. The former can be realised by standard microwaves, whereas the latter might be tailored with the laser-assisted schemes. Combining this trapping scheme with Fermi gases, this platform would open a new route towards the simulation of high-energy physics and topological insulators.

\subsubsection{Non-Abelian gauge fields and topological insulators in shaken optical lattices\cite{hauke2012non}}

\begin{figure}[htb]
\begin{center}
\resizebox{0.99\columnwidth}{!}{\includegraphics{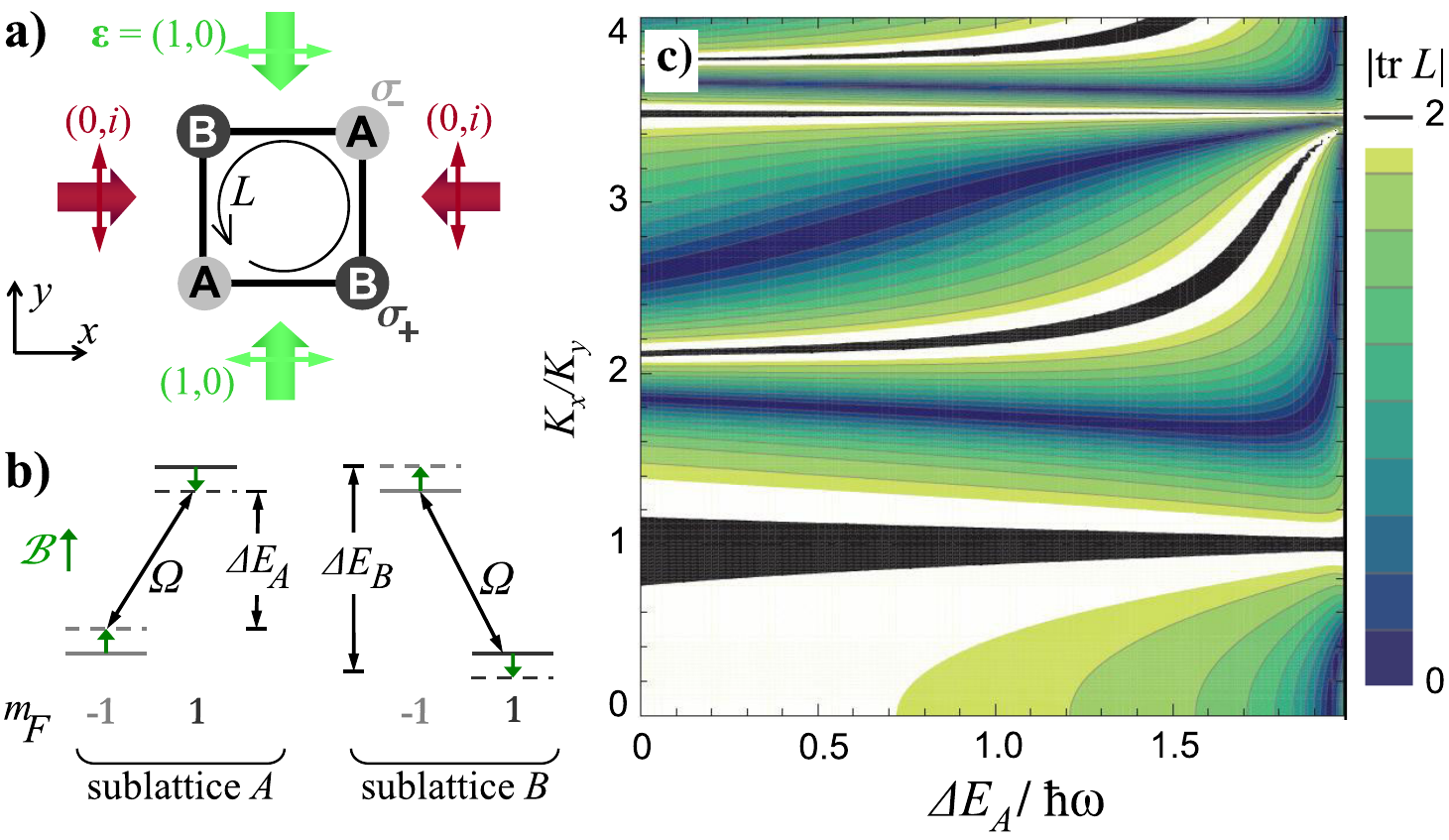}}
\caption{(color online) Non-Abelian $SU(2)$ gauge fields from lattice shaking.
(a) The optical lattice: two standing laser waves (with a phase shift of $\pi/2$ and in-plane polarisation as denoted in the figure) create a bipartite square lattice with alternating $\sigma^+$ and $\sigma^-$ polarised sites (A and B). $m_F = \pm 1$ particles feel an energy difference of $\pm \Delta E$ between A and B sites. (b) The resulting level scheme. A
constant B field realises an additional on-site energy splitting $\Delta E'$ (green arrow) such that $\Delta E_{A,B} =\sqrt{\pm\Delta E + \Delta E'}$ becomes sub-lattice dependent. In combination with a coupling $\Omega$ of both spin states realised microwave (or magnetic) fields the sub-lattice splitting gives local spin eigenbasis that differ in the two sub-lattices by $SU(2)$ rotation. With lattice shaking, the $SU(2)$ gauge transformation is converted by time-averaging into a non-trivial synthetic non-Abelian gauge field. c) Accessible Wilson loops $|$Tr $L|$ for convenient experimental parameters ($K_y/K_x$ is the relative shaking amplitude in the $y$ and $x$ directions and $\omega$ is the frequency). Deviations from 2 imply non-Abelian physics: outside the white (black) regions, $|$Tr $L| < 1.9$ ( $< 1.99$).
} 
\label{fig:fig4_hauke2012non}
\end{center}
\end{figure}

A preliminary step to quantum simulating full-fledged non-Abelian gauge theory is to consider classical non-Abelian gauge fields. It is thus crucial to devise efficient experimental strategies to achieve classical synthetic non-Abelian gauge fields for ultra-cold atoms in the bulk and in optical lattices \cite{hauke2012non}. The latter situation is especially interesting as it allows for an anomalous quantum Hall effect \cite{Goldman2009} and, in combination with strong interactions, for fractional quantum Hall states with non-Abelian anyonic excitations \cite{Burrello2010}. On the lattice, synthetic non-Abelian gauge fields (in LGT language, the Wilson operators on the links) correspond to tunnelling matrices that determine the superposition each atomic species is sent to when it tunnels to a neighbouring site. In \cite{hauke2012non} it is shown for the first time how to achieve such matrices for the $SU(2)$ gauge group from lattice shaking (for general theory of lattice shaking and Floquet driving see \cite{Eckardt2017}). For simplicity, consider a spin-dependent square lattice described by  Fig. \ref{fig:fig4_hauke2012non}, which in combination with a uniform magnetic field produces a sublattice-dependent energy splitting between the spin up and down states, e.g., $m_F=\pm 1$ hyperfine states of $^{87}$Rb. With a constant microwave beam $\Omega$ coupling the two spin states, the eigenstates of the on-site Hamiltonian in the two sublattices thus differ by an $SU(2)$ rotation. Thus, an atom tunnelling between the sublattices would experience an $SU(2)$ gauge field that is trivial: the product of the tunnelling matrices around the plaquette $L$ is the identity, i.e., its trace -the Wilson loop- is 2. Furthermore, such tunnelling is highly suppressed as out of resonance due to the energy offset between the sublattices (if it is sufficiently large).  However, the energy conservation can be restored, and thus the tunnelling, by shaking the lattice at a frequency commensurable with the energy off-sets. Such analysis can be made precise by time averaging the total Hamiltonian in the rotating frame of the driving plus the on-site Hamiltonian \cite{hauke2012non}. The main result is that for feasible experimental parameters, generic non-trivial classical $SU(2)$ gauge field configurations can be engineered, i.e., characterised by $|$Tr $L  |<2$, as shown in Fig. \ref{fig:fig4_hauke2012non}.

\subsubsection{Observation of chiral edge states with neutral fermions in synthetic Hall ribbons\cite{mancini2015observation}}

\begin{figure}
\begin{center}
\includegraphics[width=\columnwidth]{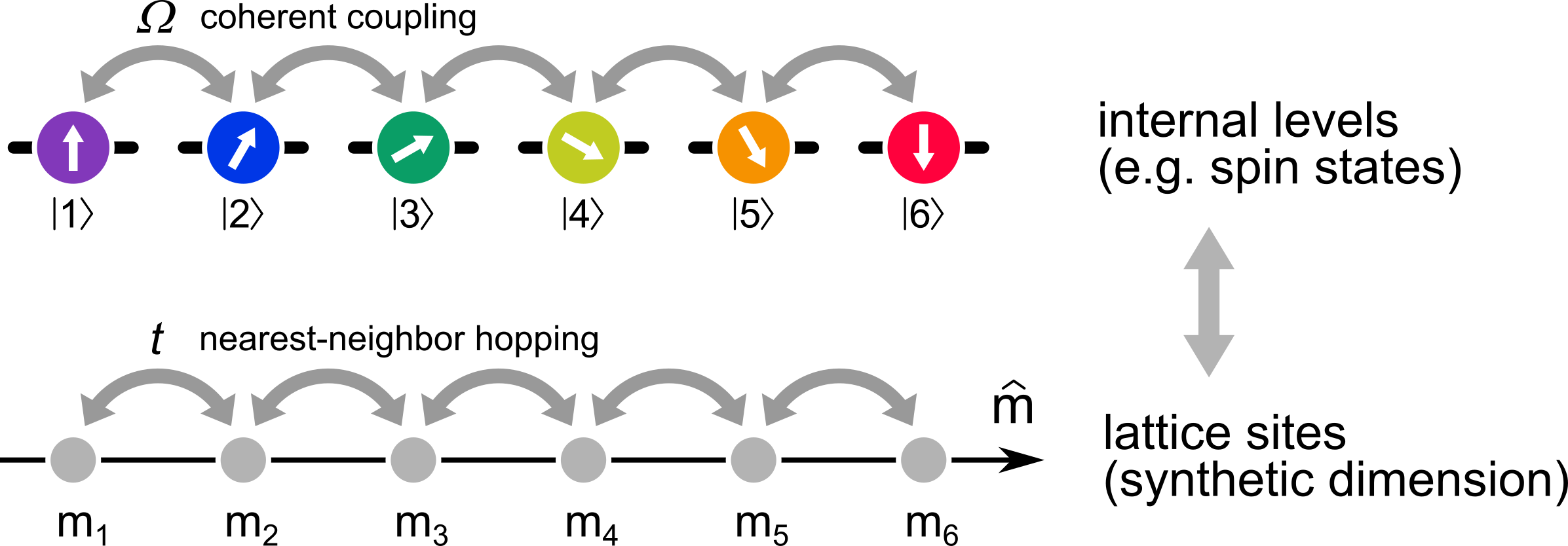}
\end{center}
\caption{A convenient approach for analog quantum simulation with ultra-cold neutral atoms relies in the concept of "synthetic dimensions": a coherent coupling $\Omega$ between internal atomic states $|i\rangle$ (induced e.g. with laser fields) mimics an effective hopping $t$ between sites with position $m_i$ of a fictitious synthetic dimension $\hat{m}$.}
\label{fig-cf-1}
\end{figure}

A powerful resource for the implementation of analog quantum simulations with ultra-cold-atomic samples is based on the manipulation of the internal atomic degrees of freedom. In this context, atoms with two valence electrons (such as alkaline-earth elements or lanthanide ytterbium) represent a convenient choice, as they provide access to several stable internal states (either nuclear or electronic), that can be initialised, manipulated with long-coherence times and read-out optically with high-fidelity. This platform is particularly suitable for implementing the concept of "synthetic dimensions", in which a manifold of internal states is mapped onto effective positions along a fictitious discretised extra-dimension, and the coherent optical coupling between the different states can be described as an effective hopping between synthetic sites (see Fig. \ref{fig-cf-1} for a sketch of the general idea).

\begin{figure}
\includegraphics[width=\columnwidth]{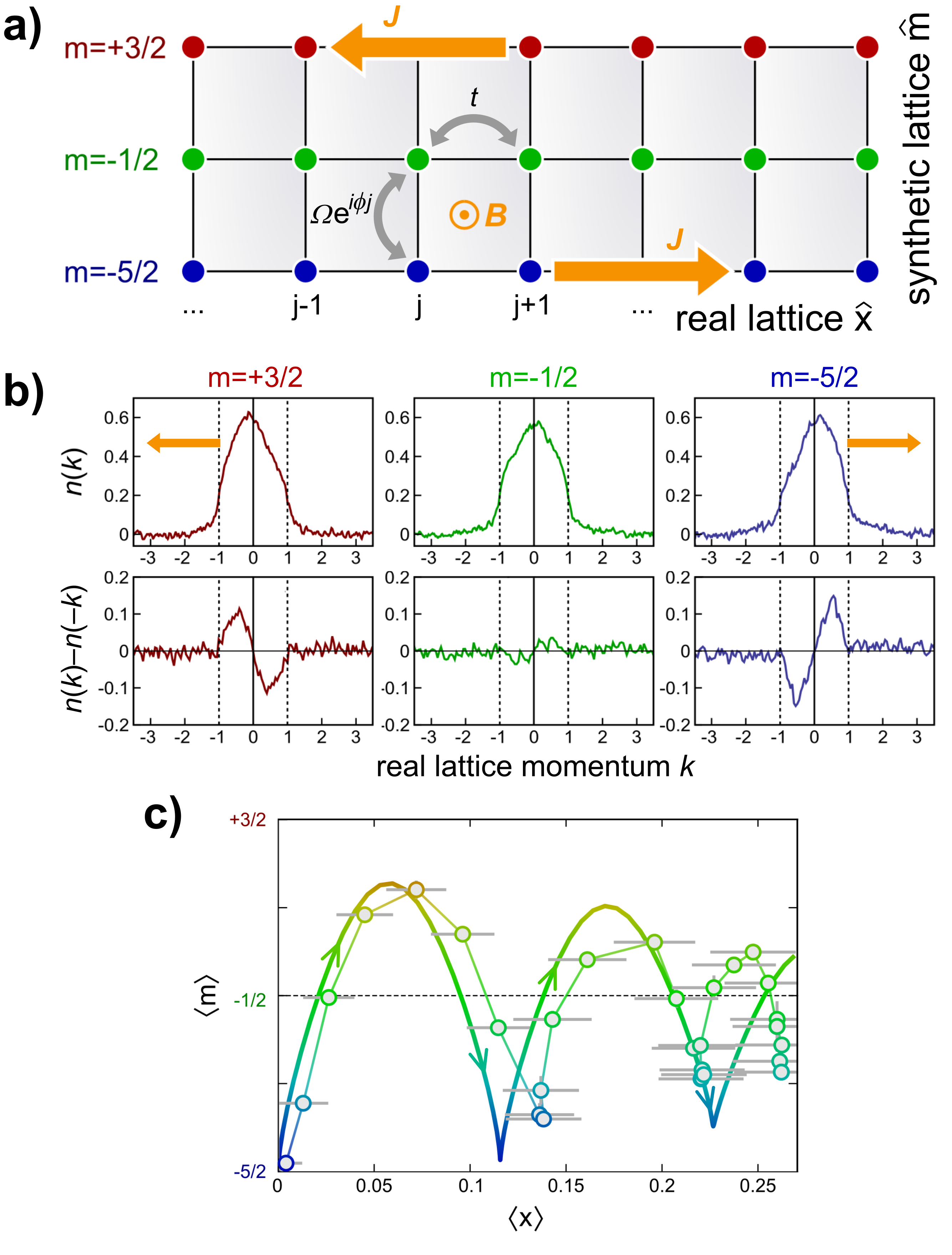}
\caption{a) Sketch of the experimental configuration employed in Ref. \cite{mancini2015observation} for the generation of classical gauge potentials. b) Measured lattice momentum distribution along the different synthetic legs of the ladder, showing the emergence of steady-state chiral edge currents $J$. c) Experimental reconstruction of the average trajectory on the synthetic strip, evidencing a non-equilibrium dynamics induced after a quench in the tunnelling. Adapted from Ref. \cite{mancini2015observation}.}
\label{fig-cf-2}
\end{figure}

This approach, initially proposed in Ref. \cite{boada2012quantumsimulation}, provided a very convenient method for the realisation of tunable background gauge potentials \cite{celi2014syntheticgauge}. This is explained in Fig. \ref{fig-cf-2}a, showing a hybrid lattice structure combining one real direction (discretised by a real optical lattice in sites with position $j$) with a synthetic direction composed by internal atomic states, depicted orthogonally to the real one. The hopping matrix element along the synthetic lattice is a complex quantity $\Omega e^{i \phi j}$, with an argument depending on the phase of the electric field inducing the transition between the internal states generating the synthetic dimension. As a consequence, hopping around a unit cell of the real and synthetic lattice can be described in terms of an effective geometric Aharonov-Bohm phase $\phi$ associated with the effect of a background synthetic magnetic field on effectively charged particles. 

This idea was experimentally realised in Ref. \cite{mancini2015observation} (and in a related experiment \cite{stuhl2015visualizingedge}), where different nuclear-spin projection states of fermionic $^{173}$Yb atoms were coupled coherently with a two-photon Raman transition, realising the scheme of Fig. \ref{fig-cf-2}a. This system is particularly suited to study edge physics, as the synthetic dimension is made up by a finite number of states/sites, resulting in a ladder geometry (with a tunable number of legs). The emergence of steady-state chiral currents at the edges of the ladder was detected with a state-dependent imaging technique (corresponding to a single-leg detection in momentum-space), evidencing a counter-propagating atomic motion on the two outer legs. This behaviour is shown in the  graphs of Fig. \ref{fig-cf-2}b for a three-leg ladder made by three nuclear-spin projection states $m=-5/2, -1/2, +3/2$. The asymmetry of the lattice momentum distribution $n(k)$ (along the real direction) provides a direct measurement of the steady-state edge currents $J$ induced in the system after an adiabatic loading. Non-equilibrium dynamics was also studied after imposing a quench on the system: after preparing the fermionic particles on a single external leg, tunnelling along the rungs was suddenly activated. The ensuing dynamics was studied by reconstructing the average trajectory of the particles on the synthetic strip: the result is shown in Fig. \ref{fig-cf-2}c, evidencing a "skipping-orbit" motion, in which the synthetic gauge field bends the centre-of-mass trajectory in a cyclotron-like fashion, with repeated bouncing at the edge of the strip producing a net motion along the edge (akin to the steady-state chiral currents discussed above).

\begin{figure}
\begin{center}
\includegraphics[width=0.85\columnwidth]{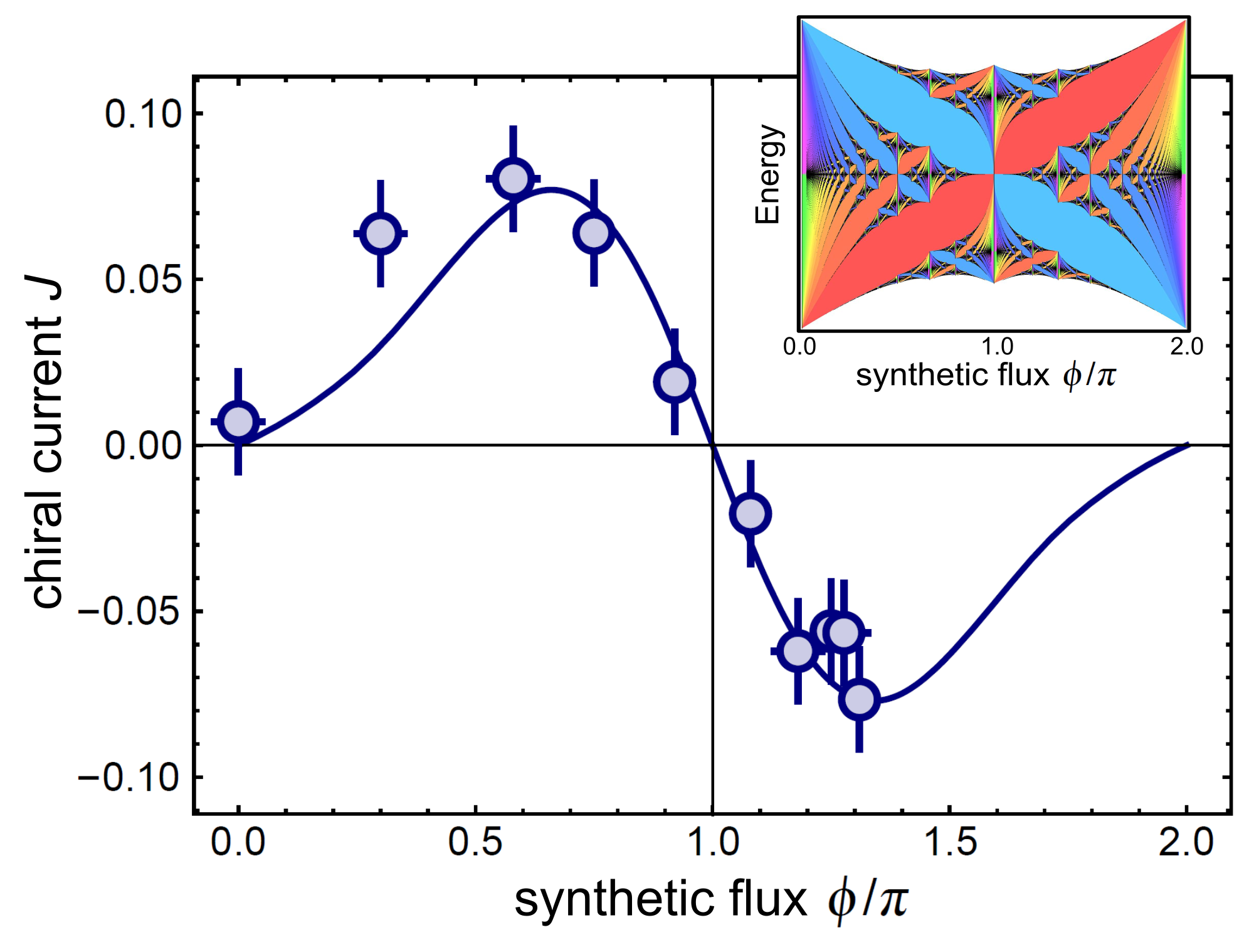}
\end{center}
\caption{The main graph shows the magnitude and direction of the chiral current $J$ vs. synthetic magnetic flux in an experimental configuration similar to that shown in Fig. \ref{fig-cf-2}b (with two legs only). The reversal of the edge current above $\phi=\pi$ flux is reminiscent of the change in sign of the Chern number for an extended two-dimensional system (plotted in the inset of the figure as a shading of the Hofstadter butterfly spectrum, with warm colours corresponding to positive Chern number and cold colours corresponding to negative Chern number). Experimental data taken from Ref. \cite{livi2016synthetic}.}
\label{fig-cf-3}
\end{figure}

Synthetic dimensions are a quite general concept, that can be adapted to different implementations. In a follow-up of that experimental work \cite{livi2016synthetic}, with a different implementation relying on the manipulation of the electronic state of $^{173}$Yb (rather than the nuclear-spin states) with a single-photon optical clock transition, the strength and direction of the chiral currents was measured as a function of the synthetic magnetic flux $\phi$, all the way from zero to above half of a quantum of flux per unit cell (a quantum of flux corresponding to $\phi=2\pi$). The measurements are summarised in Fig. \ref{fig-cf-3}, where the chiral current $J$ is plotted vs. $\phi$. It is apparent that the chiral current vanishes and then changes direction crossing the $\phi=\pi$ point. This can be explained by recalling the two underlying symmetries of the system: the time-reversal symmetry (broken by the magnetic field) imposing $J(\phi)=-J(-\phi)$ and the periodicity condition coming from the underlying lattice structure $J(\phi)=J(\phi+2\pi)$. A similar conclusion could be reached by considering the behaviour of an extended two-dimensional system, which realises the Harper-Hofstadter model describing charged particles in a two-dimensional lattice with a transverse magnetic field. The topological invariant, i.e., the Chern number, resulting from that model (shown in the inset of Fig. \ref{fig-cf-3} as a colour-coded shading of the Hofstadter butterfly spectrum) changes from positive to negative values when the $\phi=\pi$ point is crossed: this sign change is connected, by the bulk-boundary correspondence principle, to the reversal of chiral currents measured in the experimentally-realised ladder geometry.

The concept of synthetic dimensions is promising for the realisation of quantum simulators with advanced control on topological properties and site connectivities. For instance, controlling the parameters of the atom-laser interaction gives the possibility of engineering systems with periodic boundary conditions (i.e., a compactified extra-dimension) \cite{boada2015quantumsimulation} or different kinds of topological ladders. Furthermore, combining this idea with atom-atom interactions (featuring an intrinsic global $SU(N)$ symmetry in $^{173}$Yb atoms) results in intriguing possibilities for engineering synthetic quantum systems, as interactions along the synthetic dimension have an intrinsic non-local character (as different sites of the synthetic dimension correspond to the same physical position of space). While effects of atom-atom interactions have already been studied theoretically in combination with the classical gauge potential described above, with the prediction of strongly correlated states that are reminiscent of the fractional quantum Hall effect \cite{barbarino2015magnetic,taddia2017topological,strinati2017laughlin}, the application of this approach to the realisation of other kinds of gauge fields is still to be investigated.

\subsection{Quantum Simulation of Abelian Gauge Fields with ultra-cold atoms}

\subsubsection{Atomic quantum simulator for lattice gauge theories and ring exchange models\cite{buchler2005atomic}}

Reference \cite{buchler2005atomic} presents the design of a ring exchange interaction in cold-atomic gases subjected to an optical lattice using well-understood tools for manipulating and controlling such gases. The strength of this interaction can be tuned independently and describes the correlated hopping of two bosons. This design offers the possibility for the atomic quantum simulation of a certain class of strong coupling Hamiltonians and opens an alternative approach for the study of novel and exotic phases with strong correlations. A setup is discussed where this coupling term may allow for the realisation and observation of exotic quantum phases, including a deconfined insulator described by the Coulomb phase of a three-dimensional U(1) lattice gauge theory. 

\subsubsection{Cold-atom simulation of interacting relativistic quantum field theories\cite{cirac2010cold}}

Dirac fermions self-interacting or coupled to dynamic scalar fields can emerge in the low-energy sector of designed bosonic and fermionic cold-atom systems. In the reference \cite{cirac2010cold} this is illustrated with two examples defined in two space-time dimensions: the first one is the self-interacting Thirring model, and the second one is a model of Dirac fermions coupled to a dynamic scalar field that gives rise to the Gross-Neveu model. The proposed cold-atom experiments can be used to probe spectral or correlation properties of interacting quantum field theories thereby presenting an alternative to lattice gauge theory simulations. The Hamiltonians of these systems are supported on one spatial dimension. The necessary building blocks are the Dirac Hamiltonian, which describes relativistic fermions, and the interaction of fermions with themselves or with a dynamic scalar field. The presented models are exactly solvable and serve for demonstrating the ability to simulate important properties of the standard model such as dynamical symmetry breaking and mass generation with cold atoms.

\subsubsection{Confinement and lattice QED electric flux-tubes simulated with ultra-cold atoms\cite{zohar2011confinement}}

The effect of confinement is known to be linked with a mechanism of electric flux tube formation that gives rise to a linear binding potential between quarks. While confinement is a property of non-Abelian gauge theories including QCD, it has been shown that the Abelian model of compact quantum electrodynamics (cQED), also gives rise to similar phenomena. Particularly, it has been shown long ago by Polyakov that in cQED models in 2+1 dimensions, the effect of confinement persists for all values of the coupling strength, in both the strong and the weak coupling regimes, due to non-perturbative effects of instantons \cite{polyakov1977quark}. Lattice cQED in 2+1 dimensions, hence provides one of the simplest play grounds to study confinement in a setup that contains some essential properties of full fledged QCD, in a rather simple Abelian system. 

It was first suggested in \cite{zohar2011confinement} that the well-known gauge invariant Kogut-Susskind Hamiltonian, which describes cQED in 2+1-d, can be simulated by representing each link along the lattice by a localised BEC, properly tuning the atomic scattering interactions, and using external lasers. In this proposal, the cQEDs' degrees of freedom on each link are the condensate's phases which correspond to the periodic cQED vector potentials and the atomic number operator to the quantised electric field on the link. Charged sources are non-dynamical, and introduced to the model by coupling the fields to external static charges. To obtain the Kogut-Susskind Hamiltonian, it is shown that particular two- and four-body interactions between the condensates provide manifest local gauge invariance. Furthermore, to avoid the hopping processes of an ordinary Bose-Hubbard model, one introduces a four-species two-dimensional setup, Raman transitions and two-atom scattering processes in order to obtain particular 'diagonal' hopping and nonlinear interactions. It is then shown that a certain choice of parameters gives rise to gauge invariance in the low-energy sector, hence compact QED emerges. In the strong coupling limit, the atomic system gives rise to electric flux-tubes and confinement. To observe such effects one needs to measure local density deviations of the BECs. It is reasoned that the effect should persist also outside the perturbatively calculable regime.

\subsubsection{Atomic Quantum Simulation of Dynamical Gauge Fields coupled to Fermionic Matter: From String Breaking to Evolution after a Quench\cite{banerjee2012atomic}}

In \cite{banerjee2012atomic}, a quantum simulator for lattice gauge theories is proposed, where bosonic gauge fields are coupled to fermionic matter, which allows the demonstration of experiments for phenomena such as time-dependent string breaking and the dynamics after a quench. Using a Fermi-Bose mixture of ultra-cold atoms in an optical lattice, a quantum simulator is constructed for a $U(1)$ gauge theory coupled to fermionic matter. The construction is based on quantum links which realise a continuous gauge symmetry with discrete quantum variables. At low energies, quantum link models with staggered fermions emerge from a Hubbard-type model which can be quantum simulated. This allows one to investigate string breaking as well as the real-time evolution after a quench in gauge theories, which are inaccessible to classical simulation methods. While the basic elements behind the model have been demonstrated individually in the laboratory, the combination of these tools and the extension to higher dimensions remain a challenge to be tackled in future generations of optical lattice experiments. 

\begin{figure}
\begin{center}
\includegraphics[width=0.85\columnwidth]{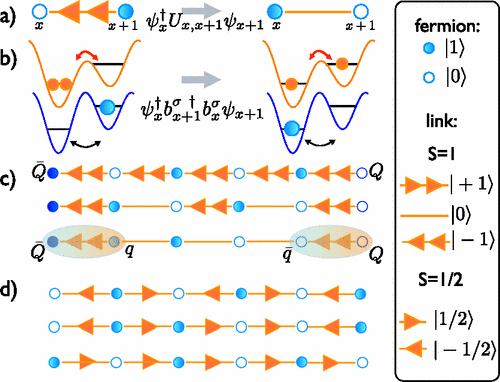}
\end{center}
\caption{(a) Correlated hop of a fermion assisted by gauge bosons consistent with Gauss' law in a QLM with spin $S=1$.  (b) Realisation of the process in (a) with bosonic and fermionic atoms in an optical superlattice. (c) Breaking of a string connecting a static quark-anti-quark pair: from an unbroken string (top), via fermion hopping (middle), to two mesons separated by vacuum (bottom). (d) From a parity invariant staggered flux state (top), via fermion hopping (middle), to the vacuum with spontaneous parity breaking. Taken from Ref. \cite{banerjee2012atomic}.}
\end{figure}

\subsubsection{Simulating Compact Quantum Electrodynamics with ultra-cold atoms: Probing confinement and non-perturbative effects\cite{zohar2012simulating}}

An alternative for simulating cQED, that relies on single atoms arranged on a lattice, (rather than small BECs), has been proposed in ref. \cite{zohar2012simulating}. In this work the idea is to use single atoms, described in terms of spin-gauge Hamiltonians, in an optical lattice, carrying $2l+1$ internal levels. As $l$ increases, it is found that the spin-gauge Hamiltonian manifests a rather fast convergence to the Kogut-Susskind cQED Hamiltonian model. This then enables the simulation in both the strong and weak regime of cQED in 2+1 dimensions. Hence the model can be used to simulate confinement effects in the non-perturbative regime, with a rather compact system and with a modest value of $l$. The case $l = 1$, corresponds to the lowest value, which is sufficient for demonstrating confinement and flux tubes between external charges, and is here explicitly constructed. This implementation with single atoms might be experimentally easier compared to the BEC approach \cite{zohar2011confinement}, as it does not rely on the overlaps of local BECs wave functions and thus involves larger Hamiltonian energy scales. 

\subsubsection{Quantum simulations of gauge theories with ultra-cold atoms: local gauge invariance from angular momentum conservation\cite{zohar2013quantum}}

Further development in realising compact QED in (1+1) and (2+1)-d is described in \cite{zohar2013quantum}, which also provides a rather systematic discussion of the structure and needs of quantum simulations of lattice gauge theory in the Hamiltonian form of Kogut and Susskind. In particular, in section IV of the article, a systematic method is described for constructing the required fermion-gauge boson interaction terms needed on the links for the particular cases involving a $U(1)$ interaction, and  a $Z(N)$ elementary interaction. The key point is that one can in fact reduce the problem of gauge invariance to that of conservation of angular momentum in elementary fermion-boson scatterings, by making a clever choice of the internal fermion and boson levels on the vertices and links. Then, while the fermion hops from one vertex to another, it interacts and scatters from bosonic species situated on the link.  By using an adequate selection of the internal fermionic and bosonic states one can then guarantee that the resulting interaction is gauge invariant. The general plaquette interaction term $\text{tr}(UUU^\dagger U^\dagger)$ has been obtained through an auxiliary particle that goes around loops, in section VI of the article. This method can be used for all gauge field interactions of the minimal form $\psi^\dagger_n U\psi_{n+1}$, with $U$ being a precise unitary, by producing such an interaction between an auxiliary particle and the physical matter. Given that that is indeed possible, it is shown how the "loop method" gives rise to plaquette interactions for the cases of U(1) and $SU(N)$ gauge fields.

\subsubsection{Quantum Spin Ice and dimer models with Rydberg atoms\cite{glaetzle2014quantum}}

Abelian gauge theories also play a rather prominent role in the theory of frustrated quantum magnets~\cite{Lacroix2010}. Refs.~\cite{glaetzle2014quantum,glaetzle2015designing} report two approaches aimed at realising the dynamics of frustrated quantum magnets with direct gauge theoretic interpretation, utilising Rydberg atoms trapped in lattices (either optical or tweezers).

\begin{figure}
\resizebox{0.99\columnwidth}{!}{\includegraphics{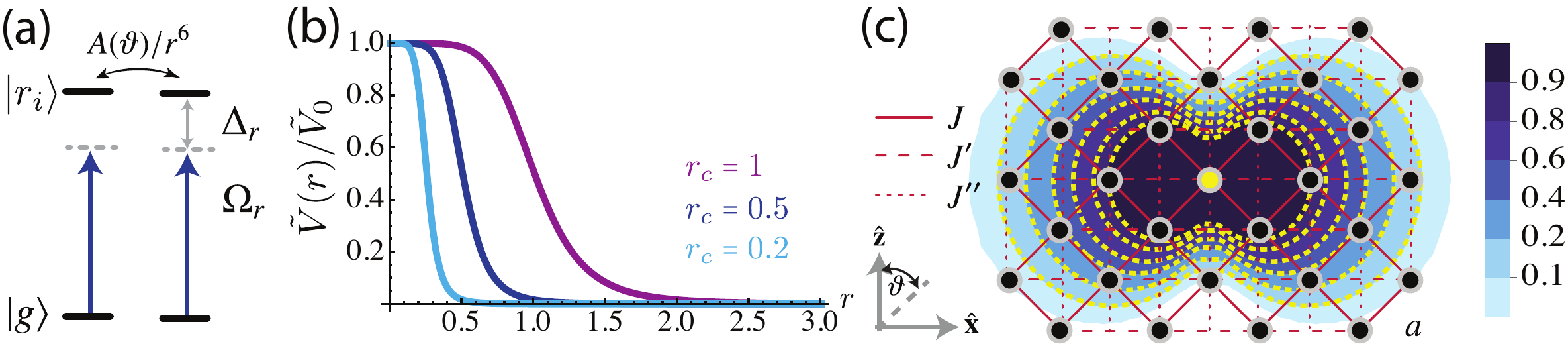}}
\caption{(a) Schematics of the energy levels (black lines) and lasers (thick solid dark-blue arrows) in the Rydberg dressing scheme. The ground state $|g\rangle$ of each atom is off-resonantly coupled to a Rydberg state $|r_i\rangle$ with a laser of Rabi frequency $\Omega_r$ and detuning $\Delta_r$. The coupling can be either direct (e.g., to p-states) or via an intermediate state. Pairwise interactions between the energetically well-isolated Rydberg states can be anisotropic, i.e. $V_{ij}(\mathbf{r})=A(\vartheta)/r^6$. (b) Typical behaviour of $V_{ij}(\mathbf{r})$ as a function of distance for different values of the Condon radius $r_c$. (c) Contour plot of the dressed ground state interaction $\tilde V_{ij}(\mathbf{r})/\tilde V_0$ between the atom in the middle (yellow circle) and the surrounding atoms (black circles) arranged on a square lattice. Figure adapted from Ref.~\cite{glaetzle2014quantum}. }
\label{fig:Ryd_Ice}
\end{figure}

Ref.~\cite{glaetzle2014quantum} introduces a toolbox to realise frustrated quantum magnets in two dimensions via Rydberg dressing. The main feature of this type of potentials, which is generated by off-resonantly coupling ground states to Rydberg states (see Fig.~\ref{fig:Ryd_Ice}a), is that, different from the bare Rydberg potential, it exhibits a plateau up to a critical distance of order of the Condon radius, as depicted in Fig.~\ref{fig:Ryd_Ice}b. This feature is already sufficient to generate gauge theory dynamics on a variety of lattices that can be decomposed in triangular unit cells, including Kagome and squagome ones. 

In addition, depending on the type of Rydberg states one is coupling to, it is possible to exploit the angular dependence of the Rydberg-Rydberg interactions (encoded in the angular dependent factor $A(\vartheta)$) to engineer Gauss' law constraints in other geometries. For the square lattice case, the resulting Hamiltonian is known as square Ice. The interactions needed to impose the corresponding constraint (the Gauss law, also known as {\it ice rule}) are both anisotropic and position dependent. This can be achieved by coupling ground state atoms to p-states: an example of the resulting interaction pattern is depicted in Fig.~\ref{fig:Ryd_Ice}c.

The resulting system dynamics, despite some additional features due to the long-range character of the Rydberg-dressing potentials, is able to reproduce the quantum ice rule, including a ground state with resonating valence-bond solid order. In addition, even in the absence of quantum fluctuations, an interesting thermal transition to an (imperfect) Coulomb phase takes place. This work has also served as a stimulus for tensor network simulations of two-dimensional systems, as reported in Ref.~\cite{tschirsich2019phase}.

In Ref.~\cite{glaetzle2015designing} (see also the related work in Ref.~\cite{Bijnen:2015aa} in the context of spin-1 models), the methods discussed above were considerably expanded. In particular, it was shown that the full dynamics within the Rydberg manifold can be exploited for the realisation of almost arbitrary spin-spin interactions, including $U(1)$ and $Z(2)$ invariant spin exchanges. Some of these terms could be additionally tuned exploiting quantum interference effects, or local AC Stark shifts. The possibility of utilising Rydberg atoms to engineer constrained models can also be directly applied to dynamics where the matter or gauge fields are integrated out explicitly. For instance, in \cite{surace2019lattice}, it was shown how recent experiments in Rydberg atoms arrays have already realised \cite{51atom} quantum simulations of gauge theories at large scales. In \cite{notarnicola2019real}, it is shown how to implement a Rydberg-atom quantum simulator to study the non-equilibrium dynamics of an Abelian $(1+1)$-D lattice gauge theory, the implementation locally codifies the degrees of freedom of a $Z(3)$ gauge field, once the matter field is integrated out by means of the Gauss local symmetries.

\subsubsection{Toolbox for Abelian lattice gauge theories with synthetic matter\cite{dutta2017toolbox}}

In \cite{dutta2017toolbox}, it is described what can be achieved by using the simplest possible implementation, taking a mixture of two Bose-gases on an optical lattice and working with lattice potentials and their time modulation (shaking). The Hamiltonian governing the system is thus the standard Bose-Hubbard Hamiltonian in which the atoms can hop between neighbouring sites around $x$  gaining an energy $J(x)$ and their interactions lead to an energy cost $U$.  The two species of bosons behave quite differently: {\it a}-bosons are strongly interacting, say in the hardcore limit, while {\it b}-bosons need to be non-interacting. Their mutual interaction is described by $U_{ab}$.

Upon fast modulation of this inter-species interaction and lattice potentials, an effective Hamiltonian is achieved (for the details of the implementation, refer to  \cite{dutta2017toolbox}), in a Floquet approximation, for the slow degrees of freedom in which the hopping of the {\it a}-bosons is modulated in amplitude and phase by the difference of occupations of the {\it b}-bosons. This means that the {\it a}-bosons behave as charged particles under a vector potential generated by the difference in occupations of {\it b}-bosons.

\begin{figure*}
\includegraphics[width=\textwidth]{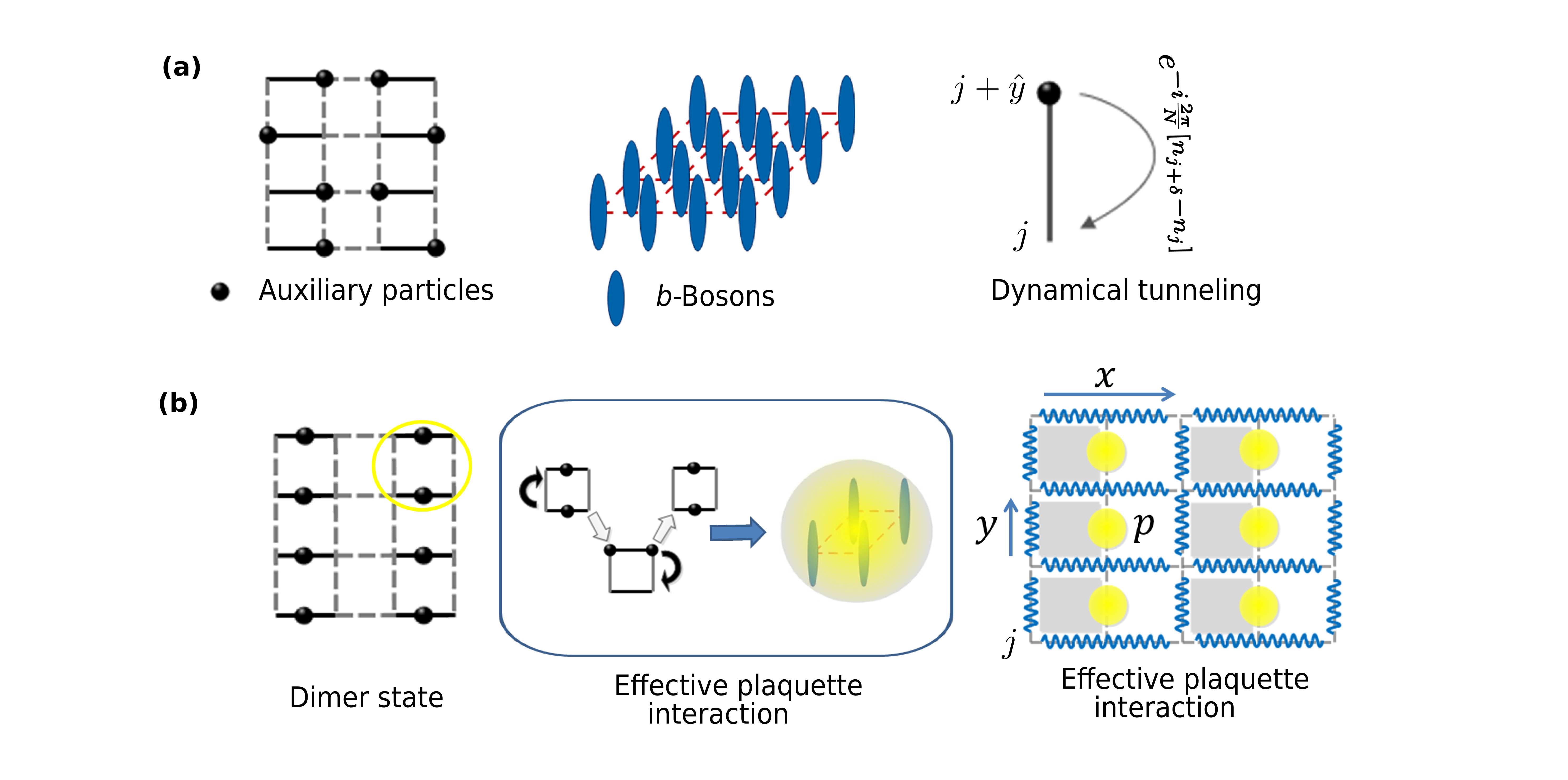}
\caption{Key ingredients - (a) Left panel: The auxiliary particles (black dots) are trapped in super-lattice geometry following the pattern in the figure. The tunnelling rate is large (small) along dark (dashed) bonds. There is one particle for every dark bond. Middle panel: A large occupation of {\it b}-bosons on every site is needed. This is achieved by trapping them along one-dimensional tubes (blue ovals) arranged in a square lattice geometry. Right panel: Upon appropriately shaking the set-up (see text for details),  the effective tunnelling of the {\it a}-bosons, at low-frequencies,  is modulated  in phase by the presence  of the {\it b}-bosons. (b) Left panel: By further increasing the tunnelling along dark bonds, the {\it a}-bosons delocalise on that bond. Middle panel: At second order in perturbation theory with respect to the weak tunnelling, the virtual processes depicted create an effective plaquette interaction for the {\it b}-bosons (yellow sphere). Right panel: Changing variable to plaquette variables, that can be thought as belonging to a coarse-grained lattice (shown by the wiggly blue lines) where the electric fields and the vector potentials live, taken from \cite{dutta2017toolbox}.}
 \label{figure1}
\end{figure*}

In this setting, an interesting scenario is obtained by considering initially the {\it a}-bosons hopping on a dimerised lattice as presented in the left-hand panel of Fig.~\ref{figure1}b where in a first approximation they  can only hop horizontally from even to odd sites with amplitude $J_a(h)$. Thus, at half-filling, there is an insulating phase where each dimerised link contains exactly one delocalised  {\it a}-boson as in Fig.~\ref{figure1}b. By switching on perturbatively both the hopping of {\it b}-bosons $J_b$ and the vertical hopping of {\it a}-bosons $J_a(v)$, applying second order perturbation theory with respect to the hopping ratios $J^2_b/J_a(h)$ and $J^2_a(v)/J_a(h)$ an effective Hamiltonian is achieved that can be described in terms of an exotic gauge theory. 

The gauge invariant variables are plaquettes construct from a coarse grained lattice, along the horizontal direction on whose links the following Hamiltonian is obtained, 
\begin{eqnarray}\label{eq:gaugeham}
{H_{\rm gauge}} &=& -2\sum_{p} \cos\hat{\mathcal{B}}_p - 2g^2 \sum_{\mathbf j} \left[ \cos\left ( \frac{4\pi}{N} \mathcal{E}_{({\mathbf j},\hat{x})}\right ) \right.\nonumber\\
&+& \left. 2\cos\left(\frac{2\pi}{N}[\mathcal{E}_{({\mathbf j},\hat{x})}-\mathcal{E}_{({\mathbf j}+\hat{y},\hat{x})}]\right)\right.\nonumber\\
&+&  \left.\cos\left(\frac{2\pi}{N}[\mathcal{E}_{({\mathbf j},\hat{y})}-\mathcal{E}_{({\mathbf j}+\hat{y},\hat{y})}]\right) \right].  
\end{eqnarray}
The $\mathcal{B}_p$ represent the standard plaquette magnetic term  and $\mathcal{E}$ is the electric field. $N$ represents the rank of the Abelian group $Z(N)$ and the $U(1)$ theory is obtained in the limit $N\to \infty$. Even in that limit, the above Hamiltonian differs from the standard two-dimensional QED Hamiltonian, since the electric field part of the Hamiltonian acts on neighbouring links, rather than on a single link as in the standard case.

\begin{figure}
\includegraphics[width=\columnwidth]{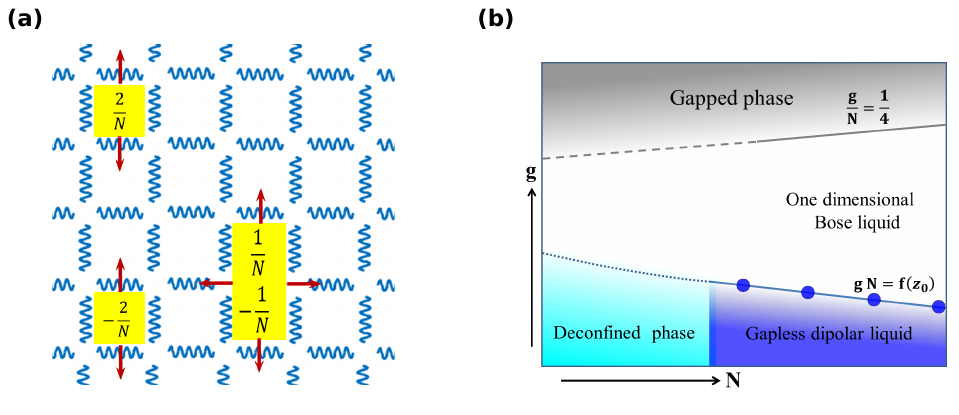}
\caption{Low-energy excitations and phase diagram. (a) At weak coupling the low-energy excitations are plaquette excitations, magnetic fluxes of strength $\pm 2/N$. Excitations can only be created in pairs inside a column and are free to move along that column. Alternatively, dipoles of magnetic fluxes involving the excitations of two adjacent plaquettes are free to move along both lattice directions.(b) Qualitative phase diagram of the Hamiltonian $H_{\rm gauge}$ in the $g$-$N$ plane. The upper shaded region denotes a gapped phase in the strong coupling limit. In the weak coupling limit, for low $N$, the system is gapped but deconfined (region shaded in light blue). In the $U(1)$ limit, the system becomes gapless and an exotic dipole liquid phase emerges (region shaded in dark blue). In the intermediate region, the system is effectively in a one-dimensional gapless Bose liquid phase\label{fig:phasediagram}, extracted from \cite{dutta2017toolbox}.}
\end{figure}

This specific pattern of the electric field implies that the low-energy sector of the model can be described by a gas of oriented magnetic dipoles rather than a gas of magnetic monopoles. As a consequence, even when the dipoles condense, they cannot screen the electric field and thus the model contains new exotic deconfined phases. The sketch of the conjectured  phase diagram is contained in Fig. \ref{fig:phasediagram}.

\subsubsection{Many-body localisation dynamics from gauge invariance\cite{brenes2018many}}

The experimental realisation of a synthetic lattice gauge theory reported in Ref.~\cite{martinez2016real} has immediately stimulated a novel interest in the real-time evolution of low-dimensional lattice gauge theories, following related works in the context of statistical mechanics models~\cite{Polkovnikov2011b}. A particularly active area of research in the latter field has been the study of disordered, interacting systems, which, under certain conditions, may fail to thermalise - a phenomenon referred to as many-body localisation (MBL)~\cite{Basko:2006hh}.

In Ref.~\cite{brenes2018many} it was shown that, in contrast to statistical mechanics models, LGTs may  display MBL even in the absence of any disorder. The feature at the basis of this phenomenon is the presence of Hilbert space sectors, which are a characteristic feature in LGTs. For Abelian theories, these subspaces are simply characterised by a given background charge distribution.

An arbitrary translationally invariant state is typically spanning several of these super-selection sectors: this implies that the resulting time evolution is actually sensitive to several, distinct background charge distributions, mimicking the time evolution of a system under an inhomogeneous potential. This does not map into uncorrelated disordered patterns, at least for the gauge group and the class of initial states considered in Ref.~\cite{brenes2018many}.

\begin{figure}
\resizebox{0.99\columnwidth}{!}{\includegraphics{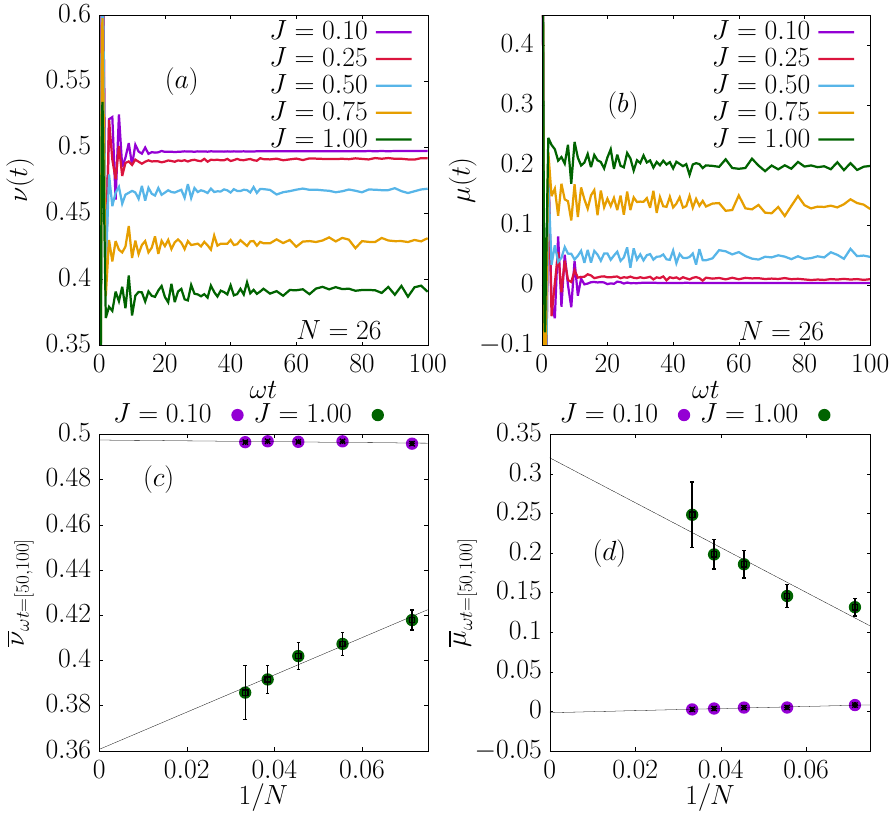}}
\caption{Dynamics following a quantum quench in the Schwinger model, starting from a state with gauge fields in an equal weight superposition of $E=(-1,0,1)$ and fermions in the bare vacuum. Top panels: overall staggered fermion occupation (a) and central staggered fermion occupation (b) for different values of the coupling strength parameter $J$, and $N = 26$. (c-d) Final time value of $\nu(t)$ and $\mu(t)$, respectively, averaged from $t=50$ to $t=100$ for different system sizes up to $N=30$ and $J=0.1, 1.0$, representative of the thermalising and non-thermalising dynamics. The absolute error due to averaging over realisations are shown as vertical bars for each point. Figure taken from Ref.~\cite{brenes2018many}.}
\label{fig:LGT_MBL}
\end{figure}

This type of dynamics was then analysed numerically using large-scale exact diagonalisation methods for the lattice Schwinger model, starting from initial states with gauge fields in an equal weight superposition of $E=(-1,0,1)$, and staggered fermions. A sample of these results is depicted in Fig.~\ref{fig:LGT_MBL}: absence of relaxation for two local observables is indicated in the upper two panels. The bottom two report a finite-size scaling analysis for a thermalising and non-thermalising case. The entanglement evolution of this MBL-type dynamics is however rather peculiar, as signalled by the evolution of the half-chain entanglement entropy: while this typically increases logarithmically with time, in the present context, a double-logarithmic increase was observed up to numerically accessible time-scales. 

A related phenomenology was also observed in theories without confinement in Ref.~\cite{Smith:2017aa}, and was also proposed as an implementation route for non-interacting models in Ref.~\cite{Smith:2018aa}

\subsubsection{Discretizing Quantum Field Theories \cite{TRS18,BTR18}}

The majority of platforms for quantum simulations: ultra-cold atoms in optical lattices, Rydberg atoms in traps or trapped ions deal with discretised versions of the considered Quantum Field Theory. As always in such cases, this can be viewed as a nuisance or an opportunity. In fact, to simulate LGT, discrete lattice models and their implementations are needed. But, studying and quantum simulating discrete versions of continuous QFT models might also lead to new fascinating physics. In \cite{BTR18} a Gross-Neveu-Wilson model was studied and its correlated symmetry-protected topological phases revealed. A Wilson-type discretisation of the Gross-Neveu model, a fermionic $N$-flavour quantum field theory displaying asymptotic freedom and chiral symmetry breaking, can serve as a playground to explore correlated symmetry-protected phases of matter using techniques borrowed from high-energy physics. A large-$N$ study, both in the Hamiltonian and Euclidean formalisms, was used and analysed. 

In \cite{TRS18} renormalisation group flows for Wilson-Hubbard matter and the topological Hamiltonian were studied. The aim was to understand the robustness of topological phases of matter in the presence of interactions, a problem that poses a difficult challenge in modern condensed matter physics, showing interesting connections to high-energy physics (see also other works facilitating quantum  simulations with ultra-cold atoms and ions: \cite{JPR16} for exploration of  interacting topological insulators with ultra-cold atoms in the synthetic Creutz-Hubbard model, \cite{MVE18} for symmetry-protected topological phases in lattice gauge theories, \cite{MVE19} for the study of the topological Schwinger model, \cite{GDG19} for $Z(N)$ gauge theories coupled to topological fermions. These connections present an analysis of the continuum long-wavelength description of a generic class of correlated topological insulators of Wilson-Hubbard type, feasible for experiments with quantum simulators. 

\subsubsection{Interacting bosons on dynamical lattices \cite{GGD18,GDG18,GDG19,CZL19}}

For ultra-cold atoms in optical lattices quantum simulations with bosons are more experimentally friendly than those with fermions. For these reasons there is a clear tendency in recent years to design and study lattice models in which bosons replace fermions. Particularly fruitful and fascinating are such models, dubbed as dynamical lattices, where some objects live on the lattice links, like in LGTs. 

Inspired by the so-called fluctuating bond superconductivity for fermionic Hubbard-like models that include quantised phonons on the links (see for instance \cite{JDC19}), a bosonic version of such models was formulated where phonons on the links are replaced by two states of a spin-$1/2$, which provides a minimal description of a dynamical lattice. These $Z(2)$ Bose-Hubbard models are extraordinarily rich and lead to bosonic Peierls transitions \cite{GGD18}, intertwined topological phases \cite{GDG18, GBG19}, symmetry-protected topological defects \cite{GDG19}, etc. Thus even in the absence of the gauge invariance, these models allow one to explore interesting strongly-correlated topological phenomena in atomic systems.

Similar models, now with exact $Z(2)$ gauge symmetry, were studied theoretically \cite{BSA18} and realised recently in a cold-atom experiment \cite{SGB19}. Peierls states and phases were also studied in models of Floquet-engineered vibrational dynamics in a two-dimensional array of trapped ions \cite{KHW19}. Recently, density-dependent Peierls phases were realised coupling dynamical gauge fields to matter \cite{Gorg:2018xyc}. 

The culmination of this effort is the paper \cite{CZL19}, in which the bosonic Schwinger model was studied and confinement and lack of thermalisation after quenches was observed. The vacuum of a relativistic theory of bosons coupled to a $U(1)$ gauge field in 1+1 dimensions (bosonic Schwinger model) was excited out of equilibrium by creating a spatially separated particle-anti-particle pair connected by an electric flux string. During the evolution, a strong confinement of the bosons is observed witnessed by the bending of their light cone, reminiscent of what was observed for the Ising model. As a consequence, for the time scales amenable to simulations, the system evades thermalisation and generates exotic asymptotic states. These states extend over two disjoint regions, an external deconfined region that seems to thermalise, and an inner core that reveals an area-law saturation of the entanglement entropy.

\subsection{Abelian Quantum Simulation with trapped ions and superconducting circuits}

\subsubsection{Quantum simulation of a lattice Schwinger model in a chain of trapped ions\cite{hauke2013quantum}}

Ref.~\cite{hauke2013quantum} represents the first attempt to identify gauge theory dynamics in trapped ion systems. From the point of view of analog quantum simulation, the main difference between atom and ion experiments resides in the type of degrees of freedom one can harness. While in former settings one is typically dealing with itinerant fermions and bosons, in the latter, the dynamics is dictated by the interactions between the ions' internal degrees of freedom (either nuclear or electronic spin) and the phonon modes generated by the ion crystal. 

The internal levels of the ions and the phonon modes are then coupled via external light fields: upon integration of the phonon degrees of freedom the resulting dynamics is then given by a spin chain, typically with $S=1/2$. Ref.~\cite{hauke2013quantum} exploits the fact that such interactions can be made spatially anisotropic. The main idea is that one can identify ions on one sub-lattice (A) as 'matter' fields, and ions on the other sub-lattice (B) as gauge fields in the quantum link formulation discussed above.

Within this setting, gauge invariance is then enforced by energy punishment. The latter is generated using a combination of two laser-assisted (e.g., Raman) interactions between qubits, that can be made spatially inhomogeneous either utilising local shifts, or by properly shaping the Rabi frequency of the light beams. The effective quantum link model dynamics is then generated in second order perturbation theory, similar to several atomic schemes. 

Here the main source of experimental imperfections is the presence of long-ranged terms in the spin-spin interactions. While the latter are gauge invariant, their presence might be detrimental to observe physical phenomena: as such, one has to find a balance between enforcing the constraint, while still keeping the effective QLM dynamics sufficiently fast. In \cite{hauke2013quantum}, this question was addressed in the context of ground state preparation: there, it is possible to find an optimal parameter regime that allows the observation of the Ising transition present in the QLM. 

Following this first work, other proposals have been presented to realise LGT dynamics in trapped ion systems. Ref.~\cite{Nath:2015to} dealt instead with two-dimensional models, mostly focusing on condensed matter realisations of $Z(2)$ quenched gauge theories which are known to exhibit topological quantum spin liquid behaviour. The main idea there was to utilise localised phonon modes generated in two-dimensional lattices either via Rydberg excitations, or by laser-pinning a subset of the trapped ions. At the few plaquette level, the model corresponds to the frustrated magnet introduced by Balents, Fisher and Girvin in Ref.~\cite{Balents:2002yi}.

\subsubsection{Superconducting Circuits for Quantum Simulation of Dynamical Gauge Fields\cite{marcos2013superconducting}}

\begin{figure}
\resizebox{0.9\columnwidth}{!}{\includegraphics{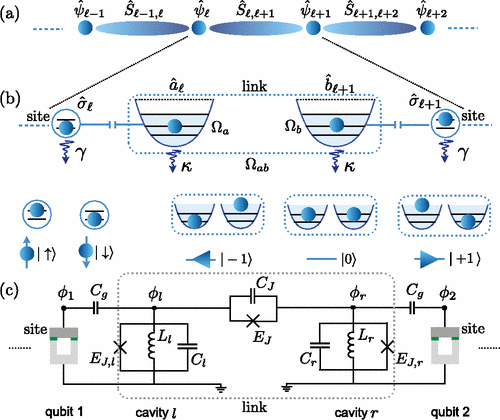}}
\caption{a) Pictorial view of a 1D quantum link model, where the operators on even (odd) sites represent matter (anti-matter) fields and the spin operators residing on each link represent the gauge fields. (b) Equivalent physical implementation, where two-level systems replace the fermionic matter fields and two oscillators with a fixed total number of excitations $N$ encode a spin $S=N/2$ on each link.(c) Superconducting-circuit implementation. Neighbouring super-conducting qubits on the sites of a 1D lattice are connected via two nonlinear LC resonators. Figure taken from Ref.~\cite{marcos2013superconducting}.}
\end{figure} 

The essential building blocks of a superconducting quantum simulator are described in~\cite{marcos2013superconducting} for dynamical lattice gauge field theories, where the basic physical effects can already be analysed with an experimentally available number of coupled superconducting circuits. This proposal analyses a one-dimensional $U(1)$ quantum link model, where superconducting qubits play the role of matter fields on the lattice sites and the gauge fields are represented by two coupled microwave resonators on each link between neighbouring sites. A detailed analysis of a minimal experimental protocol for probing the physics related to string breaking effects shows that, despite the presence of decoherence in these systems, distinctive phenomena from condensed-matter and high-energy physics can be visualised with state-of-the-art technology in small superconducting-circuit arrays.

\subsubsection{Loops and Strings in a Superconducting Lattice Gauge Simulator\cite{brennen2016loops}}

An architecture for an analog quantum simulator of electromagnetism in $2+1$ dimensions is proposed in \cite{brennen2016loops}, based on an array of superconducting fluxonium devices. The simulator can be tuned between intermediate and strong coupling regimes, and allows non-destructive measurements of non-local, space-like order and disorder parameters, resolving an outstanding gap in other proposals. Moreover, a physical encoding of the states is provided, where local electric field terms are non-trivial. The devices operate in a finite-dimensional manifold of low-lying eigenstates, to represent discrete electric fluxes on the lattice. The encoding is in the integer (spin 1) representation of the quantum link model formulation of compact U(1) lattice gauge theory.  In the article, it is shown how to engineer Gauss' law via an ancilla mediated gadget construction, and how to tune between the strongly coupled and intermediately coupled regimes. The protocol is rather robust to inhomogeneities, allowing for implementations in superconducting arrays, and numerical evidence is presented that lattice QED in quasi-$(2+1)$ dimensions exhibits confinement. Beyond ground state characterisation, the simulator can be used to probe dynamics and measure the evolution of non-local order or disorder parameters. The witnesses to the existence of the predicted confining phase of the model are provided by non-local order parameters from Wilson loops and disorder parameters from 't Hooft strings. In \cite{brennen2016loops}, it is shown how to construct such operators in this model and how to measure them non-destructively via dispersive coupling of the fluxonium islands to a microwave cavity mode. Numerical evidence was found for the existence of the confined phase in the ground state of the simulation Hamiltonian on a ladder geometry.

\begin{figure}
\resizebox{0.9\columnwidth}{!}{\includegraphics{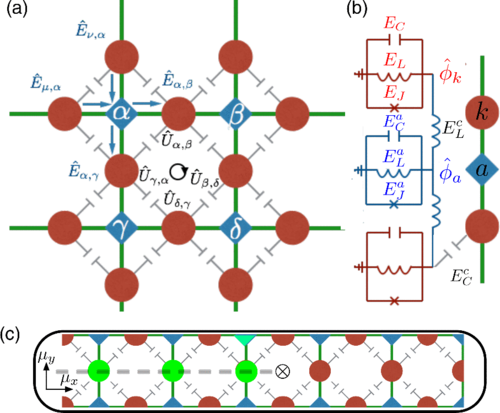}}
\caption{$U(1)$ quantum link model engineered in a fluxonium array. (a)``Electric'' $\hat{E}_{\alpha;\beta}$, and ``magnetic'  $\hat{U}_{\alpha;\beta}$, degrees of freedom are associated with links $<\alpha ; \beta>$ of a square lattice. The link degrees of freedom (red circles) are encoded in eigenstates of the fluxonia. The ancillae (blue diamonds) on vertices are inductively coupled to neighbouring link islands to mediate the Gauss constraint and plaquette interactions are obtained via link nearest-neighbour capacitive coupling. (b) Superconducting circuit elements used to build and couple components of the simulation. The link devices have local phase $\hat{\phi}_{\text{link}}$ and capacitive, inductive, and flux-biased Josephson energies $E_{C}$, $E_{L}$, and $E_{J}$, respectively, and similarly for the ancilla devices. The capacitive and inductive coupling energies are $E^{c}_{C}$ and $E^{c}_{L}$. (c) A minimal quasi-1D ``ladder'' implementation embedded in a microwave cavity (black box), in which a 't Hooft string of link fluxonia (green circles) can be measured via anancilla coupled to the cavity (green triangle). Figure taken from Ref.~\cite{brennen2016loops}.}
\end{figure} 

\subsection{Quantum Simulation of Non-Abelian Gauge Fields with ultra-cold atoms}

\subsubsection{Atomic Quantum Simulation of $U(N)$ and $SU(N)$ Non-Abelian Lattice Gauge Theories\cite{banerjee2013atomic}}

Using ultra-cold alkaline-earth atoms in optical lattices, \cite{banerjee2013atomic} constructs a quantum simulator for $U(N)$ and $SU(N)$ lattice gauge theories with fermionic matter based on quantum link models. These systems share qualitative features with QCD, including chiral symmetry breaking and restoration at non-zero temperature or baryon density. The proposal builds on the unique properties of quantum link models with rishons representing the gauge fields: this allows a formulation in terms of a Fermi-Hubbard model, which can be realised with multi-component alkaline-earth atoms in optical lattices, and where atomic physics provides both the control fields and measurement tools for studying the equilibrium and non-equilibrium dynamics and spectroscopy. 

\begin{figure}
\resizebox{0.9\columnwidth}{!}{\includegraphics{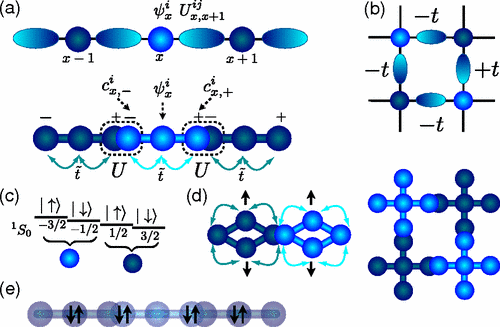}}
\caption{ a) (upper panel) $U(N)$ QLM in $(1+1)$-D with quark fields on lattice sites and gauge fields on links; (lower panel) hopping of alkaline-earth atoms between quark and rishon sites of the same shading. (b) Implementation of the QLM in rishon representation with fermionic atoms in $(2+1)$-D.(c) Encoding of the color degrees of freedom for $N=2$ in Zeeman states of a fermionic alkaline-earth atom with $I=3/2$.(d) Lattice structure to avoid the interaction in fermionic matter sites using a species-dependent optical lattice. (e) Initial state loaded in the optical lattice with a staggered distribution of doubly occupied sites for a $U(2)$ QLM with $N=2$. Figure taken from Ref.~\cite{banerjee2013atomic}.}
\end{figure} 

\subsubsection{A cold-atom quantum simulator for $SU(2)$ Yang-Mills lattice gauge theory\cite{zohar2013cold}}

A realisation of a non-Abelian lattice gauge theory, which is an $SU(2)$ Yang-Mills theory in $(1+1)$-dimensions is proposed in \cite{zohar2013cold}. The system one wants to simulate involves a non-Abelian gauge field and a dynamical fermionic matter field. As in ordinary lattice gauge theory the fermions are located at the vertices and the gauge fields on the links. Using staggered fermion methods, in the $(1+1)$-d case the Hamiltonian is equivalent, to the non-Abelian Schwinger model, with a minimal-coupling interaction of the form $\psi_n^\dagger U_n \psi_{n+1}$, involving a   $2\times 2$ unitary matrix. In order to simulate the non-Abelian $SU(2)$ Hamiltonian, this paper uses a particular realisation of the group elements and "left" and "right" generators, using a Jordan-Schwinger map, that connects harmonic oscillators (bosons) and angular momentum. Mapping $SU(N)$ to a bosonic system allows one to express the gauge fields by means of bosonic atoms in the prepotential method \cite{mathur2005harmonic}. For the $SU(2)$ group, one then needs four bosonic species for each link. Remarkably, gauge invariance arises as a consequence of conservation of angular momentum conservation and thus is fundamentally robust. However, the effective Hamiltonian obtained here is not valid beyond the strong coupling limit. 

\subsubsection{Constrained dynamics via the Zeno effect in quantum simulation: Implementing non-Abelian lattice gauge theories with cold atoms\cite{stannigel2014constrained}}

Implementing non-Abelian symmetries poses qualitatively new challenges from the theory side with respect to the Abelian case. In particular, energy punishment strategies, which are widespread for U(1) theories, are not immediately viable. The main reason is the following: consider a set of local generators $G^\alpha_x$, and a microscopic Hamiltonian of the type:
\begin{equation}
H = H_0+H_1+ \sum_x\sum_\alpha U_x^\alpha (G_x^\alpha)^2.
\end{equation}
where $H_0$ is a gauge invariant term, and $H_1$ is gauge variant. Here, differently from the Abelian case, where a single generator is considered, one deals with several constraints, that have to be satisfied in a symmetric manner - that is, $U^\alpha_x=U_x$. Typically, this requires fine-tuning, making energy punishment strategies not immediately viable. 

\begin{figure}[htb]
\begin{center}
\resizebox{0.99\columnwidth}{!}{\includegraphics{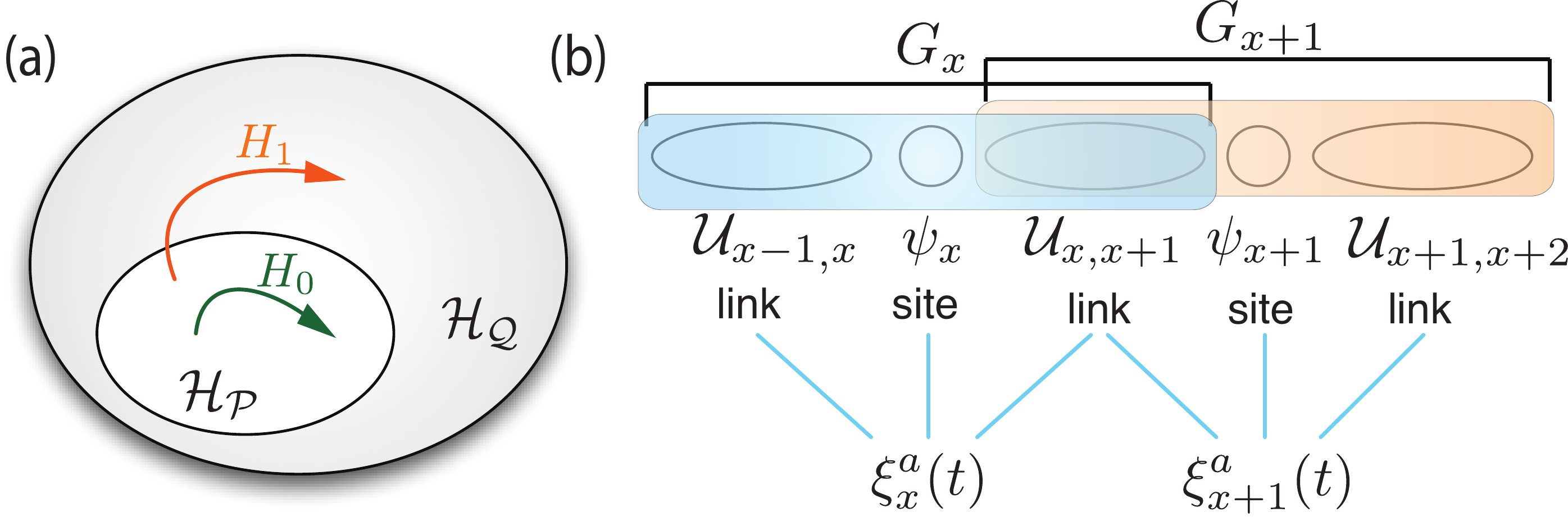}}
\caption{Dissipative protection of gauge invariance via the quantum Zeno effect. (a) The dynamics $H_0$ spans the gauge invariant subspace $\mathcal H_P$, defined by $G_{x}^{a}|\psi\rangle=0$, but gauge variant perturbations $H_1$ may drive the system into the gauge variant subspace $\mathcal H_Q$ (where $G_{x}^{a}|\psi\rangle\neq 0$). (b) The noise sources coupled to each single building block of the LGT constrain the dynamics within $\mathcal H_P$. The multi-site structure of the generators implies that the noise has to be correlated quasi-locally in space. Figure taken from Ref.~\cite{stannigel2014constrained}.
} 
\label{fig:DissLGT}
\end{center}
\end{figure}

In Ref.~\cite{stannigel2014constrained}, an alternative procedure was proposed based on quantum Zeno dynamics~\cite{Facchi:2008aa}. The basic idea is to consider a set of classical noise sources $\xi_x^\alpha(t)$ coupled to the generators of the gauge symmetry, described by the effective microscopic Hamiltonian:
\begin{equation}
H_{\mathrm{micro}} =  H_0+H_1+\sqrt{2\kappa}\sum_{x,\alpha}\xi_x^\alpha(t)G^\alpha_x
\end{equation}
In the limit of independent, white noise sources, the system dynamics is described by a master equation, whose corresponding effective Hamiltonian reads:
\begin{equation}
H_{\mathrm{eff}} =  H_0+H_1-i\kappa^2\sum_{x,\alpha}(G^\alpha_x)^2.
\end{equation}
In the large noise limit, where $\kappa$ is much larger than all microscopic scales involved in $H_1$, the effective dynamics is constrained to the gauge invariant subspace $\mathcal{H}_P$. In this limit, any coupling term between the gauge invariant subspace to the gauge variant one $\mathcal{H}_U$ is suppressed by the noise terms (see Fig.~\ref{fig:DissLGT}a). Depending on the nature of $H_1$, one can use a limited number of noise sources, with the condition that neighbouring blocks are subject to distinct noise terms, as illustrated in  Fig.~\ref{fig:DissLGT}b. From a theoretical viewpoint, this scheme can be seen as a classical analogue of the quantum Zeno effect~\cite{Facchi:2008aa}, which has also been discussed in the context of quantum many-body systems as a source of local interaction~\cite{Syassen:2008aa,Roncaglia:2010aa,Kantian:2009aa}.

The main feature of this scheme is that, different from energy penalty schemes, here the constraints are induced by just coupling to simple operator terms (there is no microscopic term $(G^\alpha_x)^2$ that shall be engineered), that can be easily controlled by means of external fields. This drastically simplifies the conditions required to achieve gauge invariant dynamics. This comes at the price of needing to realise the desired dynamics $H_0$ without the help of perturbation theory: two applications in the context of both Abelian and non-Abelian theories are discussed in Ref.~\cite{stannigel2014constrained} in the context of cold atoms in optical lattices.

\subsubsection{SO(3) "Nuclear Physics" with ultra-cold Gases\cite{rico2018so}}

An ab initio calculation of nuclear physics from QCD, the fundamental $SU(3)$ gauge theory of the strong interaction, remains an outstanding challenge. In this proposal, the emergence of key elements of nuclear physics using an $SO(3)$ lattice gauge theory as a toy model for QCD is discussed. This model is accessible to state-of-the-art quantum simulation experiments with ultra-cold atoms in an optical lattice. The phase diagram of the model is investigated, and showed that it shares some fundamental properties with QCD, most prominently confinement, the spontaneous breaking of chiral symmetry, and its restoration at finite baryon density, as well as the existence of few-body bound states, which are characteristic features of nuclear physics. The most critical step in implementing a gauge theory on a quantum simulator is to make sure that the gauge symmetry remains intact during the simulation. It is shown how the lattice gauge model, which has a non-Abelian gauge symmetry, can be realised in a quantum simulator platform by encoding the operators directly in the gauge invariant subspace, thus guaranteeing exact gauge invariance. This encoding strategy is generally applicable to the whole class of quantum link models, which are extensions of Wilson's formulation of lattice gauge theories. Quantum link models permit encoding to local spin Hamiltonians. This not only makes the implementation feasible on different platforms, such as ultra-cold atoms and molecules trapped in optical lattices, but also establishes a novel connection between non-Abelian lattice gauge theories including matter fields and quantum magnetism.

\begin{figure}[htb]
\begin{center}
\resizebox{0.99\columnwidth}{!}{\includegraphics{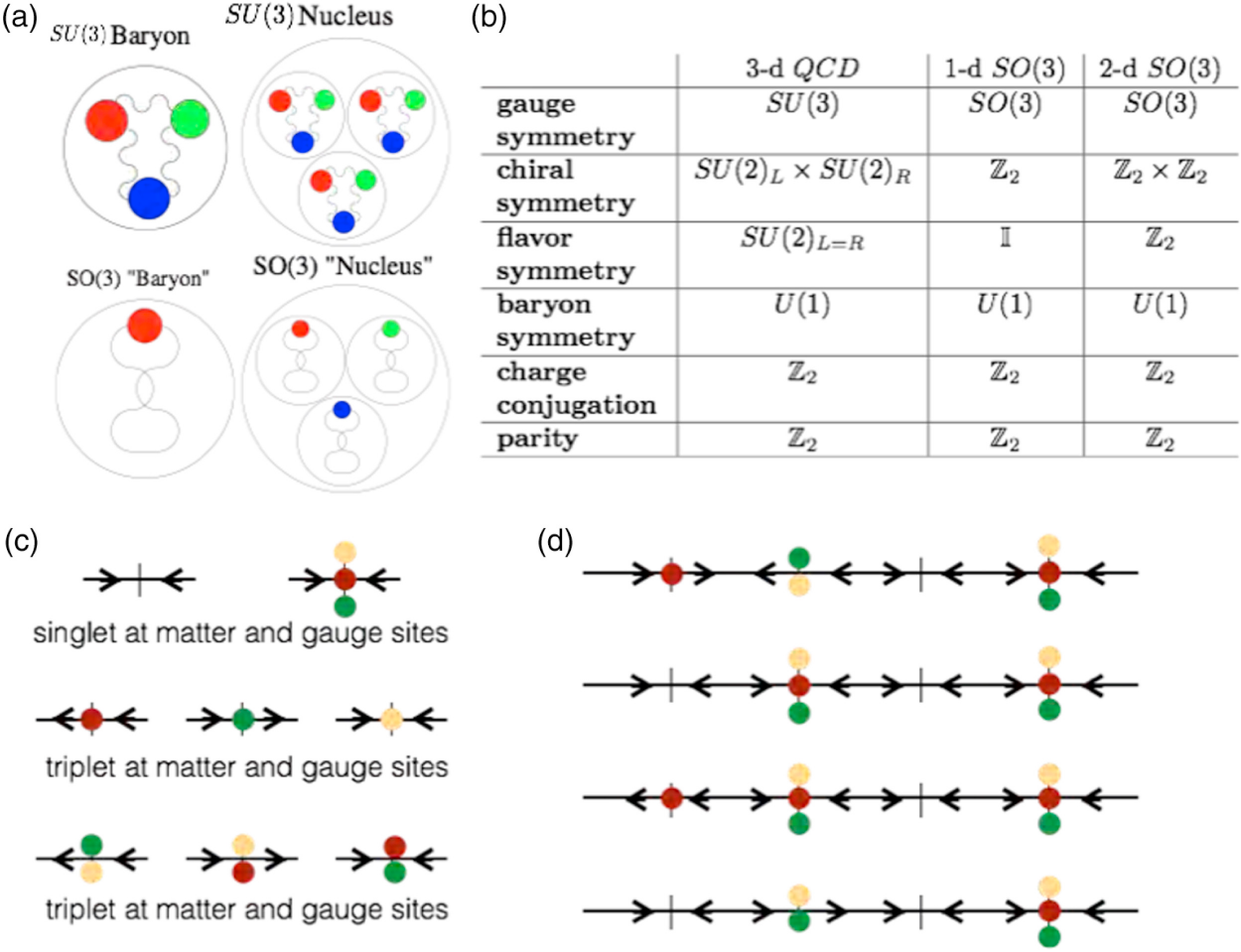}}
\caption{(a) Sketches of the different objects in $SU(3)$ and $SO(3)$ gauge theories. In both cases, three colour matter fields are present. Single baryons have different internal structures: in $SU(3)$ gauge theories, they contain three quarks, while in $SO(3)$ theories they can be formed by a single quark paired with a gluon. (b) Summary table of both local and global symmetries in the $(1+1)$-d $SO(3)$ QLM, compared with its $(2+1)$-d counterpart, and with $(3+1)$-d QCD. The model investigated here has the same baryon number symmetry as QCD, and has a non-trivial discrete chiral symmetry (which is simpler than QCD's continuous chiral symmetry). (c) Local gauge invariant states in the $SO(3)$ gauge model. Gauss' law implies that the physical subspace contains singlet states in the combined matter and gauge degrees of freedom. (d) Cartoon states for some phases of the $SO(3)$ QLM. From top to bottom: generic configuration with no order; chiral symmetry broken vacuum, where the quark population is arranged in a staggered fashion; baryon configuration, where a single baryon is created on top of the vacuum state; anti-baryon configuration, where a single anti-baryon is created on top of the vacuum state. Figure taken from Ref.~\cite{rico2018so}.}
\end{center}
\end{figure}

Concretely, it is shown in \cite{rico2018so} how $(1+1)$-d $SO(3)$ lattice gauge theories can be naturally realised using ultra-cold atoms in optical lattices. The phase diagram of these models features several paradigmatic phenomena, e.g. the presence of stable two-body bound states, phases where chiral symmetry is spontaneously broken, where the broken chiral symmetry is restored at finite baryon density, and emergent conformal invariance.  The dynamics in the gauge invariant sector can be encoded as a spin $S=3/2$ Heisenberg model, i.e., as quantum magnetism, which has a natural realisation with bosonic mixtures in optical lattices, and thus sheds light on the connection between non-Abelian gauge theories and quantum magnetism. This encoding technique has the dramatic advantage of realising gauge invariance exactly, and at the same time bypassing the complex interaction engineering which is required for non-Abelian theories. 

\section{Conclusions}

At the time of writing this review, there is a coordinated effort to study the possible applications of quantum technologies to the study of gauge theories. The global aim is to develop novel methods and techniques, namely classical and quantum hardware and software, that eventually could be applied to study open problems in different fundamental and applied fields of science ranging from materials science and quantum chemistry to astrophysics and that will impact fundamental research and our everyday life. This challenge is a highly collaborative advanced multidisciplinary science and cutting-edge engineering project with the potential to initiate or foster new lines of quantum technologies. Along the way, the scientific community that is growing around this topic, is developing a deeper fundamental and practical understanding of systems and protocols for manipulating and exploiting quantum information; at enhancing the robustness and scalability of quantum information processing; identifying new opportunities and applications fostered through quantum technologies and enhancing interdisciplinarity in crossing traditional boundaries between disciplines in order to enlarge the community involved in tackling these new challenges.

The results of this endeavour serve as benchmarks for the first generation of quantum simulators and will have far reaching consequences, e.g., in the long run this research will enable the study and design of novel materials with topological error correcting capabilities, which will play a central role in the quest for building a scalable quantum computer. In particular, in the review, the most advanced quantum simulation of a lattice gauge theory achieved so far is presented, a digital ion-trap quantum variational optimisation applied to find the ground state of the Schwinger model. Moreover, it is also reviewed how novel tensor network methods are being developed and applied to study such systems in one and two dimensions, and finally, how new directions are being proposed for quantum simulations of more complex theories.

Lattice gauge theories provide both motivation and a framework for pushing forward the interdisciplinary advancement of quantum technologies. While the quantum simulation of classical intractable aspects of QCD (such as its real-time evolution or its phase diagram at high baryon density) remain long-term goals with a potentially large impact on particle physics, a wide class of lattice gauge theories, often with applications in condensed matter physics or quantum information science, suggests itself for theoretical investigation and experimental realisation. There is a lot of interesting physics to be discovered along the way towards developing powerful hard- and software for the fast growing field of quantum simulation and computation technology.

\section{Acknowledgments}

M.C.B. is partly supported by the Deutsche Forschungsgemeinschaft (DFG, German Research Foundation) under Germany's Excellence Strategy - EXC-2111- 390814868. The research at Innsbruck is supported by the ERC Synergy Grant UQUAM, by the European Research Council (ERC) under the European Union Horizon 2020 research and innovation programme under grant agreement No. 741541, the SFB FoQuS (FWF Project No.  F4016-N23), and the Quantum Flagship PASQUANS. Florence acknowledges financial support from ERC Consolidator Grant TOPSIM, INFN project FISH, MIUR project FARE TOPSPACE and MIUR PRIN project 2015C5SEJJ. AC acknowledges support from the UAB Talent Research program and from the Spanish Ministry of Economy and Competitiveness under Contract No.FIS2017-86530-P.  JIC is partially supported by the EU, ERC grant QUENOCOBA 742102. MD is supported by the ERC under Grant No. 758329 (AGEnTh), and has received funding from the European Union Horizon 2020 research and innovation program under Grant agreement No. 817482. M.L. acknowledges support by the Spanish Ministry  MINECO  (National Plan 15 Grant: FISICATEAMO No. FIS2016-79508-P, SEVERO OCHOA No.  SEV-2015-0522, FPI), European Social Fund, Fundacio Cellex, Generalitat de Catalunya (AGAUR Grant No. 2017 SGR 1341 and CERCA/ Program), ERC AdG OSYRIS and NOQIA, and the National Science Centre, Poland-Symfonia Grant No. 2016/ 20/W/ST4/00314. S.M. acknowledges support from PASQUANS, Italian PRIN 2017 and DFG via the Twitter project. E.R. acknowledges financial support from Spanish Government PGC2018-095113-B-I00 (MCIU/AEI/FEDER, UE), Basque Government IT986-16, as well as from QMiCS (820505) and OpenSuperQ (820363) of the EU Flagship on Quantum Technologies, EU FET-Open Grant Quromorphic, and the U.S. Department of Energy, Office of Science, Office of Advanced Scientific Computing Research (ASCR) quantum algorithm teams program, under field work proposal number ERKJ333. LT is supported by the MINECO RYC-2016-20594 fellowship and the MINECO PGC2018-095862-B-C22 grant. The research at Gent was made possible through the support of the ERC grants QUTE (647905), ERQUAF (715861). J.Z. acknowledges support by PLGrid Infrastructure and by National Science Centre (Poland) under project 2017/25/Z/ST2/03029. U.-J. Wiese acknowledges funding from the Schweizerischer National fonds and from the European Research Council under the European Union Seventh Framework Programme (FP7/2007-2013)/ ERC grant agreement 339220. M.W. acknowledges support from STFC consolidated grant ST/P000681/1.     

All the authors acknowledge the participation in the EU-QUANTERA project QTFLAG.

\section{Authors contributions}
All the authors were involved in the preparation of the manuscript. All the authors have read and approved the final manuscript.

\bibliographystyle{ieeetr}
\bibliography{reviewbib}

\end{document}